# A domain specific language for data-centric infographics: technical report

Sergio España [0000-0001-7343-4270], Vijanti Ramautar [0000-0002-3744-0013], Sietse Overbeek [0000-0003-3975-200X], and Tijmen Derikx

Utrecht University, the Netherlands
`s.espana@uu.nl`, `v.d.ramautar@uu.nl`,
`s.j.overbeek@uu.nl`, `tderikx@gmail.com`

## 1 Introduction

This technical report is a companion of the paper:

> España, S., Ramautar, V., Overbeek, S., Derikx, T.:Model-driven production of data-centric infographics: an application to the impact measurement domain. RCIS 2022 (under review). Springer (2019)

That paper investigates the problems related to the conventional process to produce data-centric infographics, and redesigns this process following the model-driven paradigm. We have designed a domain-specific language to model infographics, and implemented an interpreter that generates the infographics automatically. To inform the engineering of the DSL and interpreter, we have analysed 58 infographics reporting on ethical, social and environmental (ESEA), and 10 infpgraphic design tools. We then have tested our proposal by means of modelling a sample of 10 infographics and generating them. We have also conducted a controlled experiment to assess the extent to which the generated infograhics are as visually attractive as the original ones and whether the participants can identify whether the infographic they see is original or generated.

This report complements the above-mentioned paper in the following ways:

- Section 2 presents the conceptual model that resulted from investigating the role of infographics within the impact measuremet domain.
- Section 3 provides details on the infographic analysis. In particular, it includes a list of the 58 infographics, their sources, the infographic component types that we have found in them, along with the definitions of the infographic component types.
- Section 4 provides details on the infographic design tool analysis. In particular, it provides a list of the 10 tools, the feature model that we have created for each of the tools, a table describing each of the generic features that we have identified across tools.
- Section 5 describes the sample of 10 infographics that we have used to test the DSL and the interpreter, and we have also used in a comparative experiment. In particular, it presents each pair of infographics in the sample (i.e. the original and the generated infographic), and the model that specifies the generated infographic using the DSL.
- Section 6 presents the user stories that specify the requirements for the DSL and the interpreter, tracing them to their sources and providing additional details.
- Section 7 presents instruments and data from controlled experiment in which we have compared original infographics and the corresponding ones that we have modelled and generated. In particular, we show an example of the survey used in the experiment, and the tables containing the raw data of the experiment results.

## 2  Conceptual model of infographics within the impact measurement domain

To better understand the role of data-centric infographics in the domain of impact measurement and how they are related to other relevant concepts, we have conducted a multivocal literature review [1]. We decided to include grey literature because we are not only interested in peer-reviewed scientific results, but also in how ESEA practitioners are using infographics.

The selected studies have been coded with Nvivo. Before strting the analysis, we already had some insights on what concepts are relevant withing the ESEA domain; in particular, we had earlier produced an initial version of the openESEA metamodel [2]). We used those concepts to inform our analysis and as a glossary when the terminology varies across papers. Still, we have opted for starting with an empty coding scheme and let it arise incrementally as we read and coded the studies. When a concept that seemed relevant to our purpose appeared in the text of the selected studies, we included a new node in the coding scheme. Then we checked whether such concept was present in the above-mentioned metamodel. If so, the definition and position within the metamodel (its relationships to other concepts) were taken into account. Table 1 presents and overview of results:

- **Concept**. The name of the concept.
- **Working definition**. A definition of the concept in the context of this research.
- **Occurrences**. Number of occurrences across the literature or, in other terms, the number of fragments in the selected studies that discuss this concept

**Table 1.** Concepts resulted from coding the selected papers

| Concept | Working definition | Occ |
|---|---|---|
| ESE account | The collection of data resulting  performing an ethical, social and environmental accounting (ESEA), which is the process of assessing and reporting on the social and environmental effects caused by an organisation's economic actions to particular interest groups within society and to society itself [3]. The account can be considered as measurements of the performance of an organisation on ethical, social and environmental  (ESE) topics. It can be referred also as sustainability report, non-financial report, or social balance. It always covers a given time period (often one fiscal or natural year). | 12 |
| Organisation | A social entity that is goal-directed, that is designed as a deliberately structured and coordinated activity system, and that is linked to the external environment [4]. In our context, it is the organisation to which the account data refers to. | 6 |
| Stakeholder | A person with an interest or concern in something, especially an organisation [5]. Examples of stakeholders are the CEO of a company, a given client, a neighbour living next to a manufacturing premise. | 11 |
| Stakeholder group | A category of stakeholders; organisations typically categorise them into groups, based on their conflicting needs. A stakeholder may be part of one or more stakeholder groups. Examples are consumers, clients, suppliers, shareholders, and watchdog organisations such as NGOs. | 10 |
| Accounting team | The group of people performing the ESEA. It can be an internal team (i.e. composed of members of the organisation, an external team (members from outside the organisation, or a mixed team (members from both groups). | 8 |

| Concept | Working definition | Occ |
|---|---|---|
| **Concept** | **Working definition** | **Occ** |
| ESE Topic | Ethical, social and environmental (ESE) topics represent phenomena that organisations or individuals are (or should be) concerned about, related to organisational ethics (a.k.a. governance topics), societal issues or environmental concerns | 8 |
| ESEA Method | A prescription of how an ethical, social and environmental account should be produced; among other issues, it can provide guidance on what topics should be assessed (or how to select those topics), what stakeholder groups to engage, what questions to include in the surveys to collect data for direct indicators, or how to report on the results. | 13 |
| Indicator | An indicator is a measure [6, 7] (also known as measurable element [8], or metric [9]) -i.e. a defined measurement approach and its measurement scale- that is proposed by an ESEA method as a way to assess organisational performance on ESE topics. | 6 |
| Direct indicator | An indicator that can be directly measured; for instance, by collecting its value from an enterprise information system or by asking it to a stakeholder via a survey. | 2 |
| Indirect indicator | An indicator whose value needs to be calculated from other indicators; for instance, by means of a formula that operates on other direct or indirect indicators. | 2 |
| Indicator value | The actual value of an indicator (also known as measurement result), for a given ESEA of a given company at a given point in time. | 6 |
| Network | An organisation that develops initiatives, principles or standards related to corporate social and environmental performance [10]. Organisations willing to join a network may need to meet certain performance, legal, transparency or other requirements. Examples are the B Corporations, the Economy for the Common Good, and the Spanish Network of Networks of Alternative and Solidarity Economy (REAS). | 4 |
| Document | Communication authored to report on an ESE account; that is, on the values of the indicators that report on the performance of the organisation. | 0[a] |
| Report | A document whereby an organisation communicate the results of an account to one specific stakeholder group of to the general public. Depending on who the accounting team is and who is the intended audience of the report, different types of reports can be created [3]. | 5 |
| Infographic | An infographic (portmanteau for information graphic) blends data with graphic design, helping individuals and organisations concisely communicate complex information to an audience, in a manner that can be quickly consumed and easily understood [11]. | 0[b] |
| Tool | The (software) technology that suports the application of the ESEA method, or the creation of the reporting documents. Examples of tools are a spreadsheet, a full-fledged ESEA web application, a text processor, and a graphics design application. | 2 |
| **Total** | | 95 |

a. Document is a generalisation that facilitates us the creation of the conceptual model.

b. Infographics are rarely mentioned in ESEA research. We did find them covered by many other papers in a separate literature review on infographics.

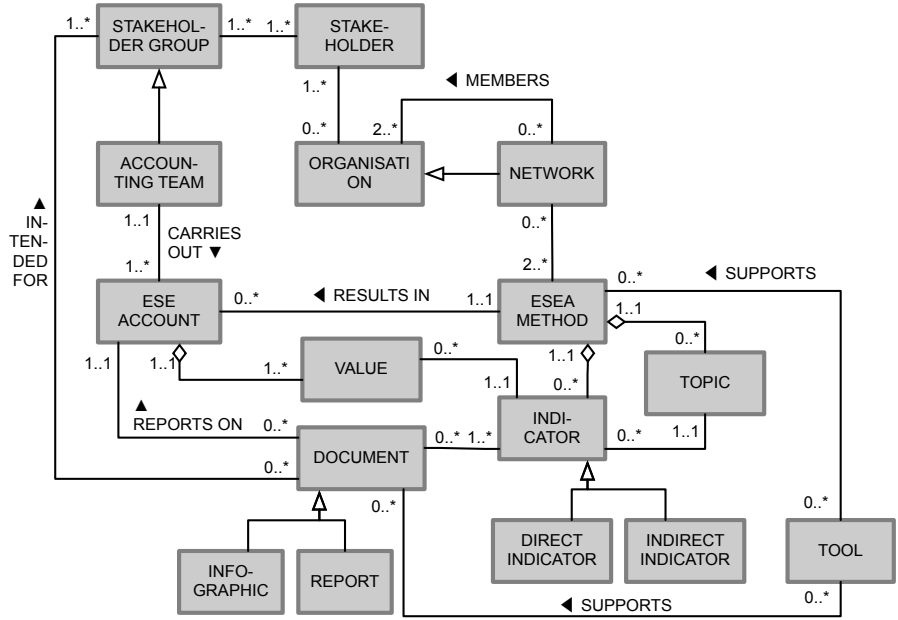

**Fig. 1.** Conceptual model of the use of infographics in the context of impact measurement; more specifically, as a means to report on the results of an ethical, social and environmental accounting process.

## 3  Analysis of infographics

We include a list of the 58 infographics and their sources (Section 3.1), we present a copy of each infographic (Section 3.2), and we report on the result of the infographic analysis and what components we found in each (Section 3.3).

### 3.1  Sources of the infographics

Table 2 enumerates the infographics that we have analysed to identify infographic component types. The columns are:

- **#**. A number identifying the infographic
- **Org**. The organisation that publishes the infographic. Therefore, the infographic presents results of the ESEA of that organisation, assessing their performance on a series of topics.
- **Year**. The year to which the data presented in the infographic refers to. This is thus the year of the ESEA.
- **URL of infographic**. The cell contains a link to the infographic, as published by the organisation. To prevent that the infographics are lost when the organisations update their websites, we also present them in section
- **URL of report**. Some infographics are part of the sustainability reports of the organisations. When that is the case, then the cell contains a link to the report.
- **S**. Whether the infographic has is part of the sample we have used to test our proposal. If the infographic is part of the sample, then the cell contains the identifier of the infographic in the sample, in the form of $I_i^O$, where *I* stands for infographic, $^O$ stands for original, and $_i$ is a sequential number. If the infographic is not part of the sample, then the cell is empty.

**Table 2.** List of infographics analysed to identify infographic component types.

| # | Org | Year | URL of infographic | URL of report | S |
|---|---|---|---|---|---|
| 1 | Nestle | 2013 | https://www.nestle.com.au/asset-library/PublishingImages/Media/News/environmental-sustainability-infographic-page.jpg | http://www.nestle.com/asset-library/documents/library/documents/corporate_social_responsibility/nestle-csv-full-report-2013-en.pdf | |
| 2 | Smithfield | 2016 | https://www.3blmedia.com/sites/www.3blmedia.com/files/styles/fmr_page_photos_blog/public/images/SFI17-022514-02_IR_Envirofograph_m_1.jpg?itok=JVzxDm5- | https://www.smithfieldfoods.com/integrated-report/2016/introduction | |
| 3 | Schroders | 2018 (Q3) | https://www.schroders.com/en/sysglobalassets/digital/insights/2018/infographics/sustainable-report-infographic/english-original/sustainable-investment-report-q3-2018-cs00637.jpg | https://www.schroders.com/en/sysglobalassets/digital/insights/2018/pdf/sustainable-investment-report/english-original/sustainable-investment-report-q3-2018.pdf | |
| 4 | Autodesk | 2015 | https://image.slidesharecdn.com/adskfy2015sustainabilityreportinfographic-150603195336-lva1-app6892/95/autodesk-sustainability-infographic-progress-report-fy2015-1-638.jpg?cb=1469490986 | http://damassets.autodesk.net/content/dam/autodesk/www/sustainability/docs/pdf/sustainability_report_2015_FINAL.pdf | $I_2^O$ |
| 5 | Wholesum Harvest | 2016 | http://www.sustainablefoodtrade.org/wp-content/uploads/2018/02/Wholesum-Harvest-INFOGRAPHIC-2016-Sustainability-Report.jpg | http://www.sustainablefoodtrade.org/wp-content/uploads/2017/09/Wholesum-Harvest-2016-Sustainability-Report.pdf | |
| 6 | Autodesk | 2014 | https://image.slidesharecdn.com/fy2014sustainabilityreportinfographic-140508175520-phpapp02/95/autodesk-sustainability-infographic-progress-report-fy2014-1-638.jpg?cb=1469490980 | https://damassets.autodesk.net/content/dam/autodesk/www/sustainability/docs/pdf/Sustainability_Report_2014.pdf | |
| 7 | Schroders | 2017 | https://www.schroders.com/en/sysglobalassets/digital/insights/2018/infographics/sustainable-report-infographic/english-original/sustainable-investment-report-annual-2017-sch43134.jpg | https://www.schroders.com/en/sysglobalassets/digital/insights/2018/pdf/sustainable-investment-report/english-original/sustainable-investment-report-annual-2017.pdf | |
| 8 | Schroders | 2019 (Q1) | https://www.schroders.com/en/sysglobalassets/digital/insights/2019/infographics/sustainability/sustainable-investment-report-q1-2019-cs1318.jpg | https://www.schroders.com/en/sysglobalassets/digital/insights/2019/pdfs/sustainability/sustainable-investment-report/sustainable-investment-report-q1-2019.pdf | |
| 9 | Alcoa | 2016 | https://www.mcshanemetalproducts.com/wp-content/uploads/2018/01/sustainability-infographic.jpg | https://www.alcoa.com/sustainability/en/pdf/archive/corporate/2016-Sustainability-Report.pdf | $I_1^O$ |
| 10 | Reas | 2017 | https://www.economiasolidaria.org/recursos/la-economia-solidaria-en-euskadi-proceso-de-auditoria-social-2017/ | NA | |
| 11 | Coca-Cola | 2016 | Local file (Page 6 of report) | https://www.coca-colacompany.com/content/dam/journey/us/en/private/fileassets/pdf/2017/2016-sustainability-update/2016-Sustainability-Report-The-Coca-Cola-Company.pdf | |
| 12 | Vanderbilt University | 2016 | https://vanderbilthustler.com/wp-content/uploads/2018/04/Summary-Infographic.png | https://www.vanderbilt.edu/sustainability/annual-sustainability-report-2016/ | |
| 13 | Vanderbilt University | 2017-2018 | https://cdn.vanderbilt.edu/vu-wp0/wp-content/uploads/sites/69/2018/10/06141552/20181105151810-hamiltcl-25921-1.png | https://www.vanderbilt.edu/sustainability/annual-sustainability-report-2017/ | $I_{10}^O$ |
| 14 | GSI: Global Salmon Iniative | 2016 | https://globalsalmoninitiative.org/files/documents/GSI-Sustainability-Report-2016-Infographic.pdf | https://globalsalmoninitiative.org/en/sustainability-report/ | |
| 15 | GSI: Global Salmon Iniative | 2017 | https://www.kingsalmon.co.nz/kingsalmon/wp-content/uploads/2018/04/Sustainability-Infographic-Image2.jpg | https://globalsalmoninitiative.org/en/sustainability-report/ | $I_6^O$ |

| #  | Org | Year | URL of infographic | URL of report | S |
|----|-----|------|--------------------|---------------|---|
| 16 | GSI: Global Salmon Iniative | 2018 | https://globalsalmoninitiative.org/files/images/Capture-ya.PNG | https://globalsalmoninitiative.org/en/sustainability-report/ | |
| 17 | Wellstone | 2014 | https://i.pinimg.com/originals/f3/eb/7e/f3eb7eb00343258decd177b550bf144e.png | https://view.publitas.com/wellstone-action/wellstone-action-2014-ar/page/1 | |
| 18 | ISTE | 2012-2013 | https://cdn.iste.org/www-root/Libraries/Images/Board%20of%20Directors/316-13-new-branding-homepag.jpg | https://www.iste.org/about/iste-story/annual-report | |
| 19 | ISTE | 2015 | http://hypsypops.com/wp-content/uploads/2017/05/iste-annual-report_2015-610x945.png | https://www.iste.org/about/iste-story/annual-report | |
| 20 | ISTE | 2016 | https://cdn.iste.org/www-root/2018-12/iste-annual-report_2016_v6_website.png | https://www.iste.org/about/iste-story/annual-report | |
| 21 | MHPM | 2010-2011 | https://images.squarespace-cdn.com/content/5bfc8dbab40b9d7dd9054f41/1550631538738-STSHYYNDP9AQD6BJILUB/MHPM%2BCSR%2BReport%2BPoster%2BFINAL%2B600px.jpg?content-type=image%2Fjpeg | N/A | |
| 22 | Trinseo | 2017 | https://mms.businesswire.com/media/20180726005012/en/669555/5/Trinseo_Sust_CSR_2017_Highlights_Slide_7.5.18.jpg | http://www.trinseo.com/sustainability | $I_9^O$ |
| 23 | Realtor | 2019 | https://www.nar.realtor/sites/default/files/2019-realtors-and-sustainability-infographic-04-18-2019-2400w-5250h.png | https://www.nar.realtor/sites/default/files/documents/2019-Sustainability-Report-04-19-2019.pdf | |
| 24 | Diageo | 2015 | https://www.diageo.com/PR1346/aws/media/1967/diageo-infographic-verdana.jpg?anchor=center&mode=crop&width=1200&heightratio=0.5625&format=jpg&quality=80&rnd=131746550340000000 | https://www.diageo.com/PR1346/aws/media/1968/2015_annual_report_-_sd_content__13_aug_2015.pdf | |
| 25 | Cook county | 2017 | http://blog.cookcountyil.gov/sustainability/wp-content/uploads/2018/01/2017Infographic.png | http://blog.cookcountyil.gov/sustainability/wp-content/uploads/2018/01/Sustainability-Report-2017_final.pdf | $I_3^O$ |
| 26 | Cook county | 2018 | http://blog.cookcountyil.gov/sustainability/wp-content/uploads/2019/01/Sustainability-Report-2018-FINAL-Infographic-768x994.png | http://blog.cookcountyil.gov/sustainability/wp-content/uploads/2019/01/Sustainability-Report-2018-FINAL.pdf | |
| 27 | Vestas | 2017-2018 | http://11thhourracing.org/wp-content/uploads/2018/09/vs11-infographicfinal-16x9-small.jpg | http://11thhourracing.org/wp-content/uploads/2018/09/vs11-sustainability-report-final.pdf | |
| 28 | ECG | 2013 | https://balance.ecogood.org/matrix-4-1-en/ecg-matrix-en.pdf | N/A | |
| 29 | Balfour Beatty | 2017 | https://www.balfourbeatty.com/media/317925/environmental-limits-infographic-2017.pdf | https://www.balfourbeatty.com/how-we-work/sustainability/sustainability-report/ | |
| 30 | GM | 2017 | https://www.smartenergydecisions.com/upload/research_+_reports/gm-operational-commitments.pdf | https://www.gmsustainability.com/ | |
| 31 | Target | 2015 | https://corporate.target.com/_media/TargetCorp/csr/images/ABV-CSR-infographic.jpg | https://corporate.target.com/_media/TargetCorp/annualreports/2015/pdfs/Target-2015-Annual-Report.pdf | |
| 32 | Lego | 2018 | https://www.lego.com/r/www/r/aboutus/-/media/aboutus/media-assets-library/annual-reports/annual-results-2018-assets/lego-annual-results-2018-infographic.pdf?l.r=-2090499682 | https://www.lego.com/r/www/r/aboutus/-/media/aboutus/media-assets-library/annual-reports/annual-results-2018.pdf?la=en-US&l.r=-1161991047 | |
| 33 | Electrolux | 2016 | https://mb.cision.com/Public/1853/2218713/8a1967c8a0960351_800x800ar.jpg | https://www.electroluxgroup.com/en/wp-content/uploads/sites/2/2017/02/electrolux-annual-report-2016.pdf | |

| # | Org | Year | URL of infographic | URL of report | S |
|---|---|---|---|---|---|
| 34 | Dow | 2017 | https://corporate.dow.com/-/media/dow/corporate/dow-corporate/sustainability/reporting/sustainability-goal-highlights.ashx?h=600&w=680&la=en-US&hash=B6E8A13AD7064A6344F7FB0680CE195633DE9B4E | https://corporate.dow.com/-/media/dow/corporate/dow-corporate/sustainability/2025-goals/pdf/dow-2017-sustainability-report.ashx | |
| 35 | Dow | 2016 | https://corporate.dow.com/-/media/dow/business-units/dow-us/pdf/science-and-sustainability/dow_2016_sustainability_highlights.ashx | https://corporate.dow.com/-/media/dow/business-units/dow-us/pdf/science-and-sustainability/dow_2016_sustainability_reportold.ashx | |
| 36 | Dow | 2013 | https://corporate.dow.com/-/media/dow/business-units/dow-us/pdf/science-and-sustainability/2013_sustainability_highlights.ashx | https://corporate.dow.com/-/media/dow/business-units/dow-us/pdf/science-and-sustainability/2013-sustainability-report.ashx | |
| 37 | Sappi | 2017 | https://sappireports.co.za/reports/sappi-gsr-2017/sites/sdr_2017/files/inline-images/7475%20SappieGR_Planet_Glance_Infographic_1366px.png | https://cdn-s3.sappi.com/s3fs-public/2017-Sappi-Annual-Integrated-Report.pdf | |
| 38 | Kroger | 2017 | http://sustainability.kroger.com/images/infographic-zhzw-2017.png | http://sustainability.kroger.com/Kroger_CSR2018.pdf | |
| 39 | Supply Chain Sustainability School | 2017-2018 | https://www.supplychainschool.co.uk/images/Impact%20Survey%20Infographic.png | https://www.supplychainschool.co.uk/documents/impact_report_2017_2018.pdf | |
| 40 | MSC | 2012-2013 | https://image.slidesharecdn.com/infographicseafoodmarketshighlights-2012-13-140513143035-phpapp02/95/seafood-market-highlights-20122013-msc-infographic-1-638.jpg?cb=1399991734 | N/A | |
| 41 | European Palm Oil Alliance | 2017 | https://www.palmoilandfood.eu/sites/default/files/Infographic%20Impact%20Report%202017.pdf | https://issuu.com/epoa/docs/impact_report_2017_2_?e=25838315/60364407 | |
| 42 | Deltec Homes | 2016 | https://x7e6w2g4.stackpathcdn.com/wp-content/uploads/2016/04/B-Corp-Infographic-website.png | N/A | |
| 43 | Zain | 2012 | https://image.slidesharecdn.com/infographics-131107062501-phpapp01/95/zain-sustainability-infographic-2011-1-638.jpg?cb=1383805629 | https://d364xagvl9owmk.cloudfront.net/media-10-4-18/media/filer_public/52/de/52def886-1e6b-4d9a-a47a-5bbceb0d7bd3/zain-sustainability-report-2012_4.pdf | |
| 44 | Crocs | 2014 | http://textilesupdate.com/wp-content/uploads/2015/03/Infographic-Image.png | N/A | $I_4^O$ |
| 45 | Crocs | 2013 | https://cdn.multichannelmerchant.com/wp-content/uploads/2014/04/crocs-sustainability-report.jpg?_ga=2.23943737.285852382.1569310452-415068144.1569310451 | N/A | |
| 46 | Quinn & Partners | 2019 | https://www.quinnandpartners.com/wp-content/uploads/2019/09/2019-07-09.QP-Impact-Poster-FY-2019.pdf | N/A | |
| 47 | Comerica | 2018 | https://www.comerica.com/content/dam/comerica/en/documents/resources/about/sustainability/Corporate-Responsibility-at-Comerica-Infographic.pdf?_ga=2.155255295.514823505.1569310748-413592400.1569310748 | https://www.comerica.com/content/dam/comerica/en/documents/reports/corporate-responsibility/comerica-corporateresponsibility-report.pdf | |
| 48 | AEG | 2017 | https://www.aegworldwide.com/sites/default/files/styles/extra_large/public/press-release/2018-02/Report%20Highlights.jpg?itok=5F0uAHIl | N/A | |
| 49 | The Home Depot | 2016 | https://corporate.homedepot.com/sites/default/files/image_gallery/THD_Sustainability_Executive-Summary_FINAL.jpg | https://corporate.homedepot.com/sites/default/files/image_gallery/PDFs/THD_0040_2016%20Responsibility%20Report_Online.pdf | $I_8^O$ |
| 50 | GM | 2010 | https://3blaws.s3.amazonaws.com/images/Corporate%20Report%20Infographic%20-%20FINAL.jpg | N/A | |

| # | Org | Year | URL of infographic | URL of report | S |
|---|---|---|---|---|---|
| 51 | Vandemoortele | 2018 | https://vandemoortele.com/sites/default/files/inline-images/infographic.JPG | https://vandemoortele.com/sites/default/files/2019-03/Sustainability_Report_2018_DIGITAL.pdf | |
| 52 | Mastercard | 2017 | https://newsroom.mastercard.com/wp-content/uploads/flickr-photos/15349501839230331664448_f8a723ae57_b.jpg | https://www.mastercard.us/content/dam/mccom/global/aboutus/Sustainability/mastercard-sustainability-report-2017.pdf | |
| 53 | Lenovo | 2018 | https://news.lenovo.com/wp-content/uploads/2018/09/Lenovo_SustainabilityReport_Infographic.png | https://www.lenovo.com/us/en/social_responsibility/2017.18-lenovo-sustainability-report.pdf | $I_7^O$ |
| 54 | Crossrail | 2016 | http://74f85f59f39b887b696f-ab656259048fb93837ecc0ecbcf0c557.r23.cf3.rackcdn.com//assets/library/image/c/original/crossrail-sustainability-highlights-2016.jpg | http://74f85f59f39b887b696f-ab656259048fb93837ecc0ecbcf0c557.r23.cf3.rackcdn.com/assets/library/document/e/original/ea753_sustrpt2016onlinefinal1.pdf | |
| 55 | PPG | 2018 | https://mms.businesswire.com/media/20190422005156/en/717120/5/2018+Sustainability+Infographic_vFINAL.jpg | http://www.sustainability.ppg.com/ | |
| 56 | ABN AMRO | 2013 | https://www.abnamro.com/en/images/Images/040_Social_Newsroom/Afwijkende_maten/Infographic_sustainability_report_2013.jpg | https://www.abnamro.com/en/images/010_About_ABN_AMRO/030_In_society/010_Sustainability/Links_en_documenten/Documenten/Rapportage_-_Sustainability_Report_2013_(EN).pdf | |
| 57 | First Horizon | 2018 | https://ml.globenewswire.com/Resource/Download/fb96cc1a-446f-46bf-9b5b-665864ead96d?size=0 | https://www.firsttennessee.com/About/Corporate-Social-Responsibility | $I_5^O$ |
| 58 | Novelis | 2015 | https://2gjjon1sdeu33dnmvp1qwsdx-wpengine.netdna-ssl.com/wp-content/uploads/2015/12/Sustainbility-Report-Infographic-Dec-2015-v7-alternate-01-751x1024.jpg | http://cdn.novelis.com/wp-content/uploads/2015/12/Novelis-2015-full-report.pdf | |

**3.2 Copies of the infographics**

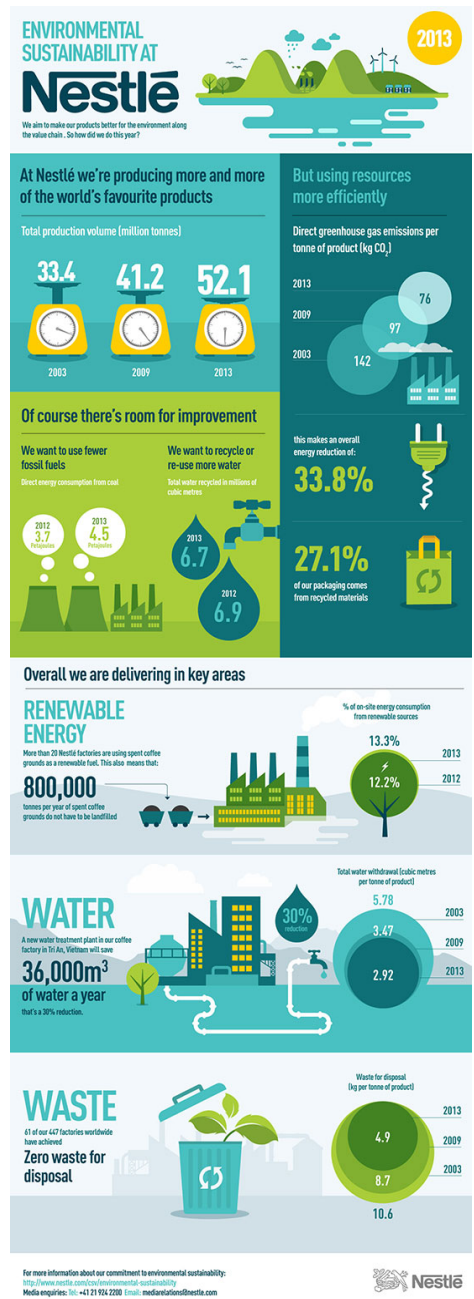

**Fig. 2.** Infographic by Nestlé (2013)

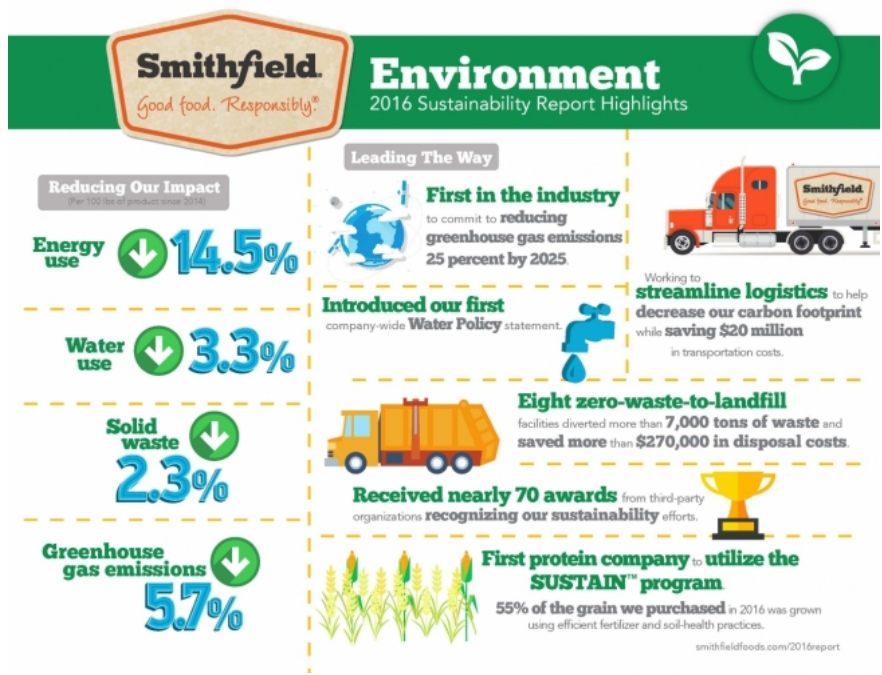

**Fig. 3.** Infographic by Smithfield (2016)

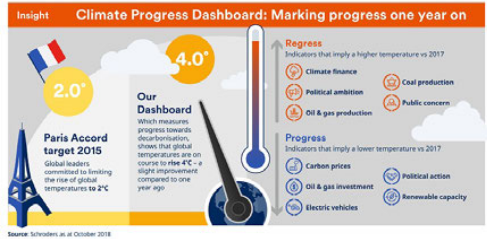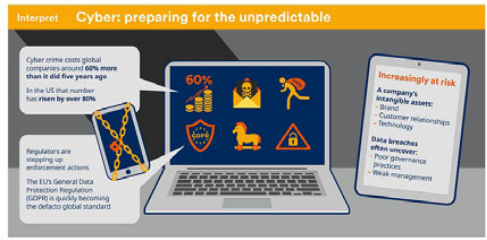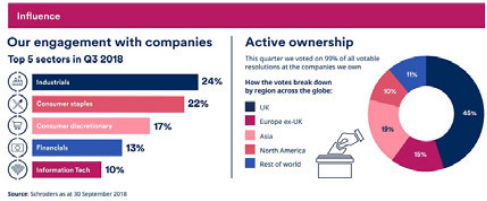

**Fig. 4.** Infographic by Schroders (2018, Q3)

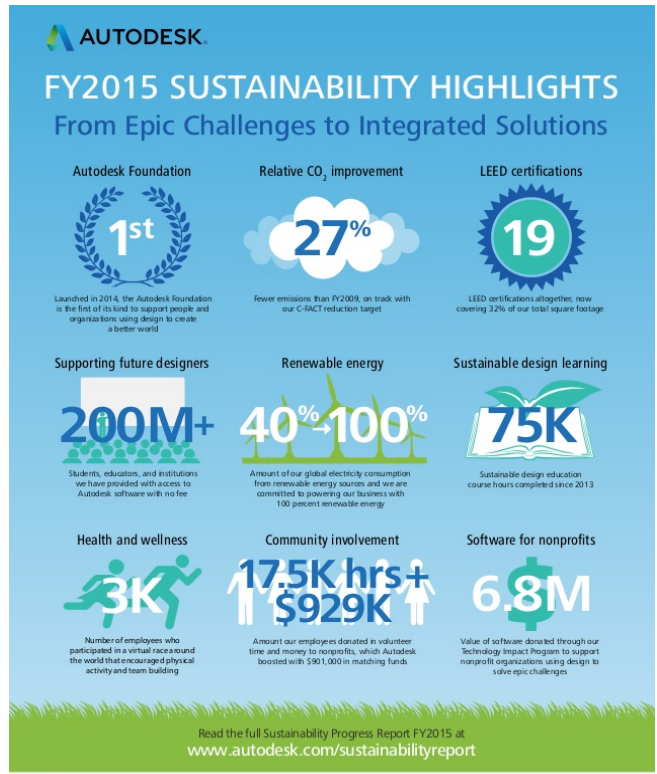

**Fig. 5.** Infographic by Autodesk (2015)

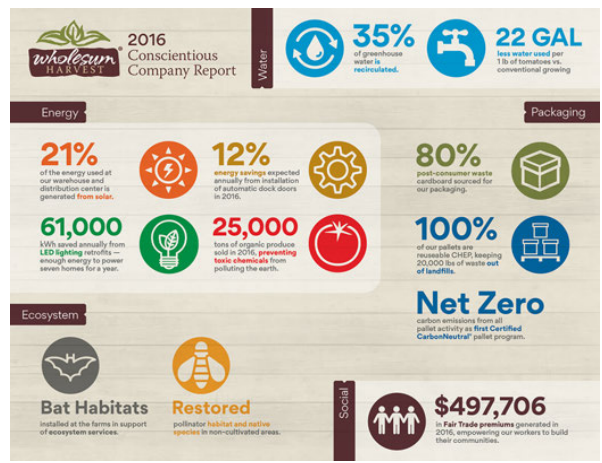

**Fig. 6.** Infographic by Wholesum Harvest (2016)

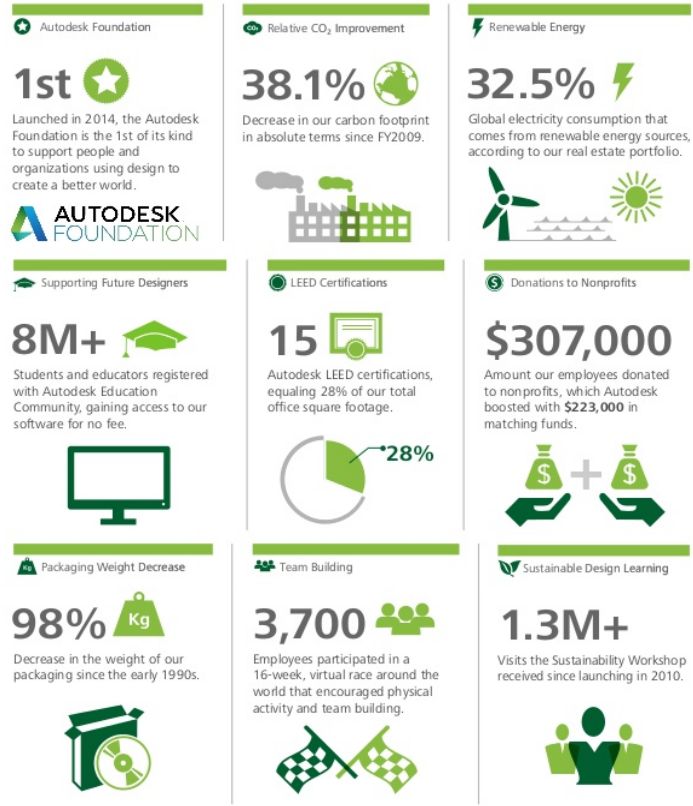

**Fig. 7.** Infographic by Autodesk (2014)

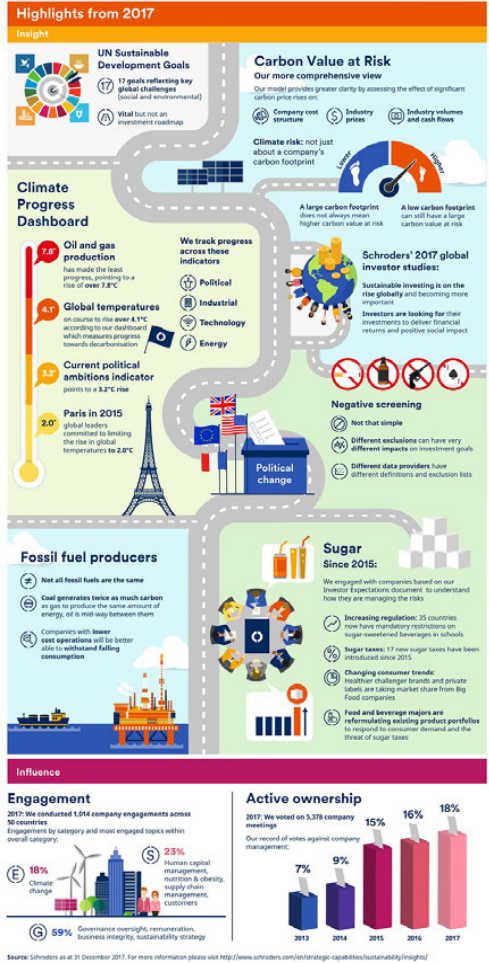

**Fig. 8.** Infographic by Schroders (2017)

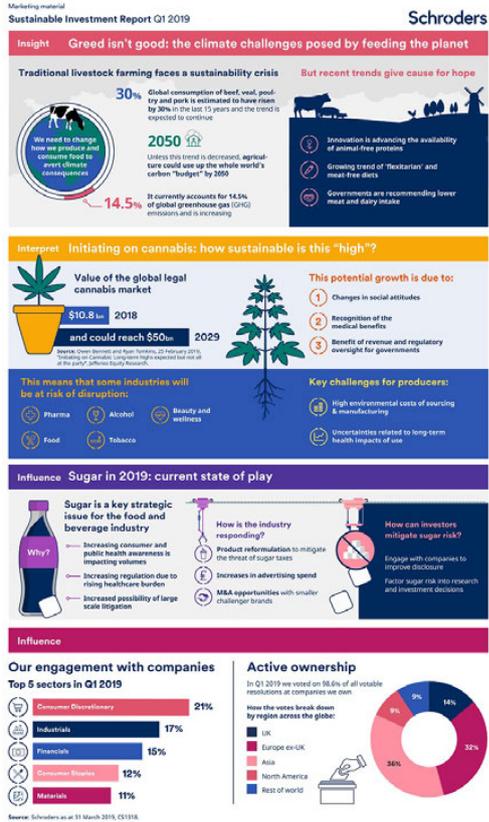

**Fig. 9.** Infographic by Schroders (2019, Q1)

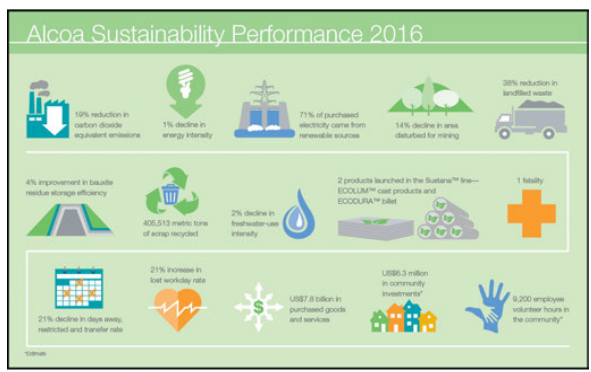

**Fig. 10.** Infographic by Alcoa (2016)

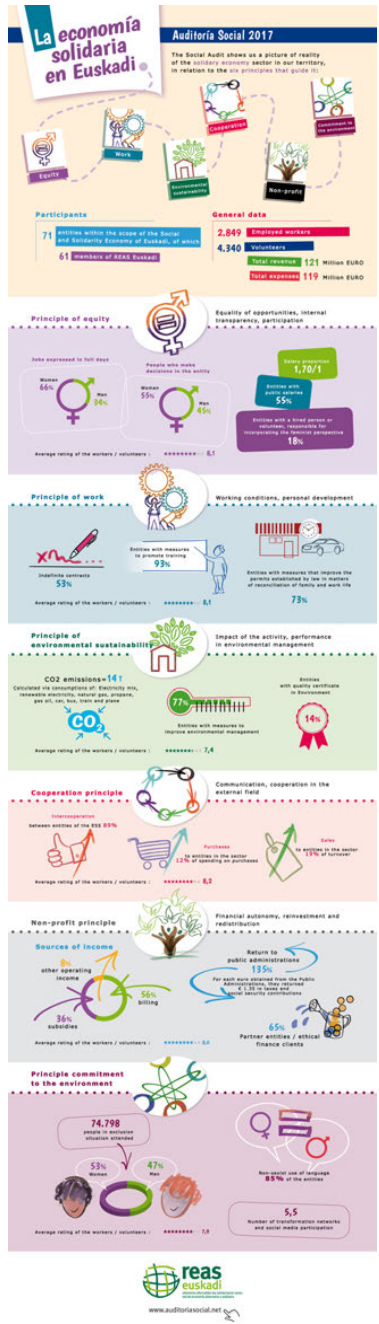

**Fig. 11.** Infographic by Reas (2017)

## 2016 SUSTAINABILITY HIGHLIGHTS

**AGRICULTURE**

**100%**
of our coffee and tea and over 50% of our lemons and beet sugar from more sustainable sources than in previous years.

**HUMAN AND WORKPLACE RIGHTS**

**89%**
of bottling partners and 90% of direct suppliers achieved compliance with our Supplier Guiding Principles.

**PACKAGING AND RECYCLING**

**60%**
of bottles and cans equivalent to what we introduced into the marketplace were refilled or recovered and recycled with our support.

**CLIMATE PROTECTION**

**14%**
estimated reduction of the $CO_2$ embedded in the "drink in your hand".*

**GIVING BACK**

**$106M+**
donated across more than 200 countries and territories.

**WATER STEWARDSHIP**

**221B**
liters of water replenished through community and watershed projects across the globe.**

**WOMEN'S ECONOMIC EMPOWERMENT**

**1.7M**
women enabled through our 5by20 initiative.

*The 14% calculation of progress toward our "drink in your hand" goal has been internally vetted using accepted and relevant scientific and technical methodologies, but those methodologies are evolving. We are working to simplify our data collection and measuring systems, and plan to have our data externally verified by an independent third party in future years. At that time, we will also revisit our estimate to ensure its accuracy and make any updates or necessary corrections, if any, to our public reporting.
**As estimated working with our third party partners.

Learn more: http://www.coca-colacompany.com/stories/2020-sustainability-goals
Track our progress: http://www.coca-colacompany.com/stories/2016-tracking-our-progress
© 2017 The Coca-Cola Company All rights reserved.

THE COCA-COLA COMPANY  6

**Fig. 12.** Infographic by Coca-Cola (2016)

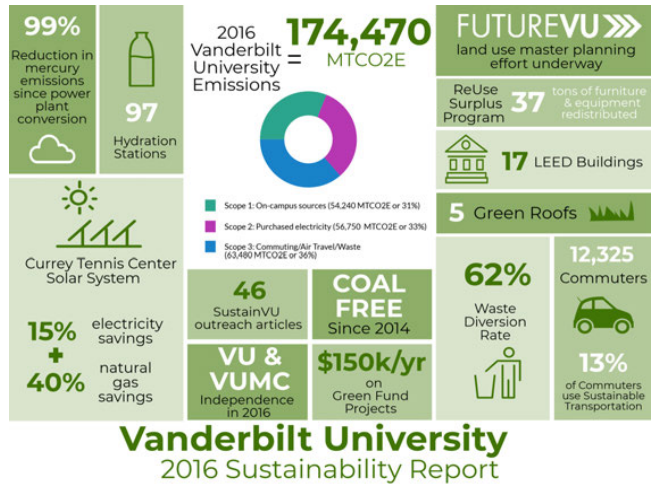

**Fig. 13.** Infographic by Vanderbilt University (2016)

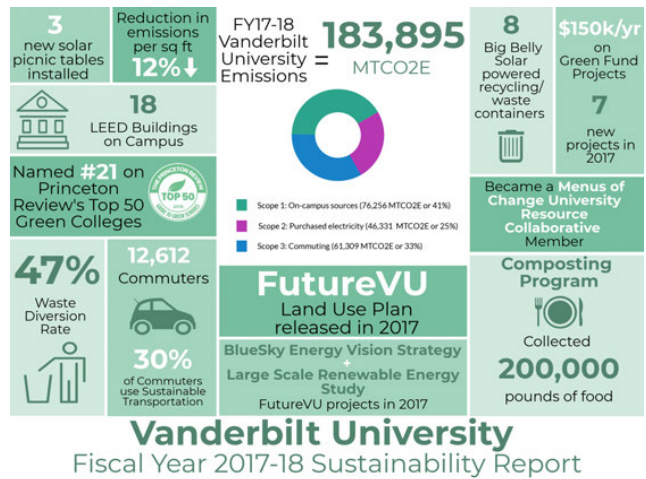

**Fig. 14.** Infographic by Vanderbilt University (2017-2018)

**Fig. 15.** Infographic by GSI: Global Salmon Iniative (2016)

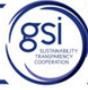

**Fig. 16.** Infographic by GSI: Global Salmon Iniative (2017)

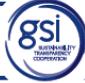

Fig. 17. Infographic by GSI: Global Salmon Iniative (2018)

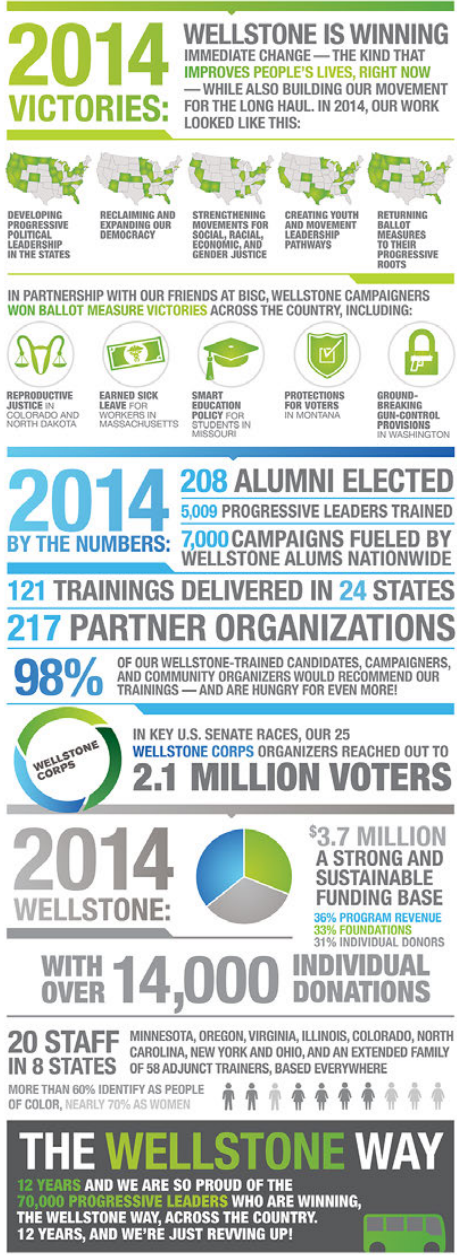

**Fig. 18.** Infographic by Wellstone (2014)

**Fig. 19.** Infographic by ISTE (2012-2013)

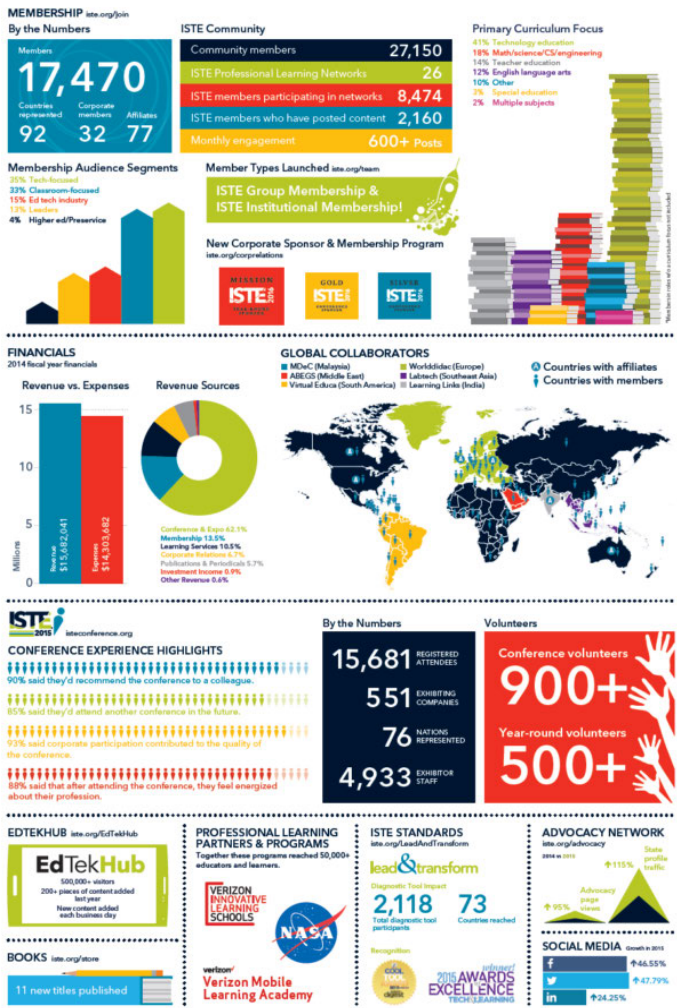

**Fig. 20.** Infographic by ISTE (2015)

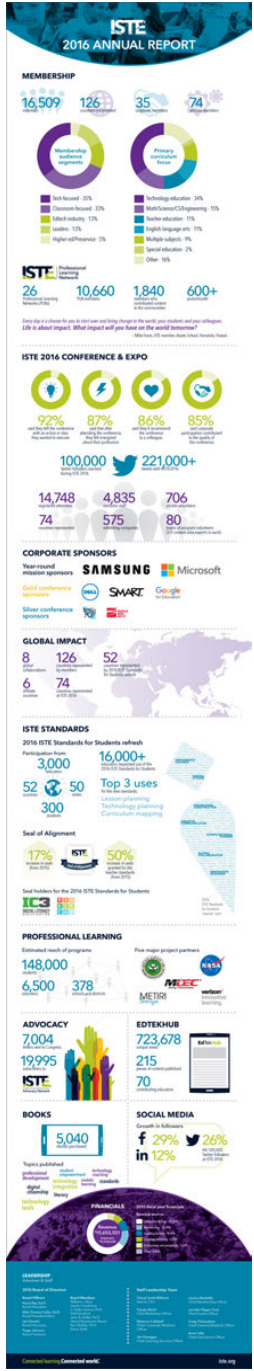

**Fig. 21.** Infographic by ISTE (2016)

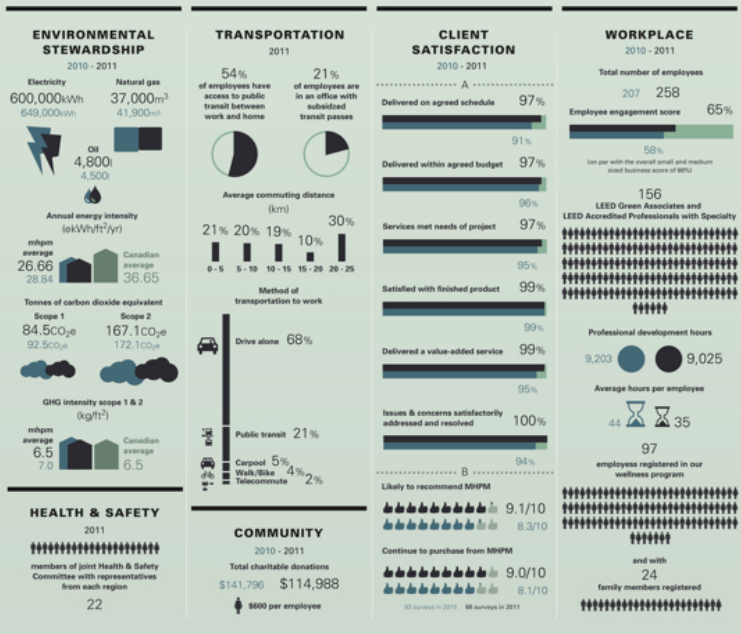

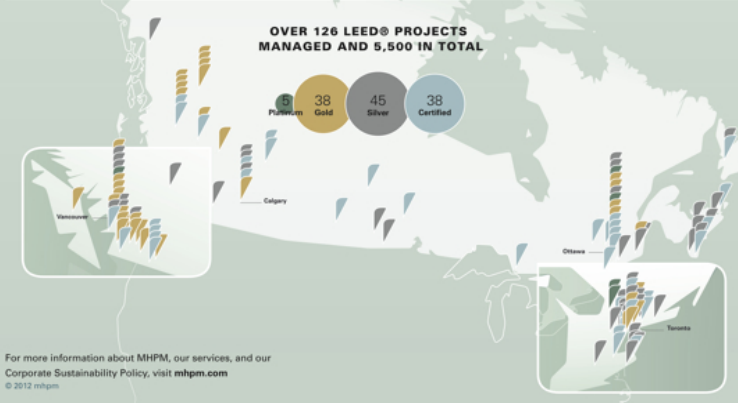

**Fig. 22.** Infographic by MHPM (2010-2011)

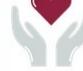

**Fig. 23.** Infographic by Trinseo (2017)

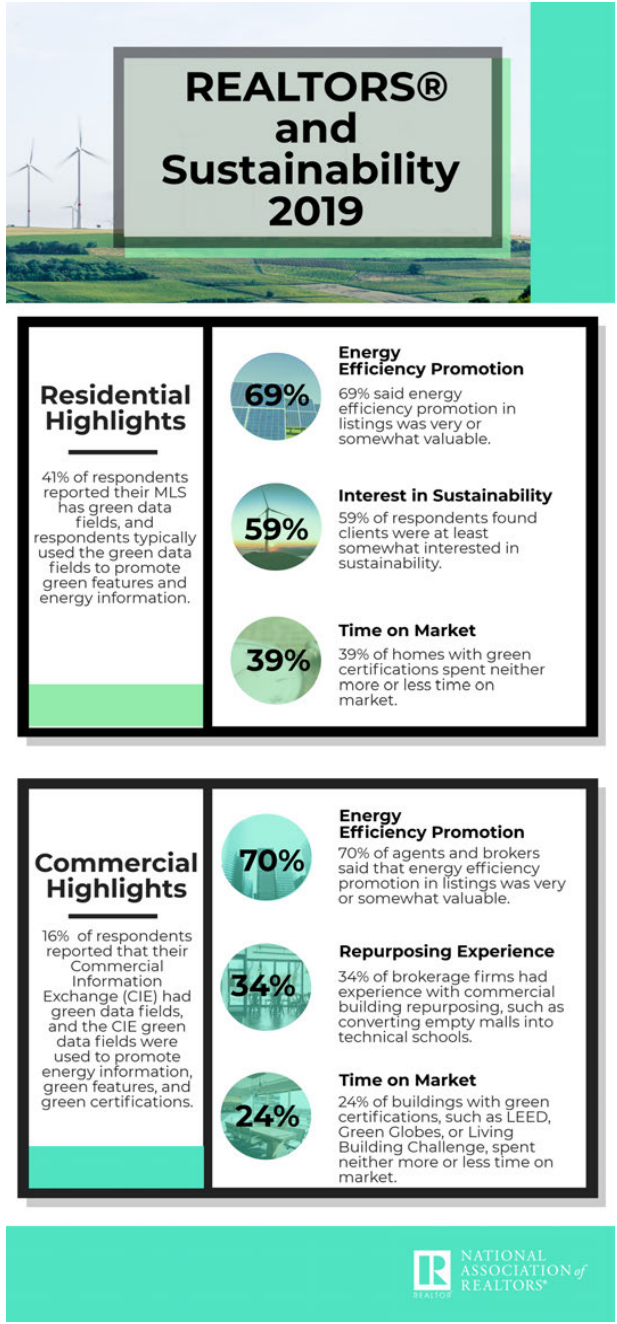

**Fig. 24.** Infographic by Realtor (2019)

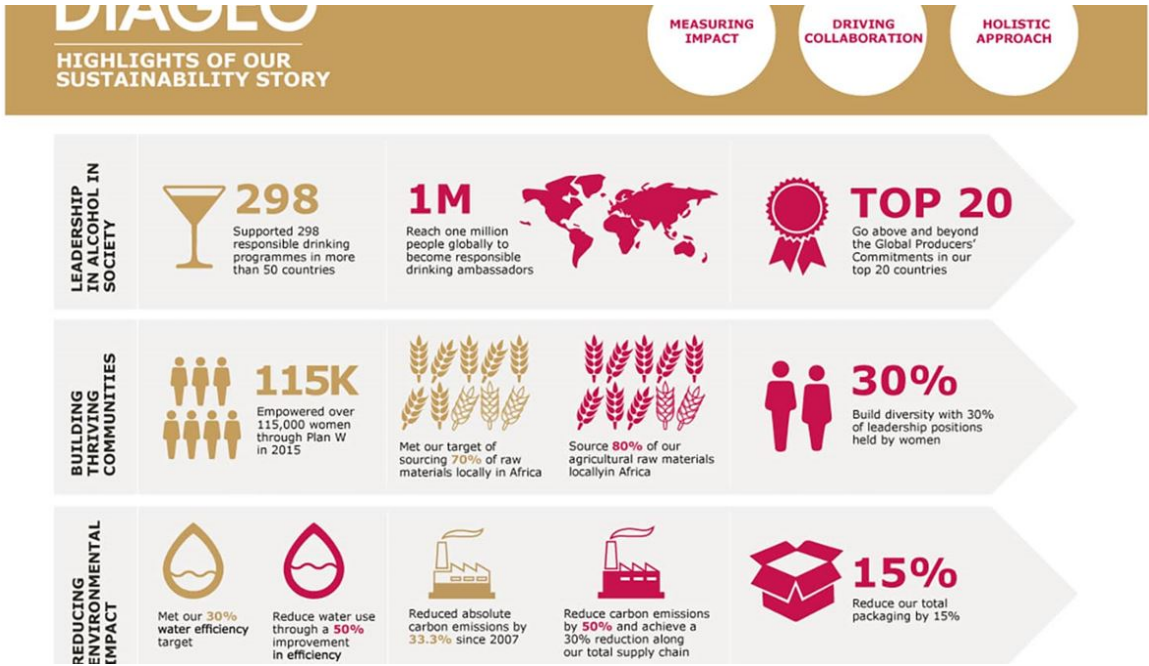

**Fig. 25.** Infographic by Diageo (2015)

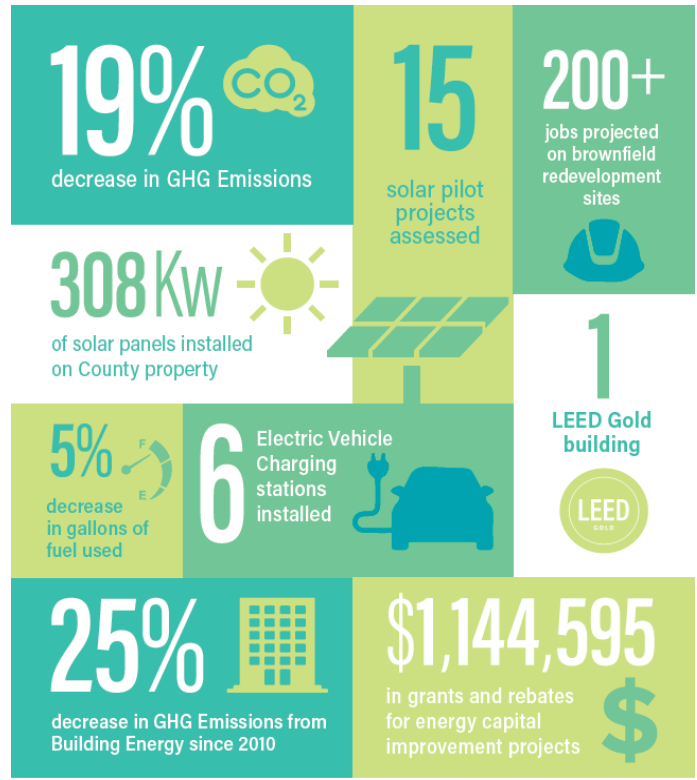

**Fig. 26.** Infographic by Cook county (2017)

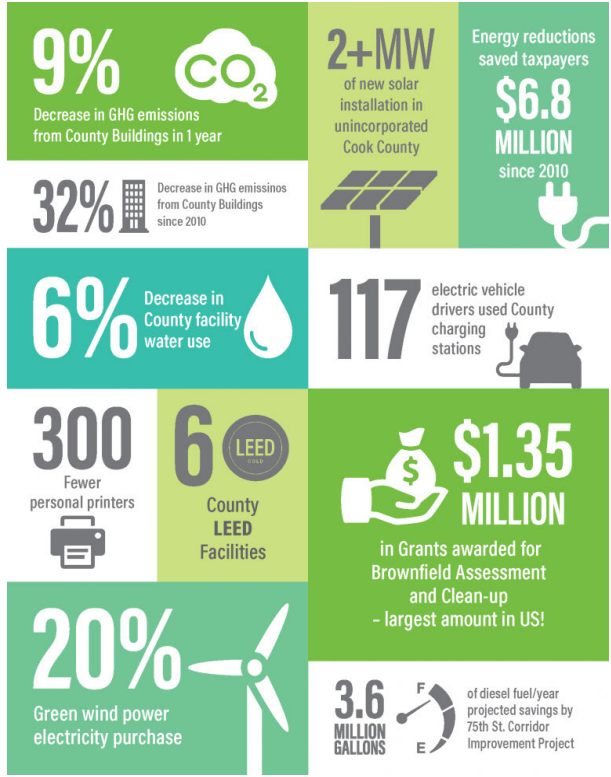

**Fig. 27.** Infographic by Cook county (2018)

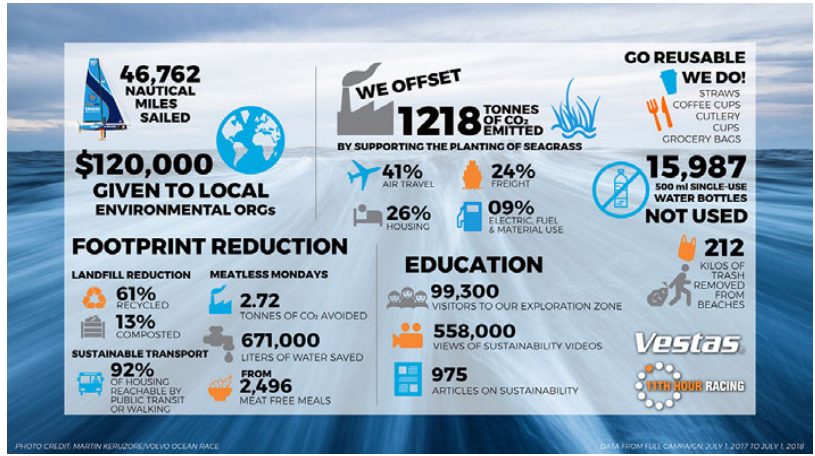

**Fig. 28.** Infographic by Vestas (2017-2018)

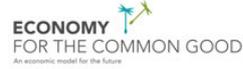

**Fig. 29.** Infographic by Economy for the Common Good (ECG) (2013)

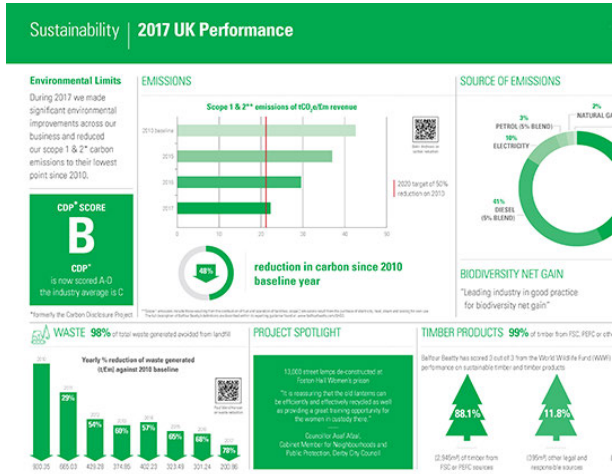

**Fig. 30.** Infographic by Balfour Beatty (2017)

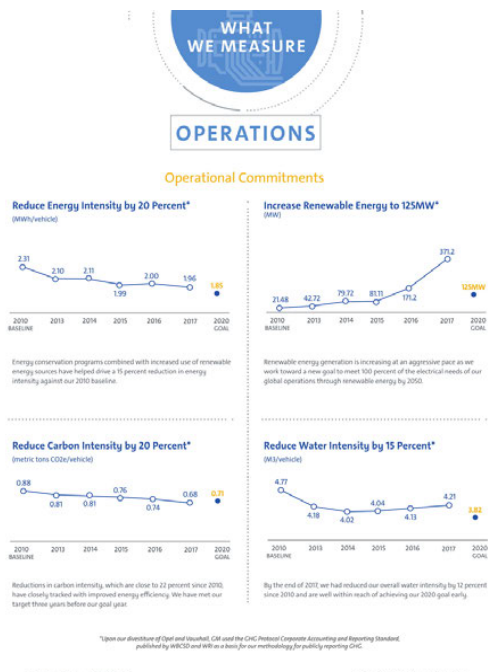
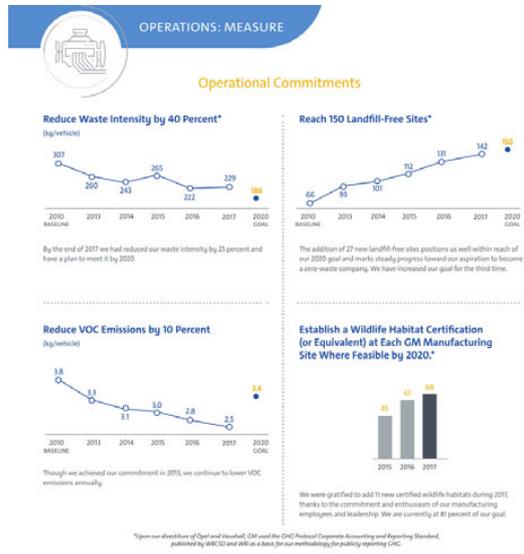

**Fig. 31.** Infographic by General Motors (GM) (2017)

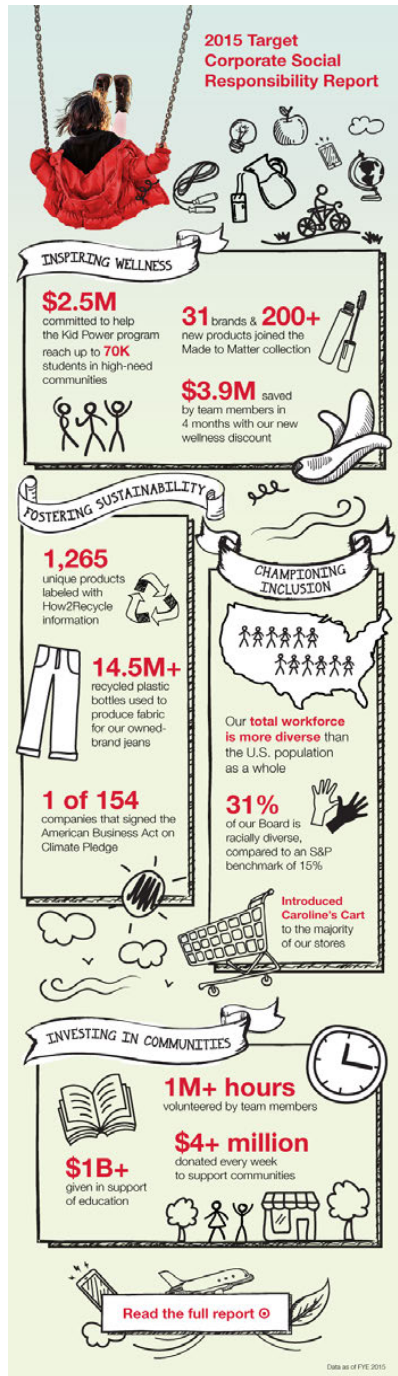

**Fig. 32.** Infographic by Target (2015)

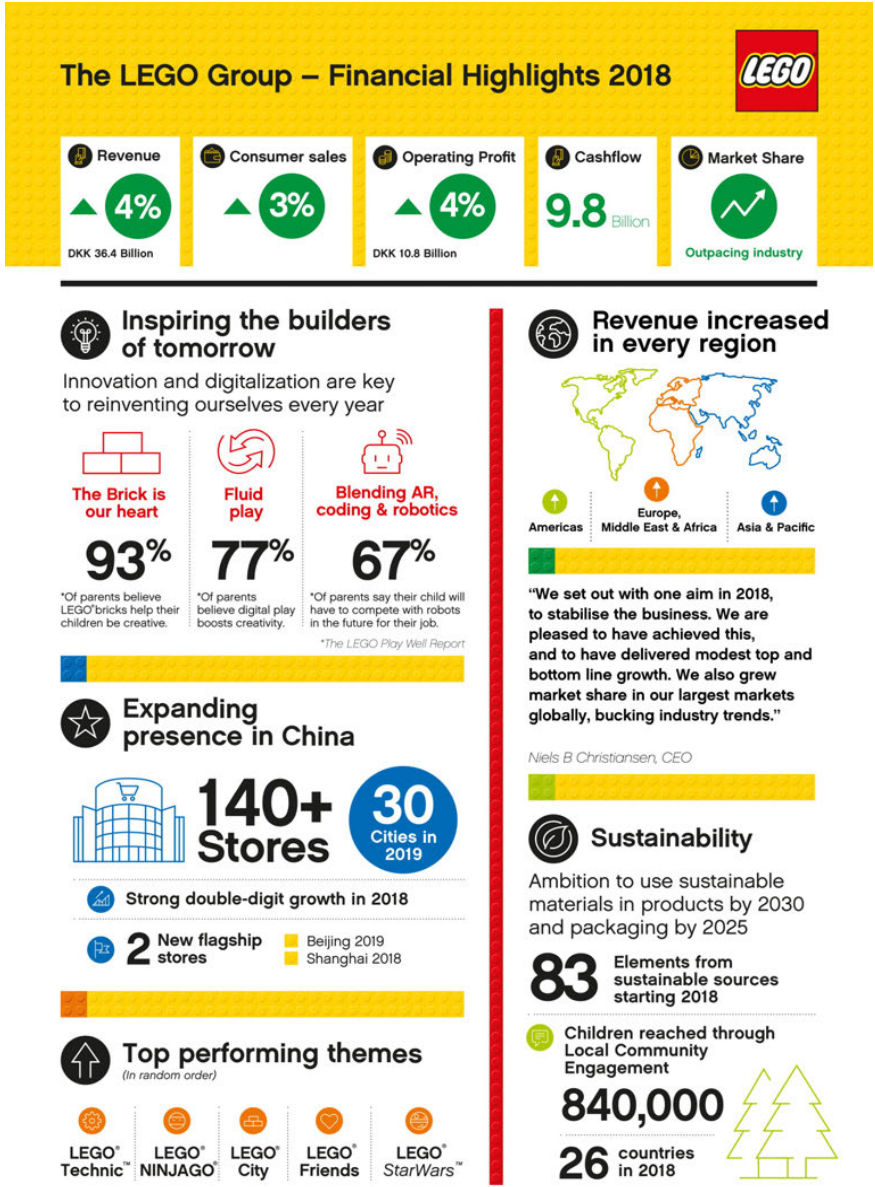

**Fig. 33.** Infographic by Lego (2018)

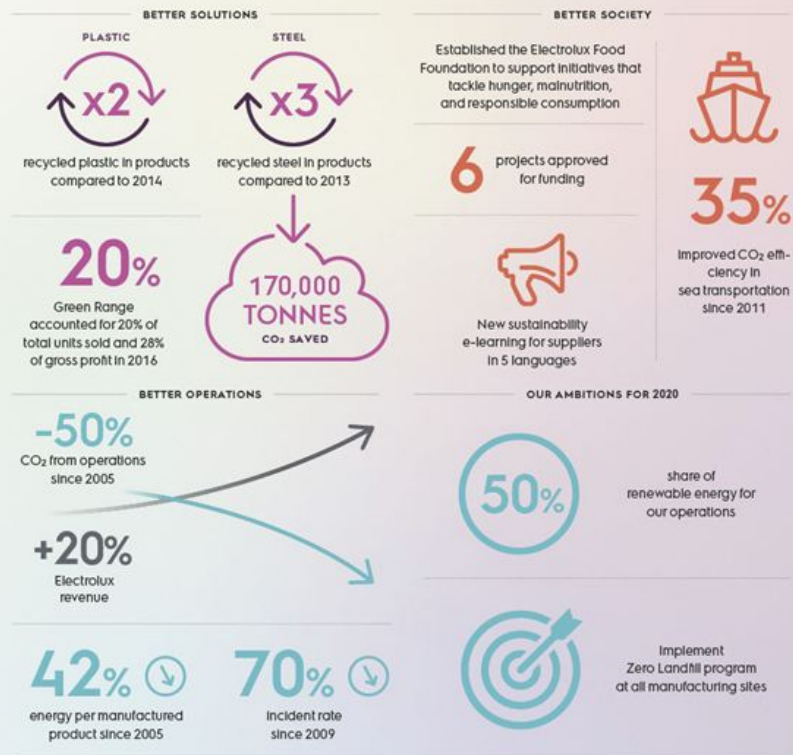

**Fig. 34.** Infographic by Electrolux (2016)

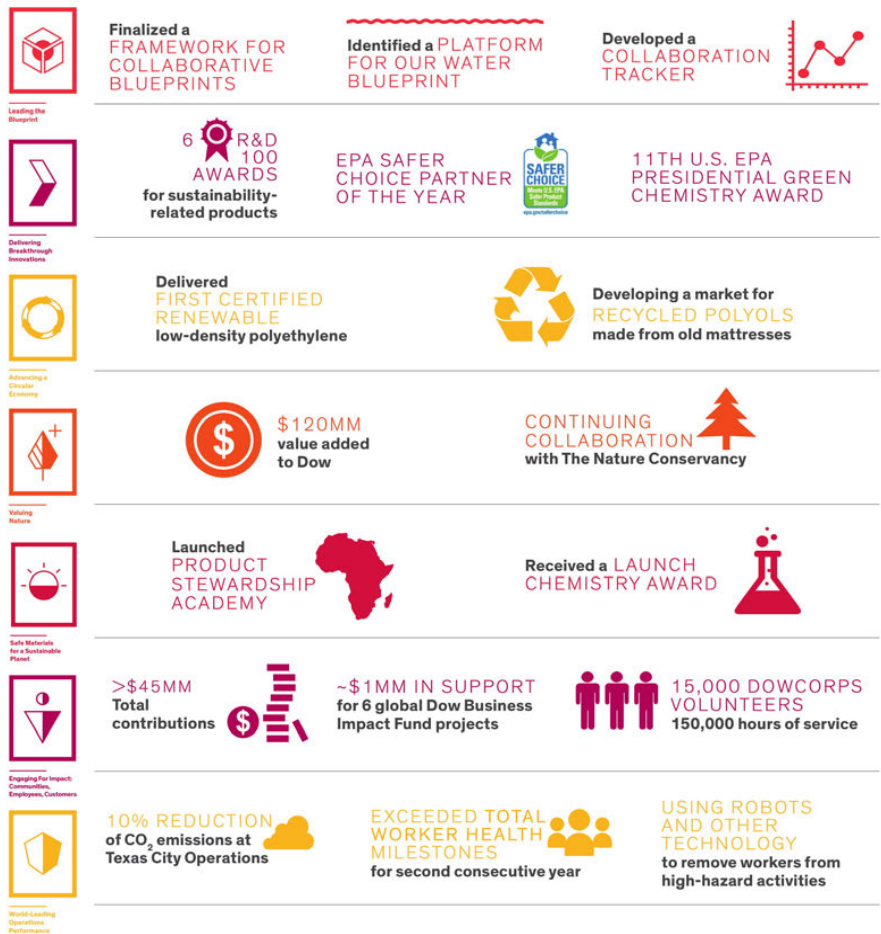

**Fig. 35.** Infographic by Dow (2017)

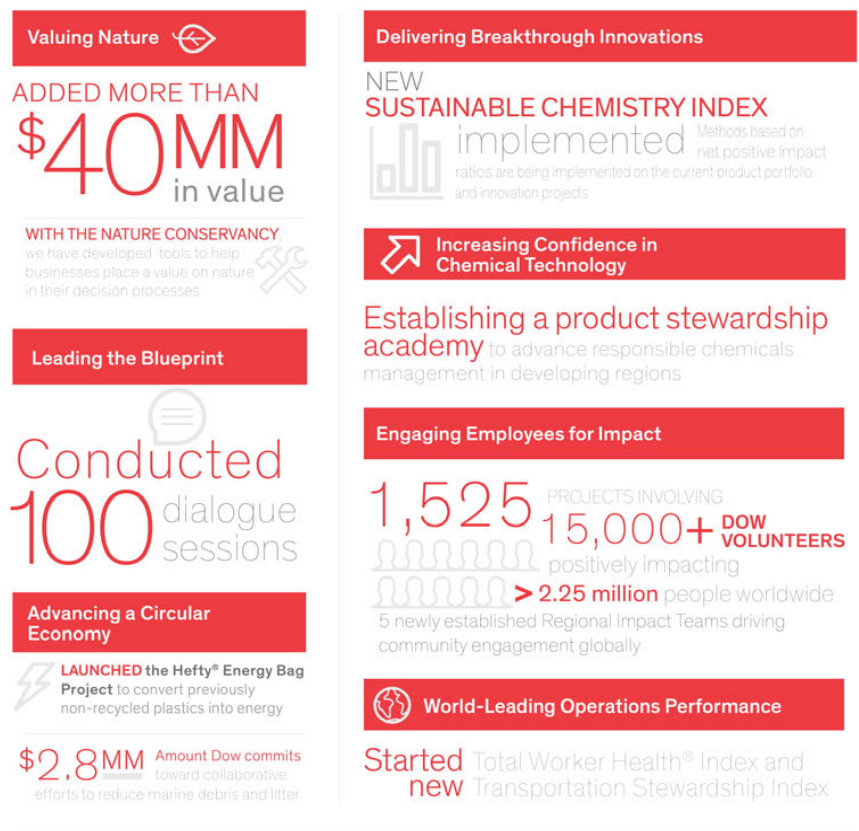

**Fig. 36.** Infographic by Dow (2016)

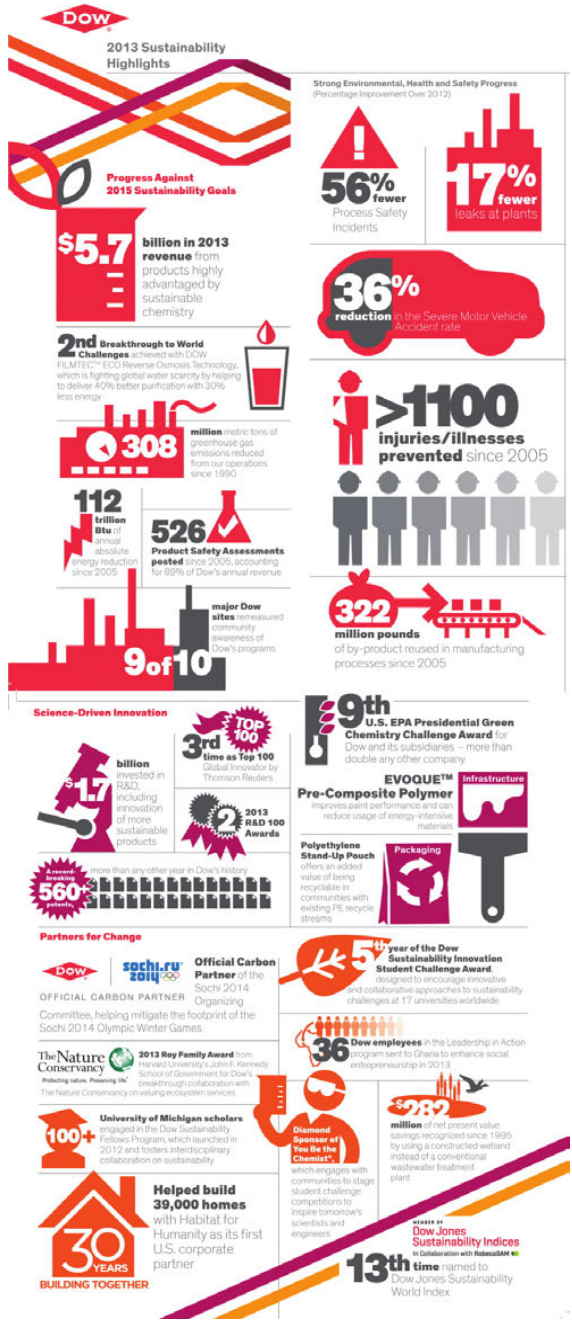

**Fig. 37.** Infographic by Dow (2013)

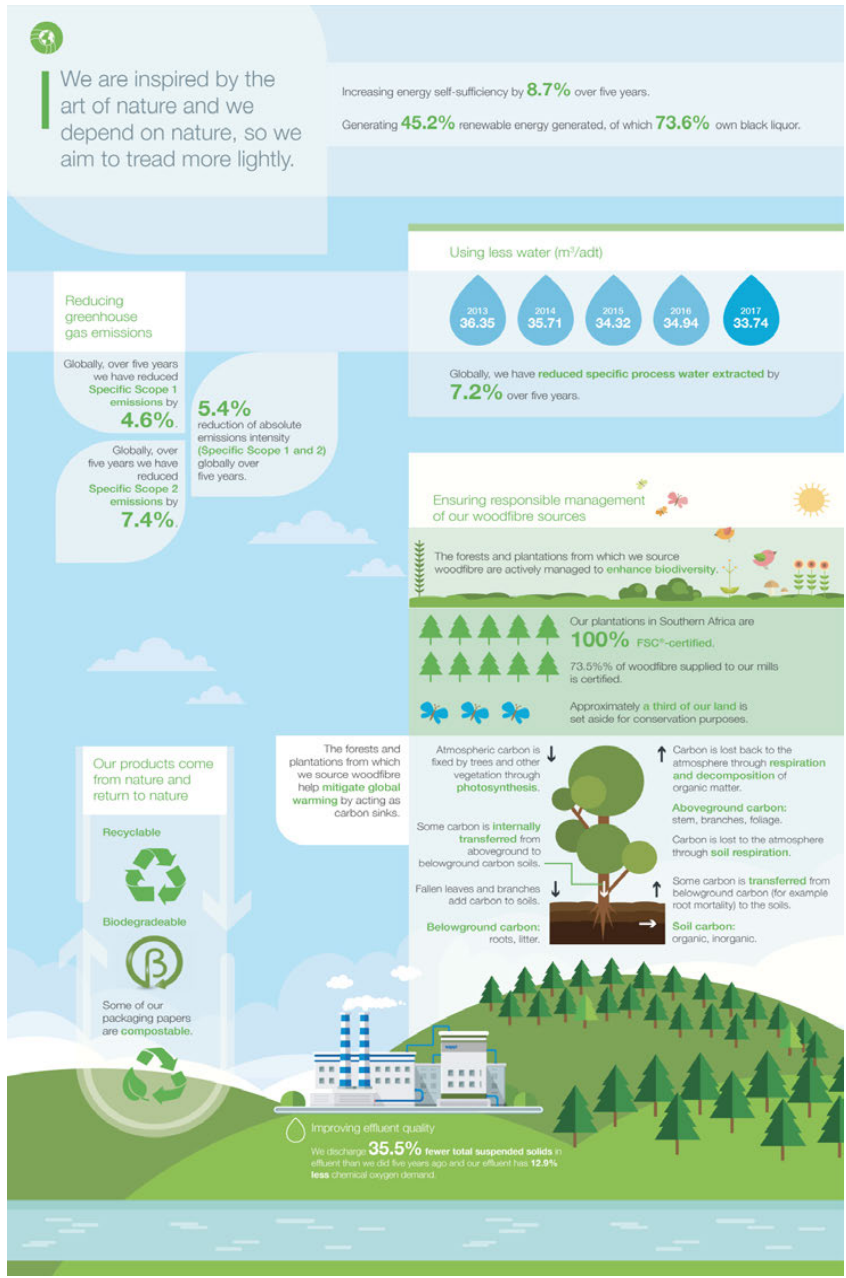

**Fig. 38.** Infographic by Sappi (2017)

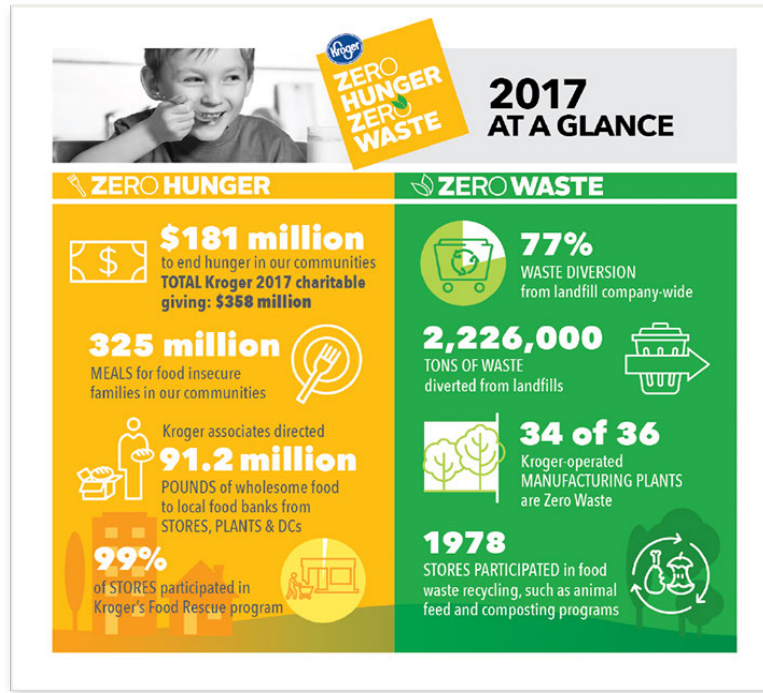

**Fig. 39.** Infographic by Kroger (2017)

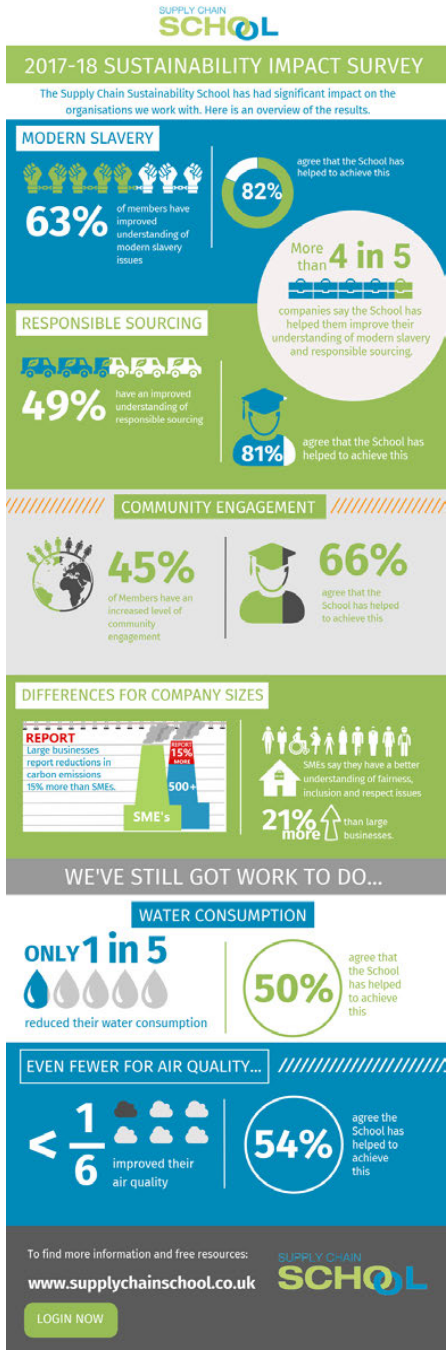

**Fig. 40.** Infographic by Supply Chain School (2017-2018)

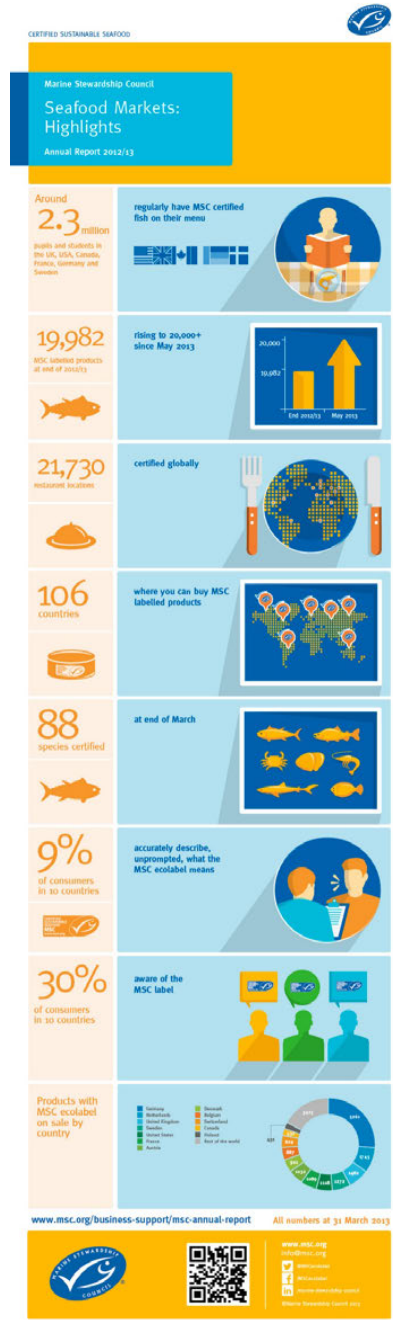

**Fig. 41.** Infographic by MSC (2012-2013)

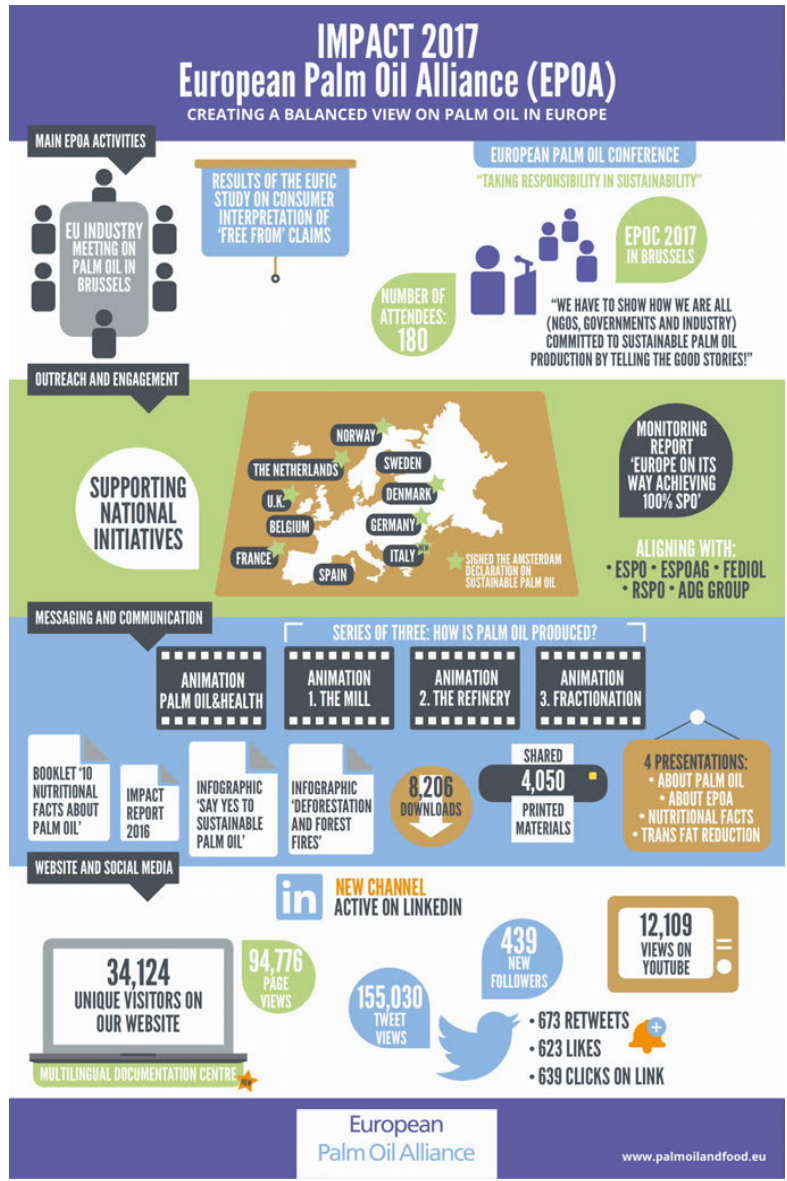

**Fig. 42.** Infographic by European Palm Oil Alliance (2017)

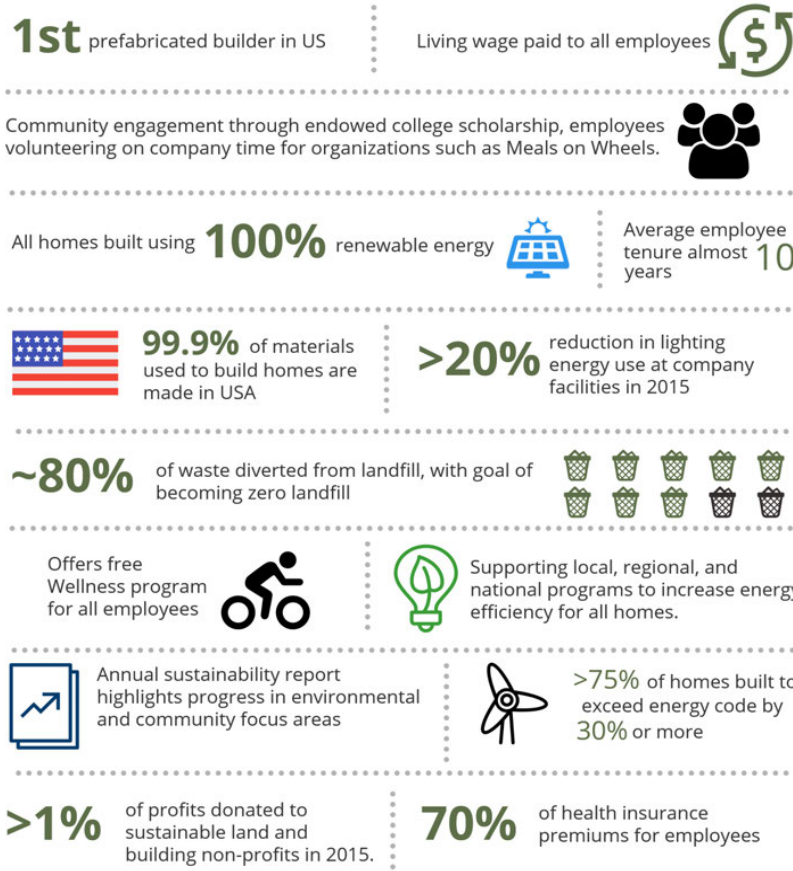

**Fig. 43.** Infographic by Deltec Homes (2016)

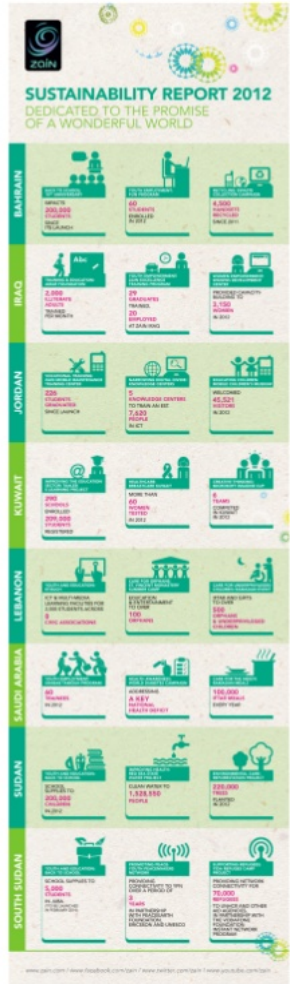

**Fig. 44.** Infographic by Zain (2012)

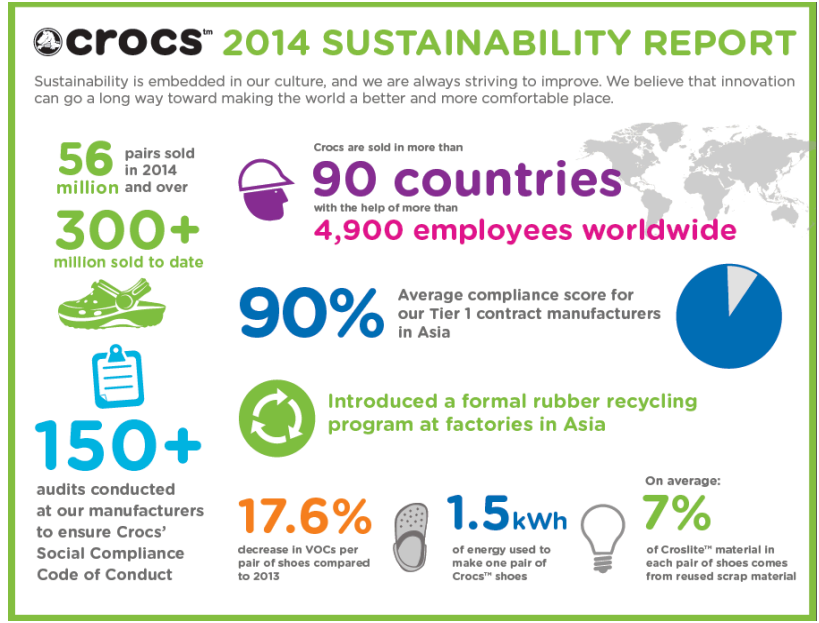

**Fig. 45.** Infographic by Crocs (2014)

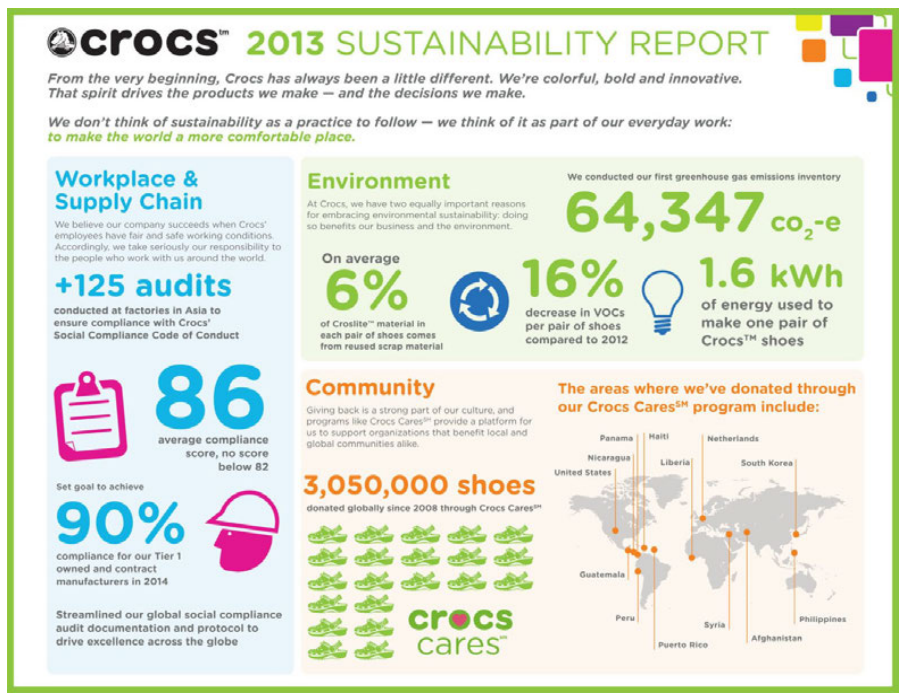

**Fig. 46.** Infographic by Crocs (2013)

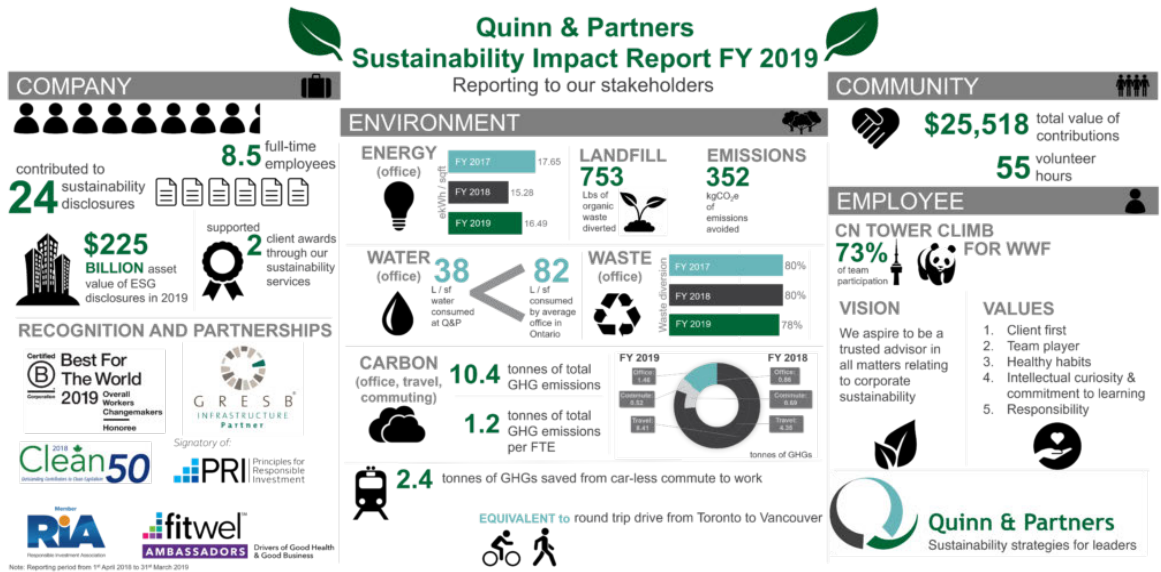

**Fig. 47.** Infographic by Quinn & Partners (2019)

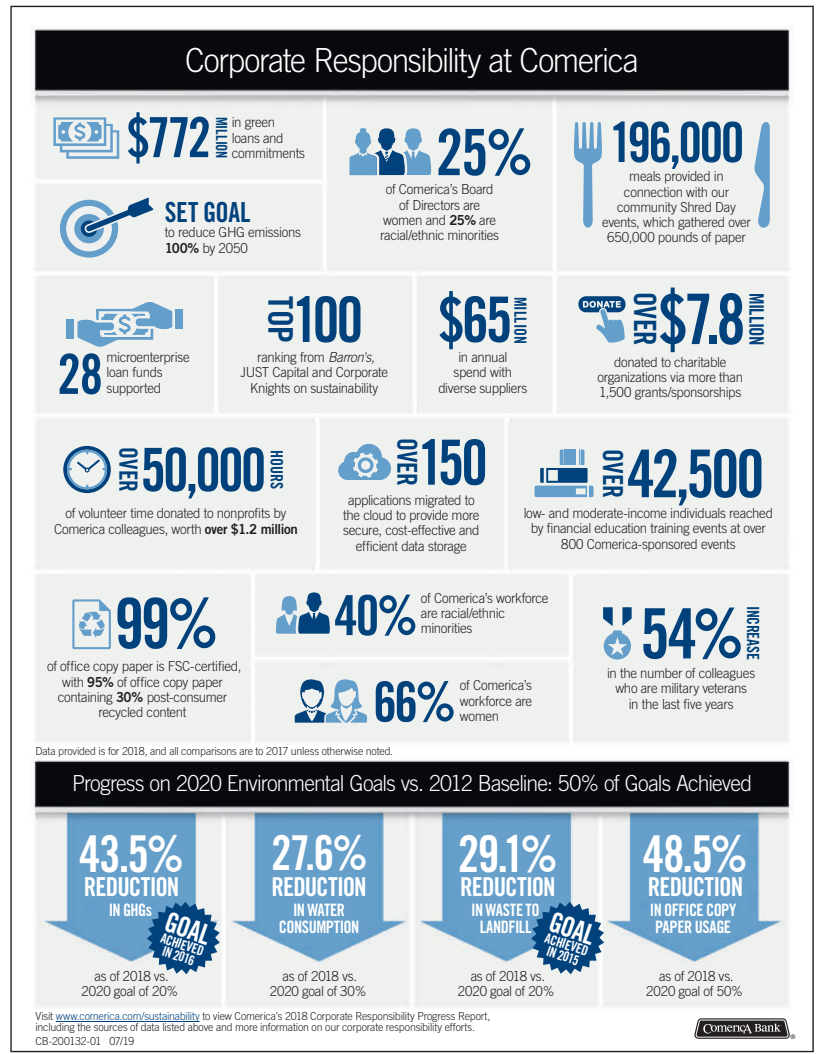

**Fig. 48.** Infographic by Comerica (2018)

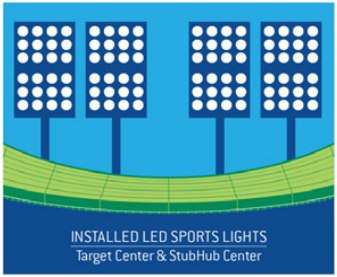
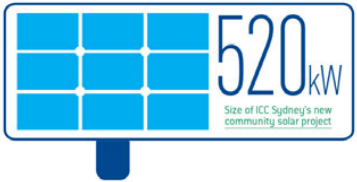
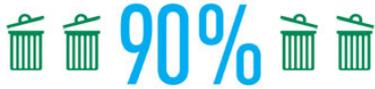
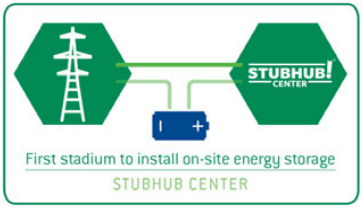
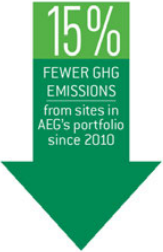
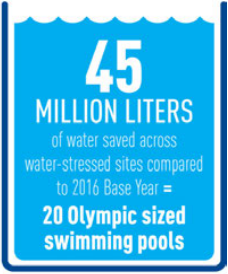

**Fig. 49.** Infographic by AEG (2017)

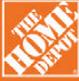

**Fig. 50.** Infographic by The Home Depot (2016)

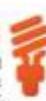
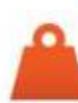

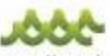

**Fig. 51.** Infographic by GM (2010)

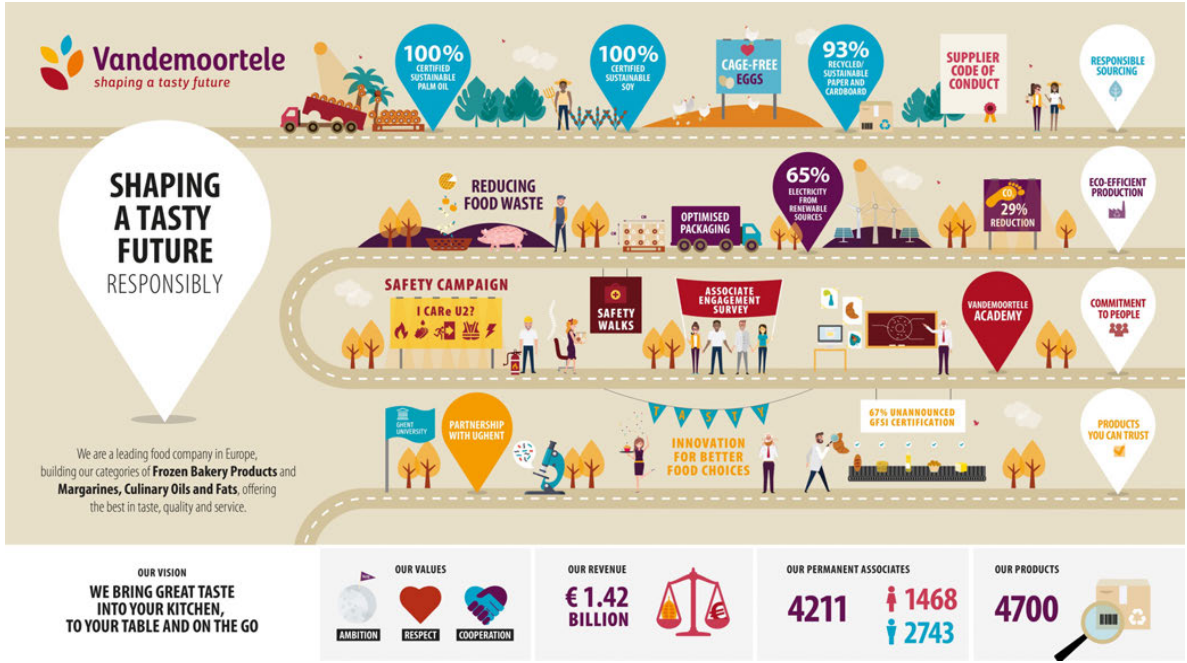

**Fig. 52.** Infographic by Vandemoortele (2018)

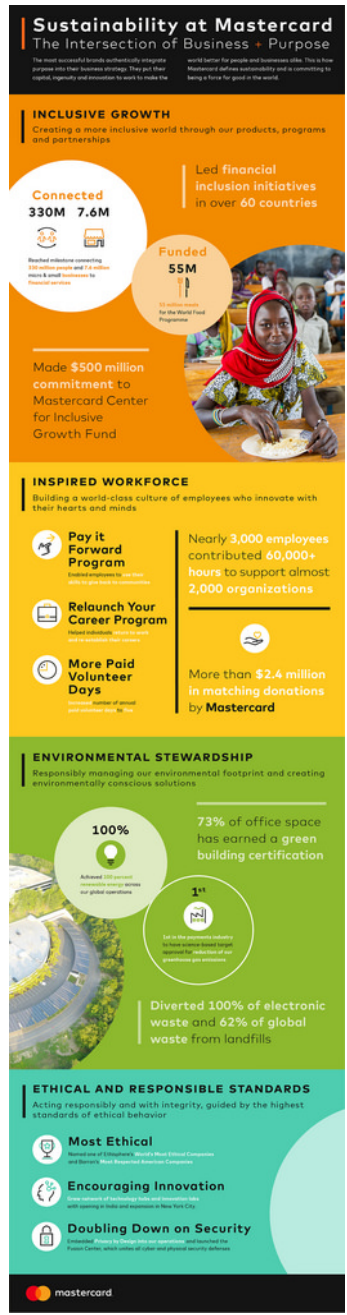

**Fig. 53.** Infographic by Mastercard (2017)

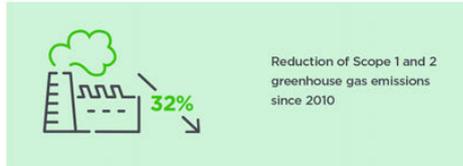

**32%** Reduction of Scope 1 and 2 greenhouse gas emissions since 2010

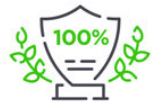

Perfect score from the Human Rights Campaign Foundation's Corporate Equality Index for two consecutive years — **100%**

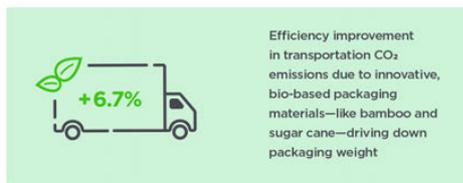

**+6.7%** Efficiency improvement in transportation $CO_2$ emissions due to innovative, bio-based packaging materials—like bamboo and sugar cane—driving down packaging weight

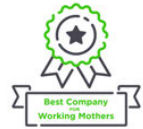

Named one of the 100 Best Companies for Working Mothers by *Working Mother* magazine

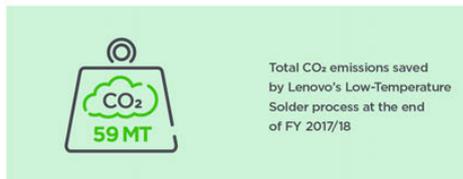

**59 MT** Total $CO_2$ emissions saved by Lenovo's Low-Temperature Solder process at the end of FY 2017/18

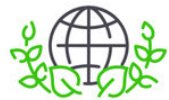

Global Week of Service directly impacted 33,000 consumers across 19 countries, benefiting minority populations and increasing access to STEM education

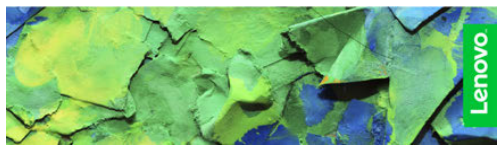

**Fig. 54.** Infographic by Lenovo (2018)

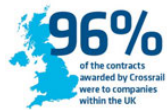
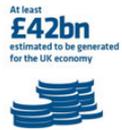
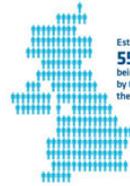
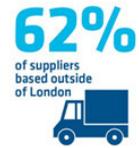
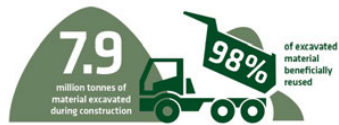
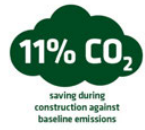
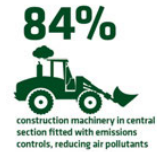
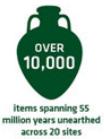
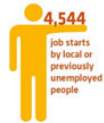
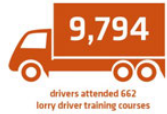
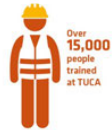
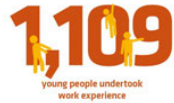
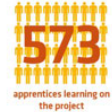

**Fig. 55.** Infographic by Crossrail (2016)

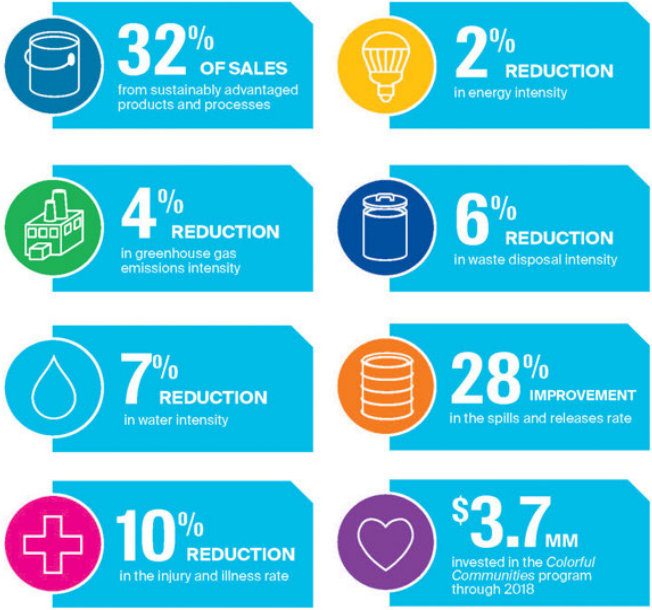

**Fig. 56.** Infographic by PPG (2018)

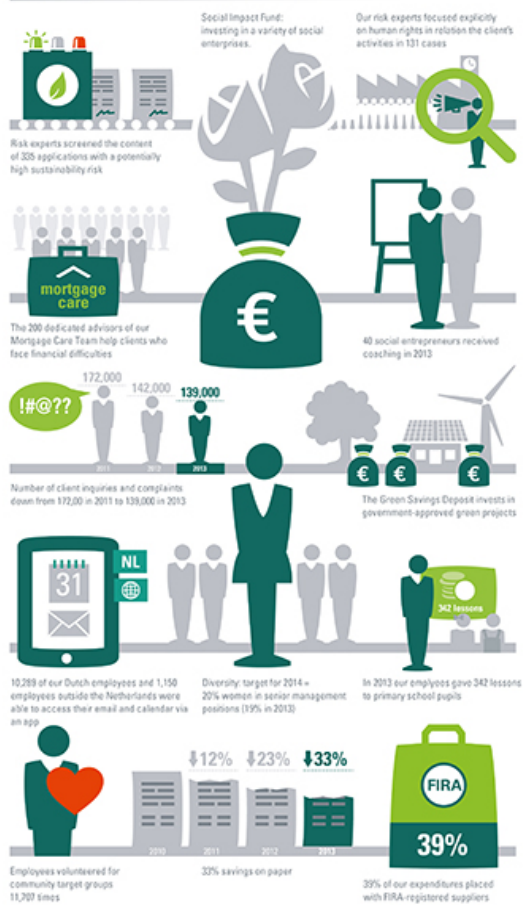

**Fig. 57.** Infographic by ABN AMRO (2013)

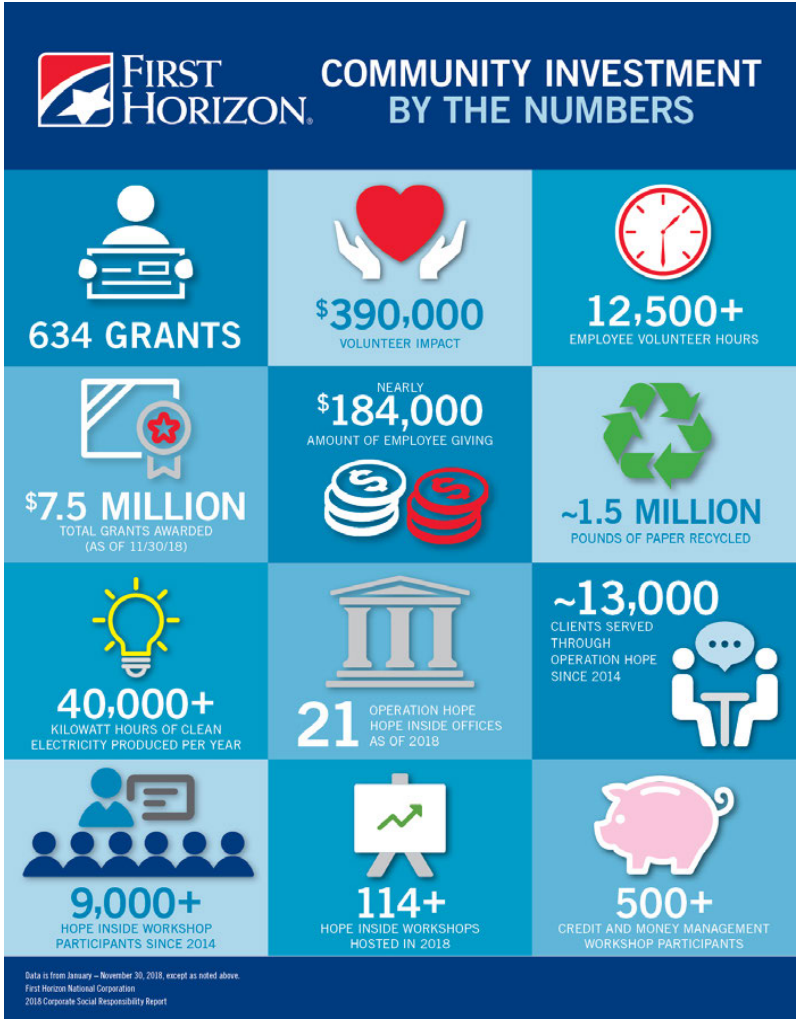

**Fig. 58.** Infographic by First Horizon (2018)

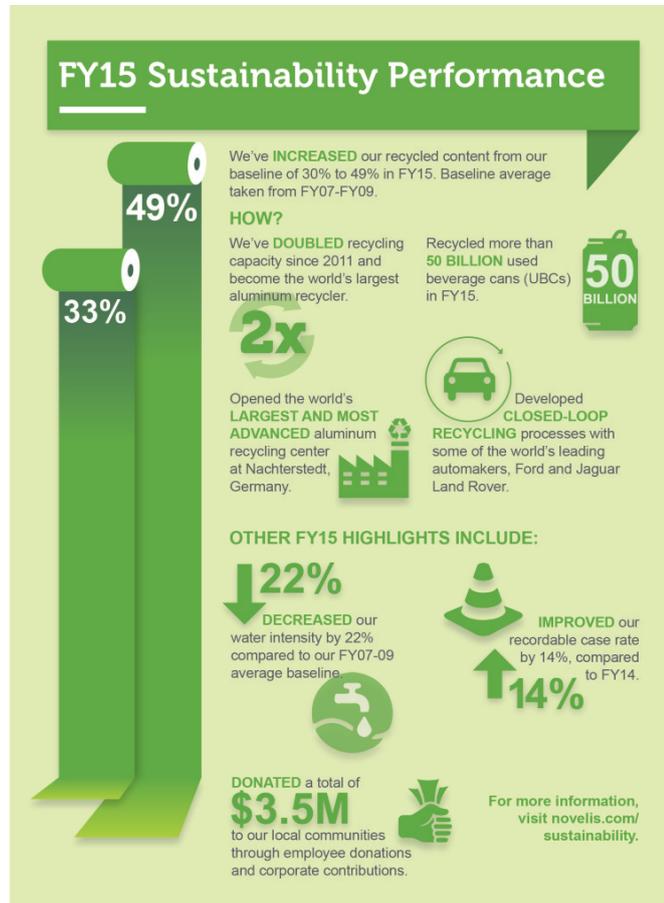

**Fig. 59.** Infographic by Novelis (2015)

### 3.3 Analysis of infographic component types

We have analysed each of the infographics in Table 2, to identify component types. Fig. 60 shows an example of how we have analysed the component architecture of a given infographic. Then Table 3 to Table 5 show the result of our analysis of the infographics, by indicating how many components of each type are present in each infographic. The definitions of infographic component types are presented in Table 6.

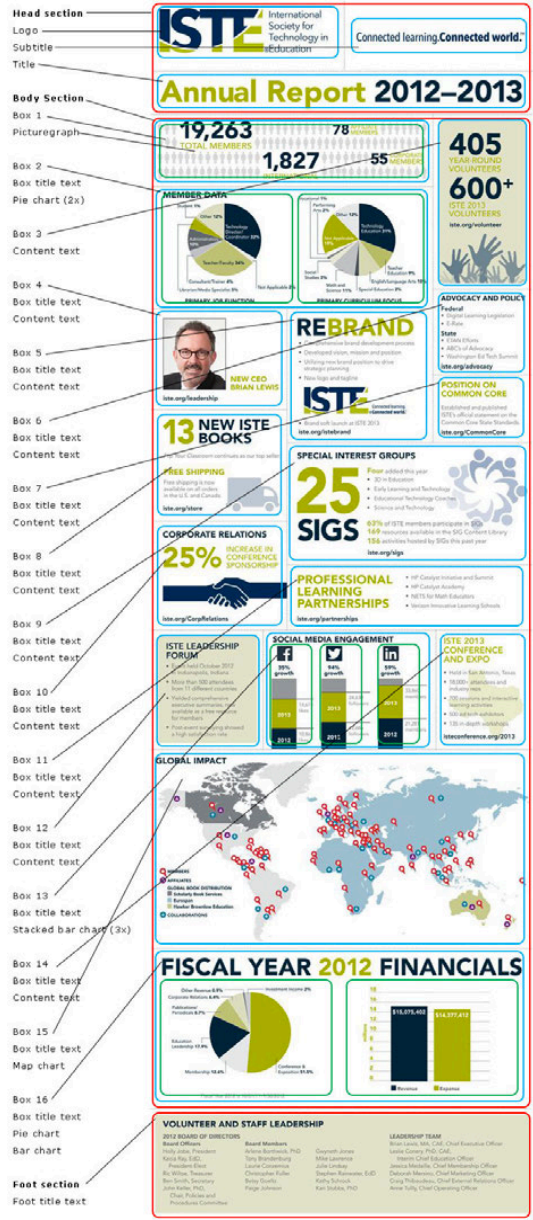

**Fig. 60.** Example of an infographic partitioned into infographic component types

**Table 3.** Analysis of infographic component types (part 1 of 3)

| Id | Name | 1 | 2 | 3 | 4 | 5 | 6 | 7 | 8 | 9 | 10 | 11 | 12 | 13 | 14 | 15 | 16 | 17 | 18 | 19 | 20 |
|---|---|---|---|---|---|---|---|---|---|---|---|---|---|---|---|---|---|---|---|---|---|
| C1 | Head section | 1 | 1 | 1 | 1 | 1 | 1 | 1 | 1 | 1 | 1 | 1 | 0 | 0 | 1 | 1 | 1 | 1 | 1 | 1 | 1 |
| C2 | Title text | 1 | 1 | 1 | 1 | 1 | 1 | 1 | 1 | 1 | 1 | 1 | 0 | 0 | 1 | 1 | 1 | 1 | 1 | 1 | 1 |
| C3 | Subtitle text | 0 | 1 | 1 | 1 | 0 | 1 | 1 | 0 | 0 | 1 | 0 | 0 | 0 | 1 | 1 | 1 | 1 | 1 | 0 | 0 |
| C4 | Introduction text | 1 | 0 | 0 | 0 | 0 | 0 | 1 | 0 | 0 | 1 | 0 | 0 | 0 | 0 | 0 | 0 | 1 | 0 | 0 | 0 |
| C5 | Logo | 1 | 1 | 1 | 1 | 1 | 1 | 1 | 1 | 0 | 0 | 0 | 0 | 0 | 1 | 1 | 1 | 0 | 1 | 1 | 1 |
| C6 | Body Section | 1 | 1 | 1 | 1 | 1 | 1 | 1 | 1 | 1 | 1 | 1 | 1 | 1 | 1 | 1 | 1 | 1 | 1 | 1 | 1 |
| C7 | Box | 6 | 10 | 8 | 9 | 5 | 9 | 9 | 11 | 15 | 17 | 8 | 14 | 13 | 7 | 7 | 7 | 6 | 16 | 16 | 12 |
| C8 | Box title text | 4 | 10 | 8 | 9 | 5 | 9 | 9 | 11 | 0 | 8 | 7 | 0 | 0 | 4 | 4 | 4 | 3 | 14 | 16 | 12 |
| C9 | Body text | 6 | 10 | 8 | 9 | 5 | 9 | 9 | 11 | 15 | 17 | 8 | 14 | 13 | 6 | 6 | 6 | 6 | 11 | 16 | 12 |
|  | *Chart* |  |  |  |  |  |  |  |  |  |  |  |  |  |  |  |  |  |  |  |  |
| C10 | Table | 0 | 0 | 0 | 0 | 0 | 0 | 0 | 0 | 0 | 0 | 0 | 0 | 0 | 1 | 1 | 1 | 0 | 0 | 0 | 0 |
| C11 | Pie chart | 0 | 0 | 1 | 0 | 0 | 0 | 0 | 2 | 0 | 4 | 0 | 1 | 1 | 0 | 0 | 0 | 0 | 3 | 1 | 7 |
| C12 | Bar chart | 0 | 0 | 1 | 0 | 0 | 0 | 1 | 1 | 0 | 0 | 0 | 0 | 0 | 0 | 0 | 0 | 0 | 1 | 3 | 0 |
| C13 | Grouped bar chart | 0 | 0 | 0 | 0 | 0 | 0 | 0 | 0 | 0 | 0 | 0 | 0 | 0 | 0 | 0 | 0 | 0 | 0 | 0 | 0 |
| C14 | Stacked bar chart | 0 | 0 | 0 | 0 | 0 | 0 | 0 | 0 | 0 | 0 | 0 | 0 | 0 | 0 | 0 | 0 | 0 | 3 | 0 | 0 |
| C15 | Picturegraph | 7 | 0 | 0 | 0 | 0 | 0 | 1 | 0 | 0 | 9 | 0 | 0 | 0 | 0 | 0 | 0 | 5 | 0 | 6 | 0 |
| C16 | Line chart | 0 | 0 | 0 | 0 | 0 | 0 | 0 | 0 | 0 | 0 | 0 | 0 | 0 | 0 | 0 | 0 | 0 | 0 | 0 | 0 |
| C17 | Map chart | 0 | 0 | 0 | 0 | 0 | 0 | 0 | 0 | 0 | 0 | 0 | 0 | 0 | 0 | 0 | 0 | 0 | 1 | 1 | 0 |
| C18 | Foot section | 1 | 1 | 1 | 1 | 0 | 1 | 1 | 1 | 1 | 1 | 1 | 1 | 1 | 1 | 1 | 1 | 1 | 1 | 0 | 1 |
| C19 | Foot title text | 0 | 0 | 0 | 0 | 0 | 0 | 0 | 0 | 0 | 0 | 1 | 1 | 1 | 1 | 1 | 1 | 1 | 1 | 0 | 1 |
| C20 | Foot text | 1 | 1 | 1 | 1 | 0 | 1 | 1 | 1 | 1 | 1 | 1 | 1 | 1 | 1 | 1 | 1 | 1 | 1 | 0 | 0 |
| C21 | Logo | 1 | 0 | 0 | 1 | 0 | 0 | 0 | 0 | 0 | 1 | 0 | 0 | 0 | 1 | 1 | 1 | 1 | 0 | 0 | 1 |
| C22 | Link to full report | 1 | 1 | 0 | 1 | 0 | 1 | 1 | 0 | 0 | 0 | 1 | 0 | 0 | 0 | 0 | 0 | 0 | 0 | 0 | 0 |
|  | **Sum** | 32 | 38 | 33 | 36 | 19 | 35 | 38 | 42 | 35 | 63 | 29 | 33 | 31 | 27 | 27 | 27 | 29 | 57 | 63 | 50 |
|  | **Count** | 13 | 11 | 12 | 12 | 7 | 11 | 14 | 11 | 7 | 13 | 9 | 7 | 7 | 13 | 13 | 13 | 13 | 15 | 11 | 11 |

**Table 4.** Analysis of infographic component types (part 2 of 3)

| Id | Name | 21 | 22 | 23 | 24 | 25 | 26 | 27 | 28 | 29 | 30 | 31 | 32 | 33 | 34 | 35 | 36 | 37 | 38 | 39 | 40 |
|---|---|---|---|---|---|---|---|---|---|---|---|---|---|---|---|---|---|---|---|---|---|
| C1 | Head section | 1 | 1 | 1 | 1 | 0 | 0 | 0 | 1 | 1 | 1 | 1 | 1 | 1 | 1 | 1 | 1 | 1 | 1 | 1 | 1 |
| C2 | Title text | 1 | 1 | 1 | 1 | 0 | 0 | 0 | 1 | 1 | 1 | 1 | 1 | 1 | 1 | 1 | 1 | 1 | 1 | 1 | 1 |
| C3 | Subtitle text | 1 | 1 | 0 | 1 | 0 | 0 | 0 | 1 | 0 | 1 | 0 | 0 | 0 | 0 | 0 | 1 | 0 | 0 | 1 | 1 |
| C4 | Introduction text | 0 | 0 | 0 | 1 | 0 | 0 | 0 | 0 | 0 | 0 | 0 | 0 | 1 | 1 | 0 | 1 | 0 | 0 | 0 | 0 |
| C5 | Logo | 1 | 0 | 0 | 0 | 0 | 0 | 0 | 1 | 0 | 0 | 0 | 1 | 0 | 1 | 1 | 1 | 1 | 1 | 1 | 1 |
| C6 | Body Section | 1 | 1 | 1 | 1 | 1 | 1 | 1 | 1 | 1 | 1 | 1 | 1 | 1 | 1 | 1 | 1 | 1 | 1 | 1 | 1 |
| C7 | Box | 7 | 7 | 10 | 9 | 9 | 11 | 5 | 1 | 8 | 9 | 4 | 7 | 4 | 7 | 7 | 24 | 7 | 2 | 7 | 8 |
| C8 | Box title text | 7 | 7 | 10 | 3 | 9 | 11 | 5 | 0 | 8 | 8 | 4 | 7 | 4 | 7 | 7 | 24 | 7 | 2 | 7 | 8 |
| C9 | Body text | 6 | 7 | 10 | 9 | 0 | 0 | 5 | 0 | 8 | 9 | 0 | 0 | 4 | 7 | 7 | 21 | 7 | 2 | 7 | 8 |
|  | *Chart* |  |  |  |  |  |  |  |  |  |  |  |  |  |  |  |  |  |  |  |  |
| C10 | Table | 0 | 0 | 0 | 0 | 0 | 0 | 0 | 1 | 0 | 0 | 0 | 0 | 0 | 0 | 0 | 0 | 0 | 0 | 0 | 0 |
| C11 | Pie chart | 2 | 0 | 0 | 0 | 0 | 0 | 0 | 0 | 2 | 0 | 0 | 0 | 0 | 0 | 0 | 0 | 0 | 2 | 1 | 1 |
| C12 | Bar chart | 0 | 0 | 0 | 0 | 0 | 0 | 0 | 0 | 2 | 1 | 0 | 0 | 0 | 0 | 0 | 0 | 0 | 0 | 0 | 0 |
| C13 | Grouped bar chart | 7 | 0 | 0 | 0 | 0 | 0 | 0 | 0 | 0 | 0 | 0 | 0 | 0 | 0 | 0 | 0 | 0 | 0 | 0 | 0 |
| C14 | Stacked bar chart | 1 | 0 | 0 | 0 | 0 | 0 | 0 | 0 | 0 | 0 | 0 | 0 | 0 | 0 | 0 | 0 | 0 | 0 | 0 | 0 |
| C15 | Picturegraph | 16 | 0 | 0 | 6 | 0 | 0 | 0 | 0 | 0 | 0 | 0 | 0 | 0 | 0 | 0 | 2 | 0 | 0 | 8 | 0 |
| C16 | Line chart | 0 | 0 | 0 | 0 | 0 | 0 | 0 | 0 | 0 | 7 | 0 | 0 | 1 | 0 | 0 | 0 | 0 | 0 | 0 | 1 |

| Id | Name | 21 | 22 | 23 | 24 | 25 | 26 | 27 | 28 | 29 | 30 | 31 | 32 | 33 | 34 | 35 | 36 | 37 | 38 | 39 | 40 |
|---|---|---|---|---|---|---|---|---|---|---|---|---|---|---|---|---|---|---|---|---|---|
| C17 | Map chart | 1 | 0 | 0 | 0 | 0 | 0 | 0 | 0 | 0 | 0 | 0 | 0 | 0 | 0 | 0 | 0 | 0 | 0 | 0 | 0 |
| C18 | Foot section | 1 | 1 | 1 | 0 | 1 | 0 | 1 | 0 | 0 | 0 | 1 | 0 | 1 | 0 | 1 | 1 | 0 | 0 | 1 | 1 |
| C19 | Foot title text | 0 | 0 | 0 | 0 | 0 | 0 | 0 | 0 | 0 | 0 | 0 | 0 | 0 | 0 | 0 | 1 | 0 | 0 | 0 | 1 |
| C20 | Foot text | 1 | 0 | 0 | 0 | 0 | 0 | 1 | 0 | 0 | 0 | 1 | 0 | 0 | 0 | 1 | 0 | 0 | 0 | 1 | 1 |
| C21 | Logo | 0 | 1 | 1 | 0 | 1 | 0 | 0 | 0 | 0 | 0 | 0 | 0 | 1 | 0 | 0 | 1 | 0 | 0 | 1 | 1 |
| C22 | Link to full report | 1 | 0 | 0 | 0 | 0 | 0 | 0 | 0 | 0 | 0 | 1 | 0 | 0 | 0 | 0 | 0 | 0 | 0 | 1 | 1 |
| | Sum | 55 | 27 | 35 | 32 | 21 | 23 | 18 | 7 | 31 | 38 | 14 | 18 | 18 | 26 | 28 | 79 | 26 | 12 | 39 | 36 |
| | Count | 16 | 9 | 8 | 9 | 5 | 3 | 6 | 7 | 8 | 9 | 8 | 6 | 9 | 8 | 10 | 12 | 8 | 8 | 14 | 15 |

**Table 5.** Analysis of infographic component types (part 3 of 3)

| Id | Name | 41 | 42 | 43 | 44 | 45 | 46 | 47 | 48 | 49 | 50 | 51 | 52 | 53 | 54 | 55 | 56 | 57 | 58 |
|---|---|---|---|---|---|---|---|---|---|---|---|---|---|---|---|---|---|---|---|
| C1 | Head section | 1 | 0 | 1 | 1 | 1 | 1 | 1 | 1 | 1 | 1 | 1 | 1 | 1 | 1 | 1 | 1 | 1 | 1 |
| C2 | Title text | 1 | 0 | 1 | 1 | 1 | 1 | 1 | 1 | 1 | 1 | 1 | 1 | 1 | 1 | 1 | 1 | 1 | 1 |
| C3 | Subtitle text | 1 | 0 | 1 | 0 | 0 | 1 | 0 | 0 | 0 | 0 | 0 | 1 | 0 | 0 | 0 | 1 | 0 | 0 |
| C4 | Introduction text | 0 | 0 | 0 | 1 | 1 | 0 | 0 | 0 | 0 | 0 | 1 | 1 | 0 | 0 | 0 | 0 | 0 | 0 |
| C5 | Logo | 0 | 0 | 1 | 1 | 1 | 0 | 0 | 0 | 1 | 1 | 1 | 0 | 0 | 0 | 1 | 0 | 1 | 0 |
| C6 | Body Section | 1 | 1 | 1 | 1 | 1 | 1 | 1 | 1 | 1 | 1 | 1 | 1 | 1 | 1 | 1 | 1 | 1 | 1 |
| C7 | Box | 5 | 14 | 8 | 8 | 3 | 12 | 16 | 6 | 3 | 4 | 5 | 4 | 6 | 3 | 8 | 13 | 12 | 8 |
| C8 | Box title text | 5 | 0 | 8 | 8 | 3 | 11 | 16 | 6 | 3 | 4 | 5 | 4 | 0 | 3 | 8 | 13 | 12 | 3 |
| C9 | Body text | 5 | 0 | 0 | 8 | 3 | 11 | 16 | 4 | 3 | 4 | 5 | 4 | 6 | 3 | 8 | 0 | 0 | 0 |
| | *Chart* | | | | | | | | | | | | | | | | | | |
| C10 | Table | 0 | 0 | 0 | 0 | 0 | 0 | 0 | 0 | 0 | 0 | 0 | 0 | 0 | 0 | 0 | 0 | 0 | 0 |
| C11 | Pie chart | 0 | 0 | 0 | 1 | 0 | 1 | 0 | 0 | 0 | 0 | 0 | 0 | 0 | 0 | 0 | 0 | 0 | 0 |
| C12 | Bar chart | 0 | 0 | 0 | 0 | 0 | 2 | 0 | 0 | 0 | 0 | 0 | 0 | 0 | 0 | 0 | 0 | 0 | 0 |
| C13 | Grouped bar chart | 0 | 0 | 0 | 0 | 0 | 0 | 0 | 0 | 0 | 0 | 0 | 0 | 0 | 0 | 0 | 0 | 0 | 0 |
| C14 | Stacked bar chart | 0 | 0 | 0 | 0 | 0 | 0 | 0 | 0 | 0 | 0 | 0 | 0 | 0 | 0 | 0 | 0 | 0 | 0 |
| C15 | Picturegraph | 0 | 1 | 0 | 0 | 0 | 0 | 0 | 0 | 0 | 0 | 0 | 0 | 0 | 0 | 0 | 2 | 0 | 0 |
| C16 | Line chart | 0 | 0 | 0 | 0 | 0 | 0 | 0 | 0 | 0 | 0 | 0 | 0 | 0 | 0 | 0 | 0 | 0 | 0 |
| C17 | Map chart | 0 | 0 | 0 | 0 | 1 | 0 | 0 | 0 | 0 | 0 | 0 | 0 | 0 | 0 | 0 | 0 | 0 | 0 |
| C18 | Foot section | 1 | 0 | 1 | 0 | 0 | 1 | 1 | 0 | 0 | 1 | 0 | 1 | 1 | 0 | 1 | 0 | 1 | 1 |
| C19 | Foot title text | 0 | 0 | 0 | 0 | 0 | 0 | 1 | 0 | 0 | 0 | 0 | 0 | 0 | 1 | 0 | 0 | 0 | 0 |
| C20 | Foot text | 1 | 0 | 1 | 0 | 0 | 0 | 1 | 0 | 0 | 1 | 0 | 0 | 0 | 1 | 0 | 1 | 0 | 0 |
| C21 | Logo | 1 | 0 | 0 | 0 | 0 | 1 | 1 | 0 | 0 | 0 | 0 | 1 | 1 | 0 | 1 | 0 | 0 | 0 |
| C22 | Link to full report | 0 | 0 | 0 | 0 | 0 | 0 | 1 | 0 | 0 | 1 | 0 | 0 | 0 | 1 | 0 | 0 | 1 | 0 |
| | Sum | 22 | 16 | 23 | 30 | 15 | 43 | 56 | 19 | 13 | 19 | 20 | 19 | 17 | 12 | 33 | 32 | 30 | 16 |
| | Count | 10 | 3 | 9 | 9 | 9 | 11 | 11 | 6 | 7 | 10 | 8 | 10 | 7 | 6 | 12 | 7 | 8 | 7 |

**Table 6.** Definitions of the infographic component types

| Id | Name | Description |
|---|---|---|
| C1 | Head section | The head section is the top part of the infographic often displaying the infographic title, subtitle, introduction text and logo. |
| C2 | Title text | Title text refers to the title of the infographic and the title is often found in the head section. |
| C3 | Subtitle text | Subtitle text refers to the subtitle of the infographic. Often displayed below the title text. |
| C4 | Introduction text | Introduction text refers to text found in the head section, which may spreads over multiple lines. |

| Id | Name | Description |
|---|---|---|
| C5 | Logo | Logo refers to an image representing the logo of the company, which often was present in the head section. |
| C6 | Body section | The body section is the middle part of the infographic displaying the text, data and chart elements. |
| C7 | Box | A box is used to demarcate a certain width and height on the infographic. Within a box, it is possible to display text, data or chart elements. |
| C8 | Box title text | Box title text refers to the title of the box and may not spread over multiple lines. |
| C9 | Body text | Body text refers to the text displayed in a box and is allowed to spread over multiple lines. |
|  | Chart | A chart is a graphical representation of data and multiple forms of charts exist. |
| C10 | Table | An arrangement of facts and figures, usually in rows and columns, that makes information easy to understand. |
| C11 | Pie chart | A pie chart is a circular statistical graph, which is divided into slices to illustrate numeric proportion. |
| C12 | Bar chart | A bar chart is a way of visualising categorical data with rectangular bars with heights/lengths proportional to the value they represent. |
| C13 | Grouped bar chart | A grouped bar chart differs from a normal bar chart since multiple bars are used for each categorical group. |
| C14 | Stacked bar chart | A stacked bar chart differs from a normal bar chart since stacked bar charts are displaying different groups on top of each other. |
| C15 | Picturegraph A | picturegraph represents data by using symbols and pictures. Variations in picturegraphs are unlimited while every symbol or picture may be used for representing data. |
| C16 | Line chart | A line chart is displaying information as a series of data points connected by straight line segments. |
| C17 | Map chart | A map chart allows the visualisation of spatial relationships in data by indicating data on a geographical map. |
| C18 | Foot section | The foot section refers to the bottom part of an infographic often displaying a foot title text, a foot text, logo and a link to the full report. |
| C19 | Foot title text | Foot title text refers to the title of the foot section and is not allowed to spread over multiple lines. |
| C20 | Foot text | Foot text refers to the text displayed in the foot section and is allowed to spread over multiple lines. |
| C21 | Logo | Logo refers to an image representing the logo of the company, which was less often present in the foot section. |
| C22 | Link to full report | Is there an URL present on the infographic linking to the full sustainability report or to additional information. |

## 4 Infographic design tool analysis

This section provides a list of the 10 tools (Table 7). Table 8 describes each of the generic features that we have identified across tools. We also provide the feature model that we have created for each of the tools (Fig. 61 to Fig. 70).

Table **7.** List of the 10 infographic design tools whose features we have analysed

| Name | URL | Description |
|---|---|---|
| BeFunky | https://www.befunky.com | BeFunky is an online and mobile photo editor launched in 2007. Only the online tool is analysed in this research. It enables the user to edit photos, add photo effects, create photo collages and also create infographics. |

| Name | URL | Description |
|---|---|---|
| Canva | https://www.canva.com | Canva is a simplified graphic-design tool website, founded in 2012. It uses a drag-and-drop format and provides the user with an extensive amount of photographs, vectors, graphics and fonts.[1] |
| Easel.ly | https://www.easel.ly | Easel.ly is a web-based infographic platform used for representing and conveying ideas, concepts, reports, processes and projects through visual forms that are easy to follow and inspiring.[2] |
| Infogram | https://infogram.com | Infogram is a web-based data visualization and infographics platform. It allows people to make digital charts, infographics and maps. Infogram offers an intuitive WYSIWYG editor with the possibility to publish, embed or share the created infographics.[3] |
| Lucidpress | https://www.lucidpress.com | Lucidpress is a web-based desktop publishing software application developed by Lucid Software, which developed Lucidchart. It can be used to create flyers, brochures, infographics, posters, magazines and presentations. |
| Mind the Graph | https://mindthegraph.com | Mind the Graph is an infographic maker for scientists, particularly focused on Life Sciences and Health. It's a useful tool to create figures for papers, infographics, slides and presentations for classes.[4] |
| Piktochart | https://piktochart.com | Piktochart is a web-based infographic application which allows users without intensive experience as graphic designers to easily create infographics and visuals using themed templates[5]. |
| Snappa | https://snappa.com | Snappa is a cloud-based graphics editor for social media, personal, and marketing purposes. This makes creating visual content easy without the use of complex tools like those in Photoshop and other similar image editing applications.[6] |
| Venngage | https://venngage.com | Venngage is a free online infographic tool and their goal is to make design more accessible for everyone. Venngage lets users easily drag and drop icons and text widgets directly onto their canvas.[7] |
| Visme | https://www.visme.co | Visme is an intuitive online platform with a mission to allow anyone to easily visualize their ideas into engaging presentations, infographics, animations and more. It's entirely based on HTML5 and runs on any browser with simple easy to use drag and drop functions.[8] |

---

[1] https://en.wikipedia.org/wiki/Canva
[2] https://reviews.financesonline.com/p/easel-ly/
[3] https://en.m.wikipedia.org/wiki/Infogram
[4] https://www.producthunt.com/alternatives/mind-the-graph-2
[5] https://en.wikipedia.org/wiki/Piktochart
[6] https://reviews.financesonline.com/p/snappa/
[7] https://www.youtube.com/user/VenngageInc/about
[8] https://www.linkedin.com/company/visme/about/

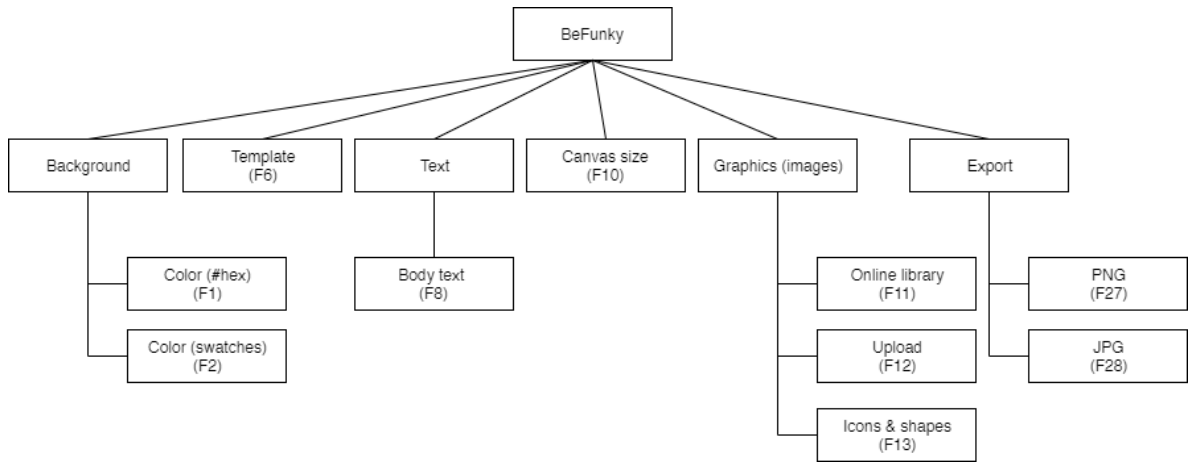

**Fig. 61.** Feature diagram of BeFunky

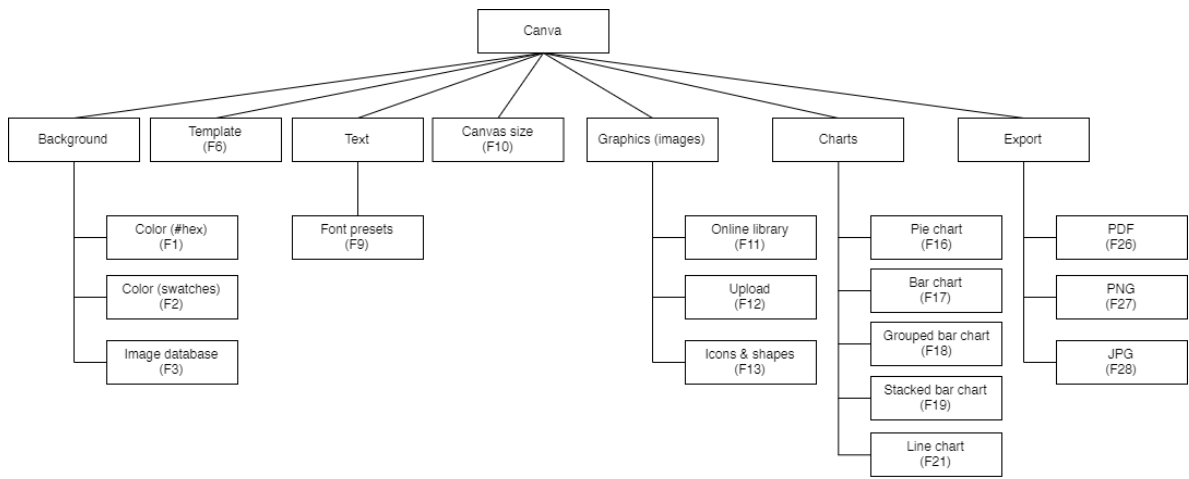

**Fig. 62.** Feature diagram of Canva

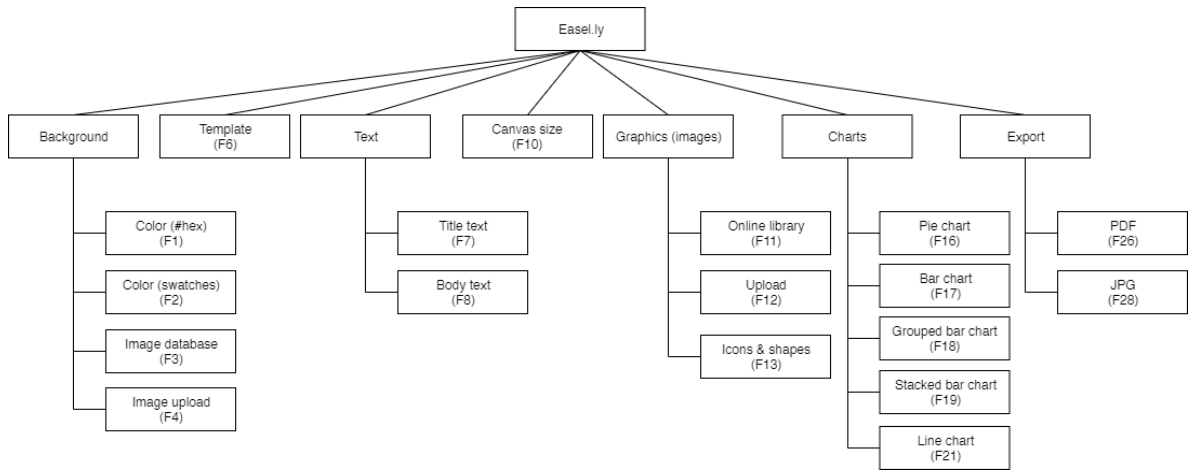

**Fig. 63.** Feature diagram of Easel.ly

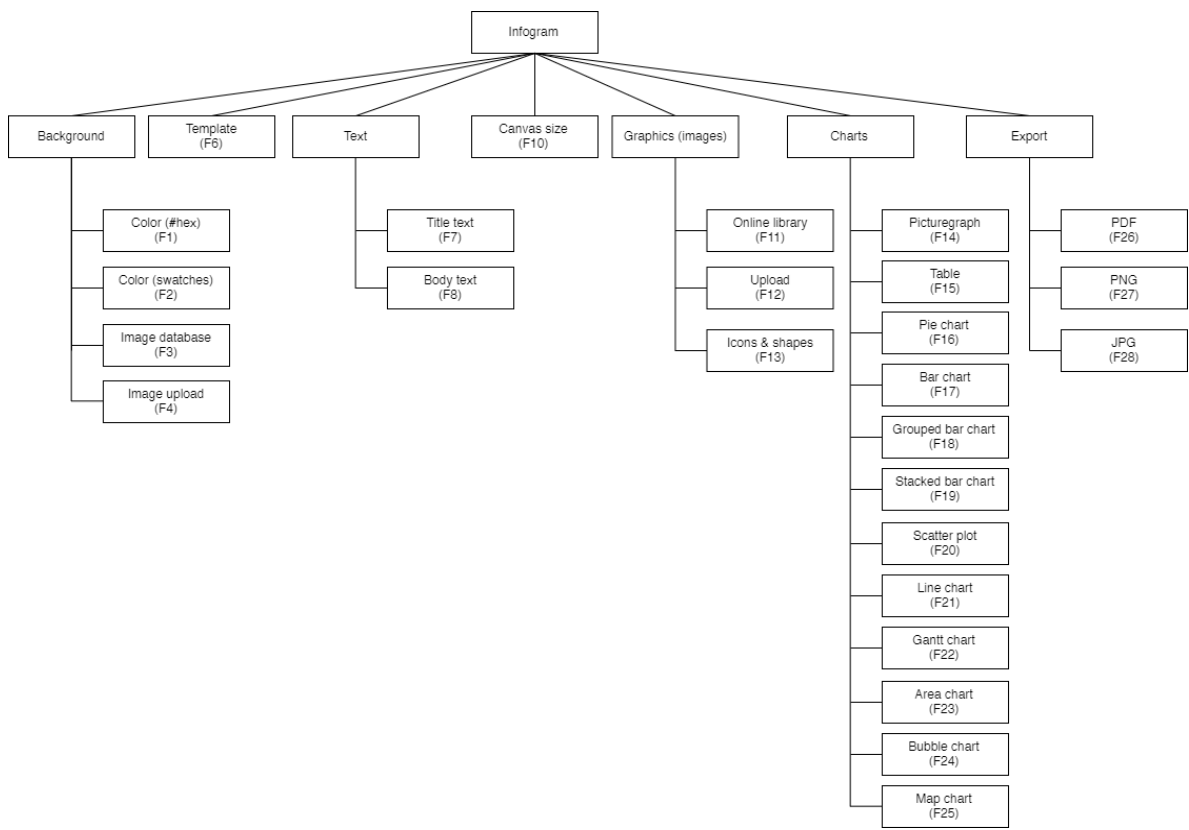

**Fig. 64.** Feature diagram of Infogram

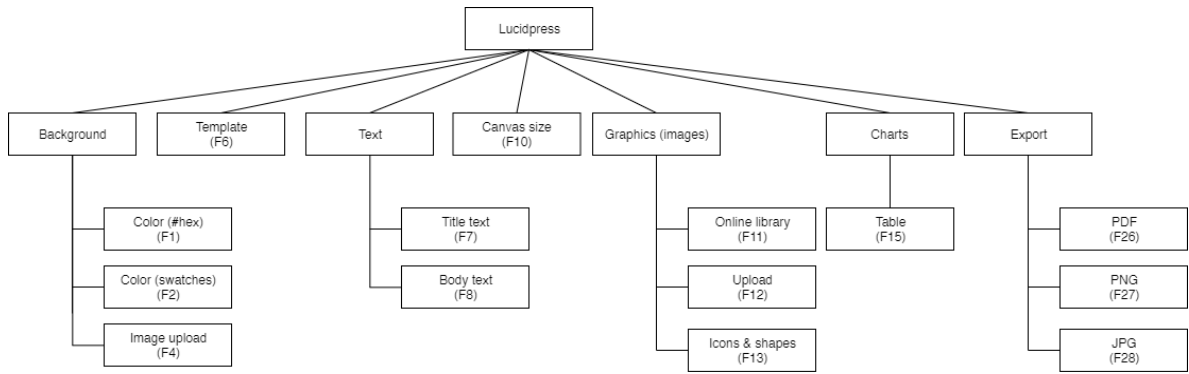

**Fig. 65.** Feature diagram of Lucidpress

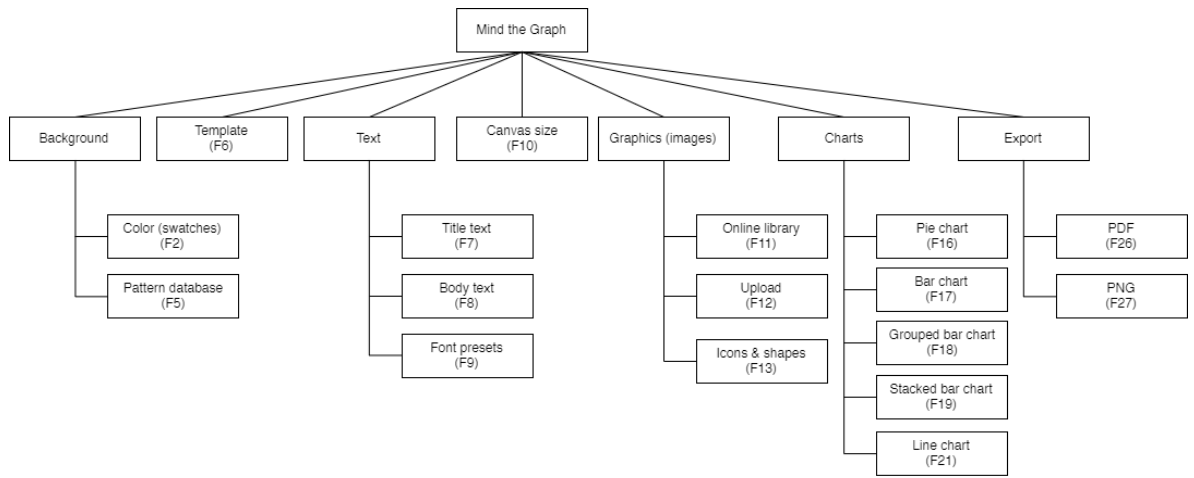

**Fig. 66.** Feature diagram of Mind the Graph

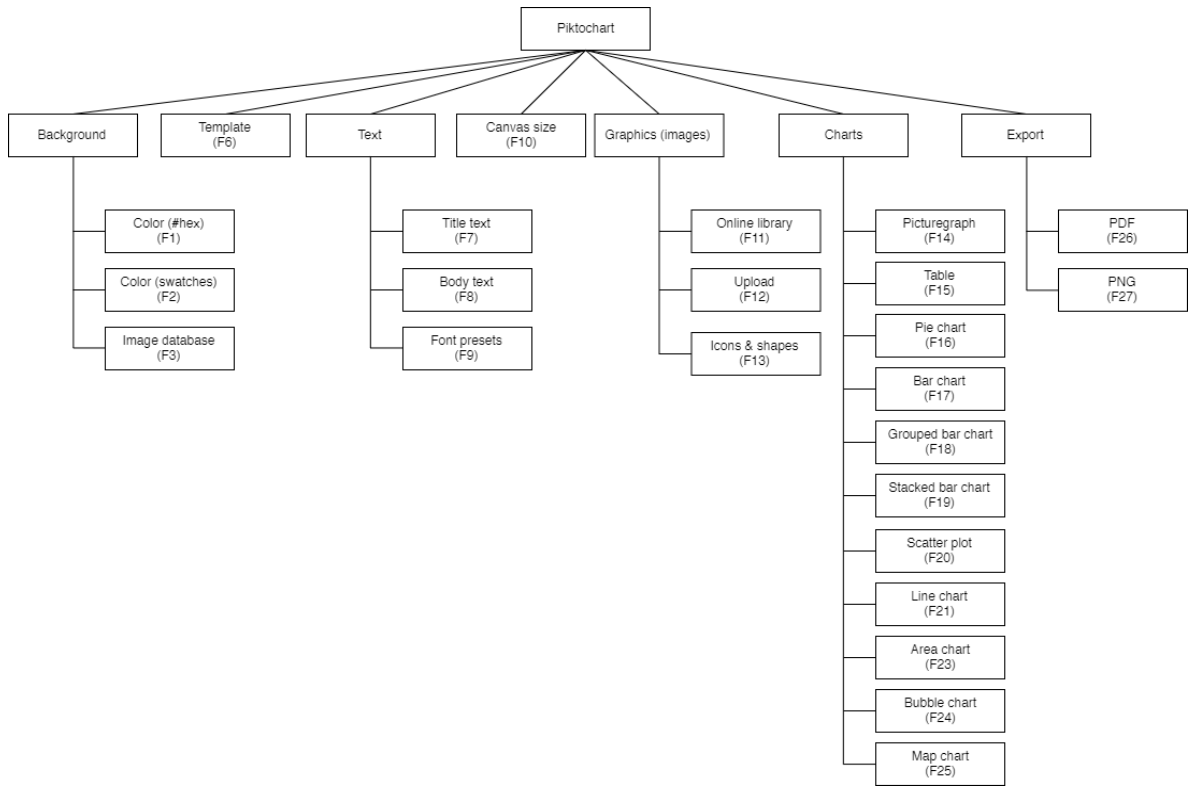

**Fig. 67.** Feature diagram of Piktochart

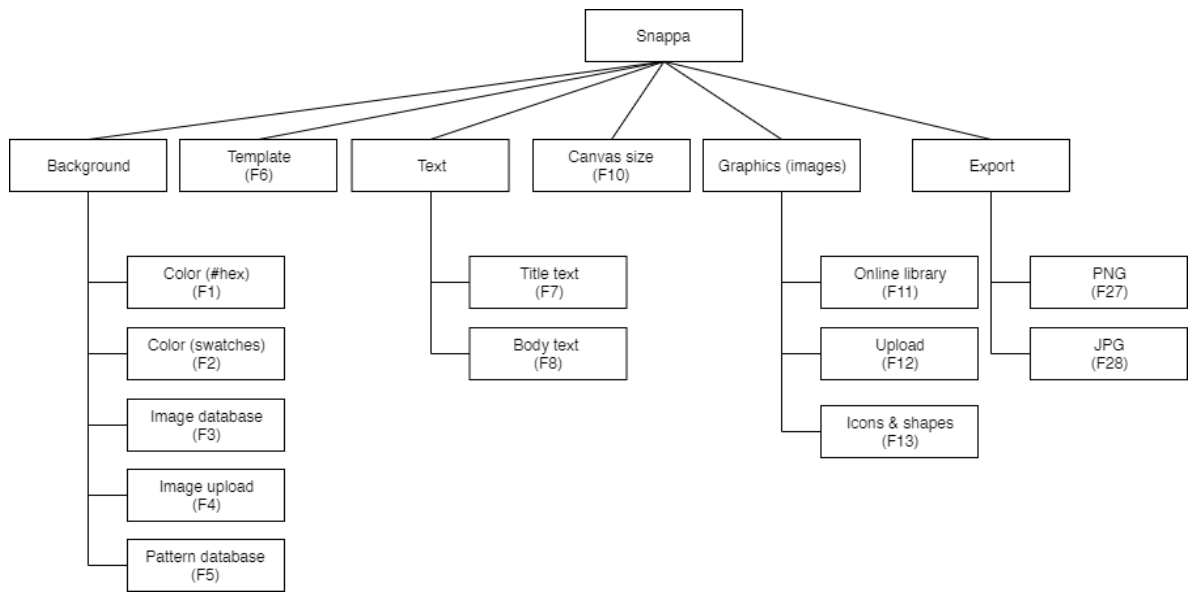

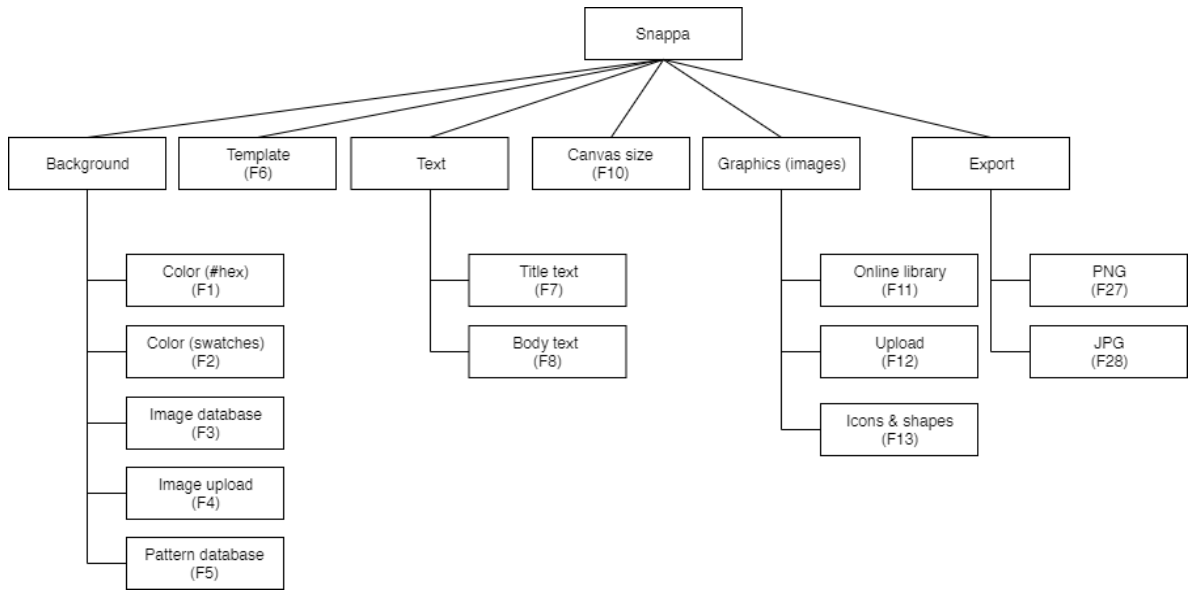

**Fig. 68.** Feature diagram of Snappa

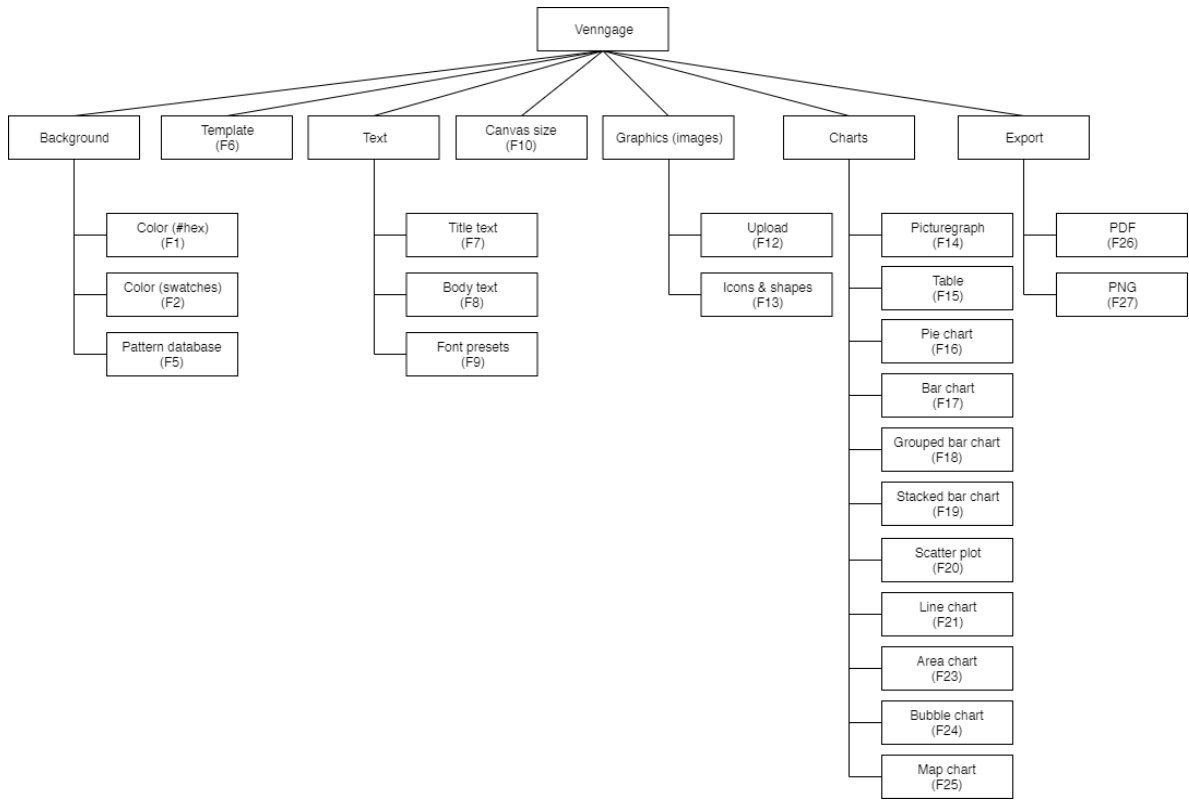

**Fig. 69.** Feature diagram of Venngage

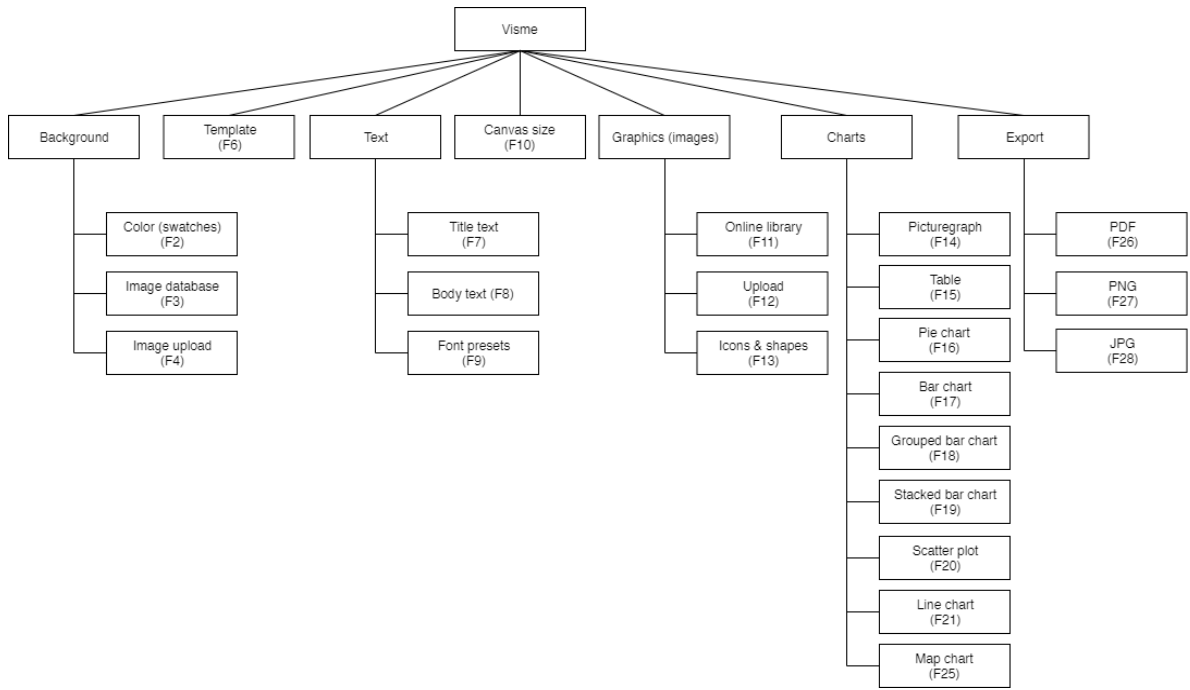

**Fig. 70.** Feature diagram of Visme

**Table 8.** Generic features found across several tools

| Id | Name | Description | Screenshot | NT |
|---|---|---|---|---|
| FG1 | *Background* | The part of the infographic representing what lies behind objects in the foreground. Five different features (F2 to F6) were identified to create the background of an infographic. | 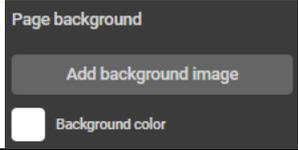 | 10 |
| F1 | Color #hex | The ability to define the background with a color hex code that describes the composition of a certain color in a specific color space, usually RGB. For example, the color hex code for white is #FFFFFF; and at the opposite end is black #000000. | 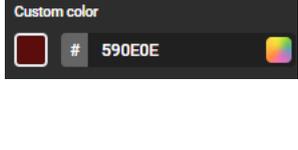 | 8 |
| F2 | Color swatch | The ability to define the background with sample colors, as displayed in the screenshot. | 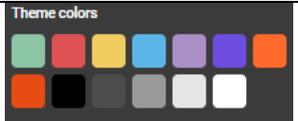 | 10 |

| Id | Name | Description | Screenshot | NT |
|---|---|---|---|---|
| F3 | Image database | The ability to select and use a sample image as the background of the infographic. | 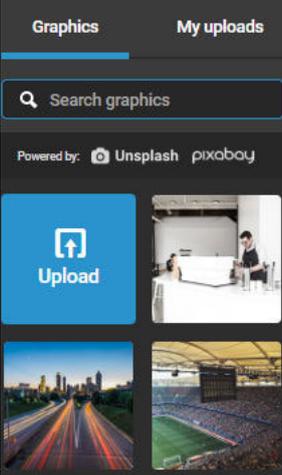 | 6 |
| F4 | Image upload | The ability to use an uploaded image as the background of the infographic. | 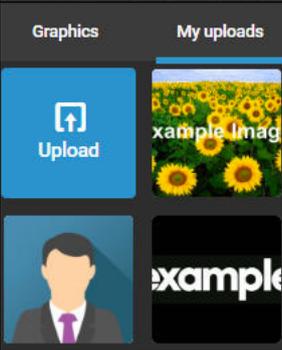 | 5 |
| F5 | Pattern database. | The ability to select a pattern and use this as the background of the infographic | 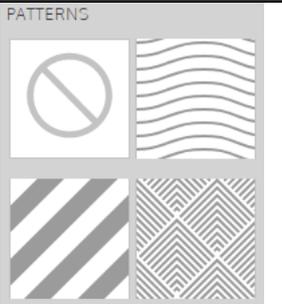 | 3 |

| Id | Name | Description | Screenshot | NT |
|---|---|---|---|---|
| F6 | Template | The ability to select a blank template or a pre-defined template to start with when designing an infographic. | 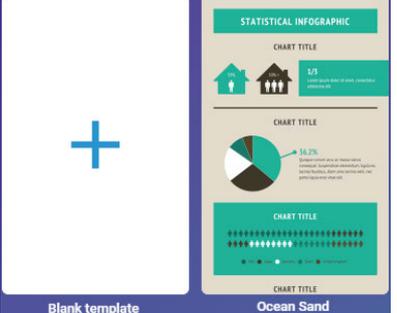 | 10 |
| FG3 | Text | The ability to place a text box and change its settings, such as font, font-size, color, text-box size and alignment among other settings. Three different features (F7 to F9) were identified to insert and use text on the infographic. | 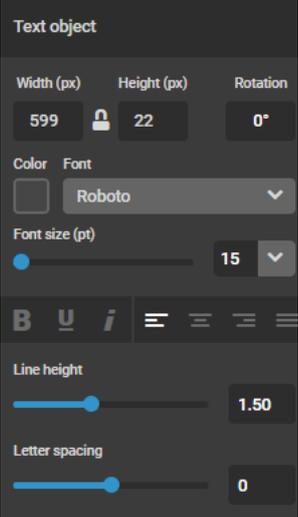 | 10 |
| F7 | Title text | The ability to place a text box with pre-set settings so it will display a title text. | 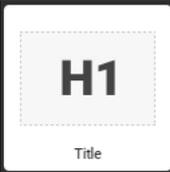 | 8 |
| F8 | Body text | The ability to place a text box with pre-set settings so it will display a body text. | 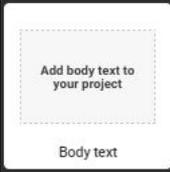 | 9 |
| F9 | Font presets | The ability to place additional different text boxes with pre-set settings other than F9 and F10. | 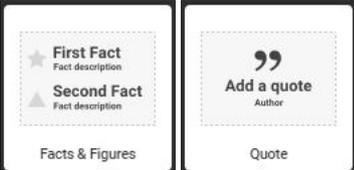 | 6 |

| Id | Name | Description | Screenshot | NT |
|---|---|---|---|---|
| F10 | Canvas size | The ability to adjust the canvas size according to the user's preference. This might be done with pre-set sizes like A3 or A4, or by filling in a number of pixels. | 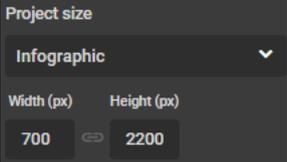 | 10 |
| FG5 | Images | The ability to use images on the infographic. The user is able to resize and rotate the images on the infographic. Three different features (F11 to F13) were identified to select and use images on the infographic. | 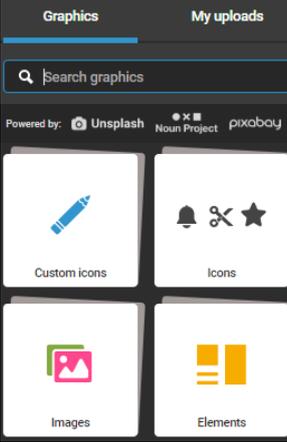 | 10 |
| F11 | Online library | The ability to select an image from a database and use this on the infographic. | 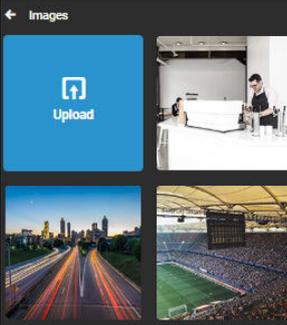 | 10 |
| F12 | Upload | The ability to upload an image from your computer and use this on the infographic. | 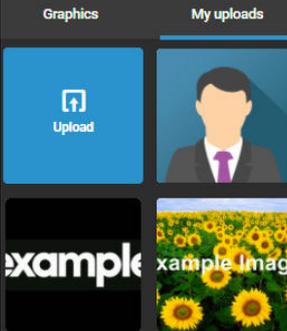 | 10 |

| Id | Name | Description | Screenshot | NT |
|---|---|---|---|---|
| F13 | Icons & shapes | The ability to select an icon or shape from a database and use this on the infographic. | 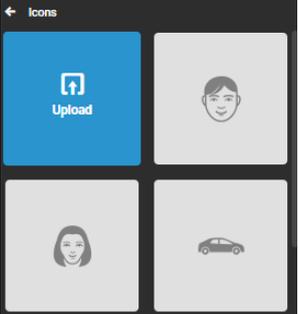 | 10 |
| | Charts | The ability to insert a chart to the infographic. Nine different features (F14 to F25) were identified to insert and use charts on the infographic. This is a feature generalisation and not a group. | 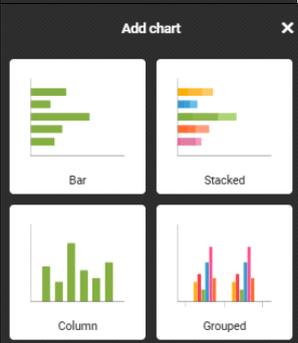 | 8 |
| F14 | Picturegraph | The ability to display picturegraphs on the infographic. This feature may also be called 'pictorial' or 'picture chart'. | 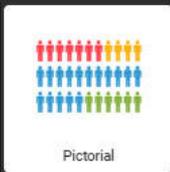 | 4 |
| F15 | Table | The ability to display tables on the infographic. | 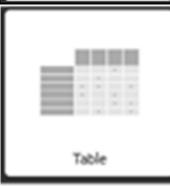 | 5 |
| F16 | Pie chart | The ability to display pie charts on the infographic. | 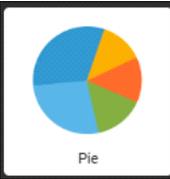 | 7 |
| F17 | Bar chart | The ability to display bar charts on the infographic. | 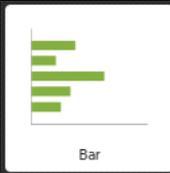 | 7 |

| Id | Name | Description | Screenshot | NT |
|---|---|---|---|---|
| F18 | Grouped bar chart | The ability to display grouped bar charts on the infographic. | 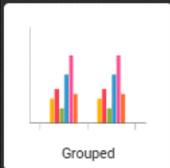 | 7 |
| F19 | Stacked bar chart | The ability to display stacked bar charts on the infographic. | 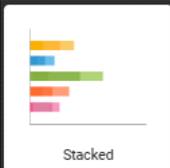 | 7 |
| F20 | Scatterplot | The ability to display scatterplots on the infographic. | 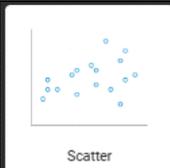 | 4 |
| F21 | Line chart | The ability to display line charts on the infographic. | 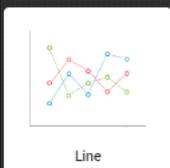 | 7 |
| F22 | Gantt chart | The ability to display Gantt charts on the infographic. | 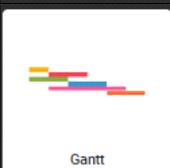 | 1 |
| F23 | Area chart | The ability to display area charts on the infographic. | 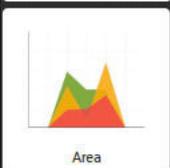 | 3 |
| F24 | Bubble chart | The ability to display bubble charts on the infographic. | 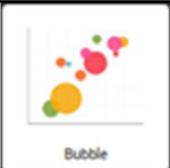 | 3 |
| F25 | Map chart | The ability to display map charts on the infographic. | 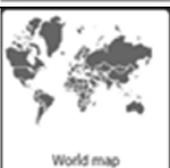 | 4 |

| Id | Name | Description | Screenshot | NT |
|---|---|---|---|---|
| FG7 | Export | The ability to export the infographic from the tool so you will obtain a downloadable copy. This feature may also be called 'publish' or 'download'. Three different features (F26 to F28) were identified to export the infographic. | 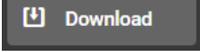 | 10 |
| F26 | PDF | The ability to export the infographic from the tool to a PDF format. | 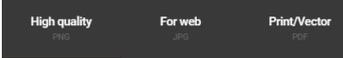 | 8 |
| F27 | PNG | The ability to export the infographic from the tool to a PNG format. | 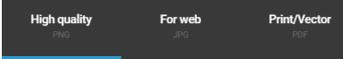 | 9 |
| F28 | JPG | The ability to export the infographic from the tool to a JPG format. | 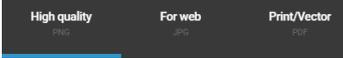 | 7 |

# 5 Sample of 10 infographics

We have sampled 10 of the infographics analysed in Section 3. For each, we have created a model and generated the infographic automatically using the openESEA tool. This has allowed us to test the DSL and the interpreter. This section presents each pair of infographics in the sample; that is the original (at the left in Fig. 71 to Fig. 80) and the generated infographic (at the right of those figures). We also present the models that specify the generated infographics using the DSL, in Table 9 to Table 18. To facilitate the independent testing of our tool by third parties, we have encoded the data with its real values, instead of referencing indicators from an ESEA method.

## 5.1 Alcoa

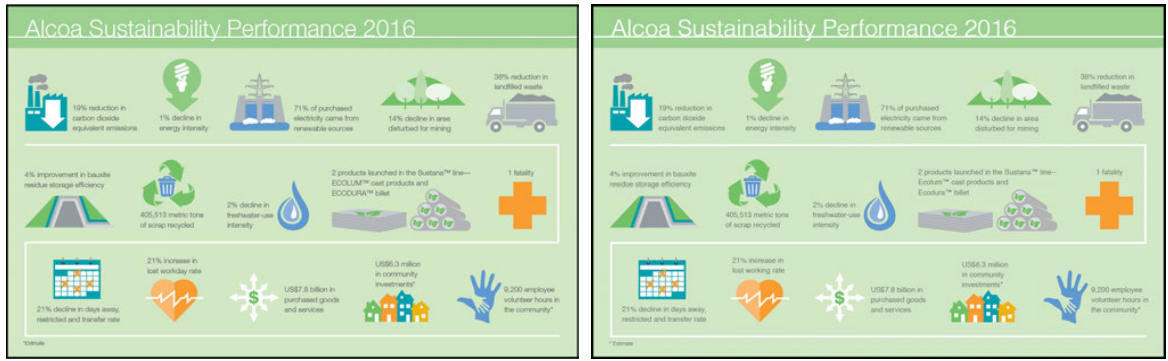

**Fig. 71.** Original infographic by Alcoa (left) and its corresponding generated infographic (right), using the specification in Table 9.

**Table 9.** Specification of the infografic shown at the right of Fig. 71

```
bgcolor: black
bgsize: 1200x747
foot: off
head:
```

```
  position: 5x5
  size: 1190x87
  bgcolor: 9dd191
  title:
    font: 49px Helvetica Light
    value: Alcoa Sustainability Performance 2016
    position: 42x67
    color: white
    maxwidth: 1200
  subtitle: off
box1:
  bgcolor: d1e7c6
  position: 5x92
  size: 1190x650
box2:
  position: 5x67
  size: 1190x3
  bgcolor: white
box3:
  size: 1119x4
  position: 40x295
  bgcolor: white
box4:
  size: 4x201
  position: 1155x299
  bgcolor: white
box5:
  size: 1119x4
  position: 40x500
  bgcolor: white
box6:
  size: 4x197
  position: 40x504
  bgcolor: white
box7:
  size: 1119x4
  position: 40x697
  bgcolor: white
image1:
  size: 89x122
  position: 40x145
  src: "https://i.imgur.com/m5XwK2f.png"
image2:
  size: 81x113
  position: 328x114
  src: "https://i.imgur.com/FkvpI9u.png"
image3:
  size: 127x125
  position: 468x143
  src: "https://i.imgur.com/8ay9Uid.png"
image4:
  size: 182x85
  position: 783x137
  src: "https://i.imgur.com/CDD1JcA.png"
image5:
  size: 152x85
  position: 1004x185
  src: "https://i.imgur.com/M5qB8XL.png"
image6:
  size: 188x70
  position: 38x402
  src: "https://i.imgur.com/GOY0wXy.png"
image7:
  size: 113x104
  position: 284x326
  src: "https://i.imgur.com/kk2wx0Y.png"
image8:
  size: 69x108
  position: 568x369
  src: "https://i.imgur.com/8ityt1l.png"
image9:
  size: 294x92
  position: 681x388
  src: "https://i.imgur.com/7smmmGS.png"
image10:
  size: 102x103
```

```
    position: 1031x372
    src: "https://i.imgur.com/sAL59lZ.png"
image11:
    size: 103x92
    position: 97x531
    src: "https://i.imgur.com/JE2aN8c.png"
image12:
    size: 123x92
    position: 291x578
    src: "https://i.imgur.com/0WLVYVJ.png"
image13:
    size: 106x105
    position: 466x565
    src: "https://i.imgur.com/3bbjgNf.png"
image14:
    size: 141x71
    position: 748x600
    src: "https://i.imgur.com/QJ9qPUp.png"
image15:
    size: 94x114
    position: 944x559
    src: "https://i.imgur.com/LxudDi4.png"
text1:
    font: 15px Helvetica Regular
    color: 6a766a
    position: 141x224
    lineheight: 21
    maxwidth: 145
    value: "19% reduction in carbon dioxide equivalent emissions"
text2:
    font: 15px Helvetica Regular
    color: 6a766a
    position: 323x245
    lineheight: 21
    maxwidth: 120
    value: "1% decline in energy intensity"
text3:
    font: 15px Helvetica Regular
    color: 6a766a
    position: 605x224
    lineheight: 21
    maxwidth: 145
    value: "71% of purchased electricity came from renewable sources"
text4:
    font: 15px Helvetica Regular
    color: 6a766a
    position: 802x246
    lineheight: 21
    maxwidth: 145
    value: "14% decline in area disturbed for mining"
text5:
    font: 15px Helvetica Regular
    color: 6a766a
    position: 1022x158
    lineheight: 21
    maxwidth: 145
    value: "38% reduction in landfilled waste"
text6:
    font: 15px Helvetica Regular
    color: 6a766a
    position: 39x363
    lineheight: 21
    maxwidth: 195
    value: "4% improvement in bauxite residue storage efficiency"
text7:
    font: 15px Helvetica Regular
    color: 6a766a
    position: 281x448
    lineheight: 21
    maxwidth: 142
    value: "405,513 metric tons of scrap recycled"
text8:
    font: 15px Helvetica Regular
    color: 6a766a
    position: 463x427
    lineheight: 21
```

```
      maxwidth: 100
      value: "2% decline in freshwater-use intensity"
    text9:
      font: 15px Helvetica Regular
      color: 6a766a
      position: 683x359
      lineheight: 21
      maxwidth: 295
      value: "2 products launched in the Sustana,Ñ¢ line-- Ecolum,Ñ¢ cast products and           Ecodura,Ñ¢
billet"
    text10:
      font: 15px Helvetica Regular
      color: 6a766a
      position: 1054x359
      lineheight: 21
      maxwidth: 145
      value: "1 fatality"
    text11:
      font: 15px Helvetica Regular
      color: 6a766a
      position: 65x647
      lineheight: 21
      maxwidth: 185
      value: "21% decline in days away, restricted and transfer rate"
    text12:
      font: 15px Helvetica Regular
      color: 6a766a
      position: 295x544
      lineheight: 21
      maxwidth: 120
      value: "21% increase in lost working rate"
    text13:
      font: 15px Helvetica Regular
      color: 6a766a
      position: 584x606
      lineheight: 21
      maxwidth: 145
      value: "US$7.8 billion in purchased goods and services"
    text14:
      font: 15px Helvetica Regular
      color: 6a766a
      position: 774x553
      lineheight: 21
      maxwidth: 105
      value: "US$6.3 million in community investments*"
    text15:
      font: 15px Helvetica Regular
      color: 6a766a
      position: 1044x606
      lineheight: 21
      maxwidth: 125
      value: "9,200 employee volunteer hours in the community*"
    text16:
      font: 11px Helvetica Regular
      color: 6a766a
      position: 38x718
      value: "* Estimate"
```

## 5.2 Autodesk

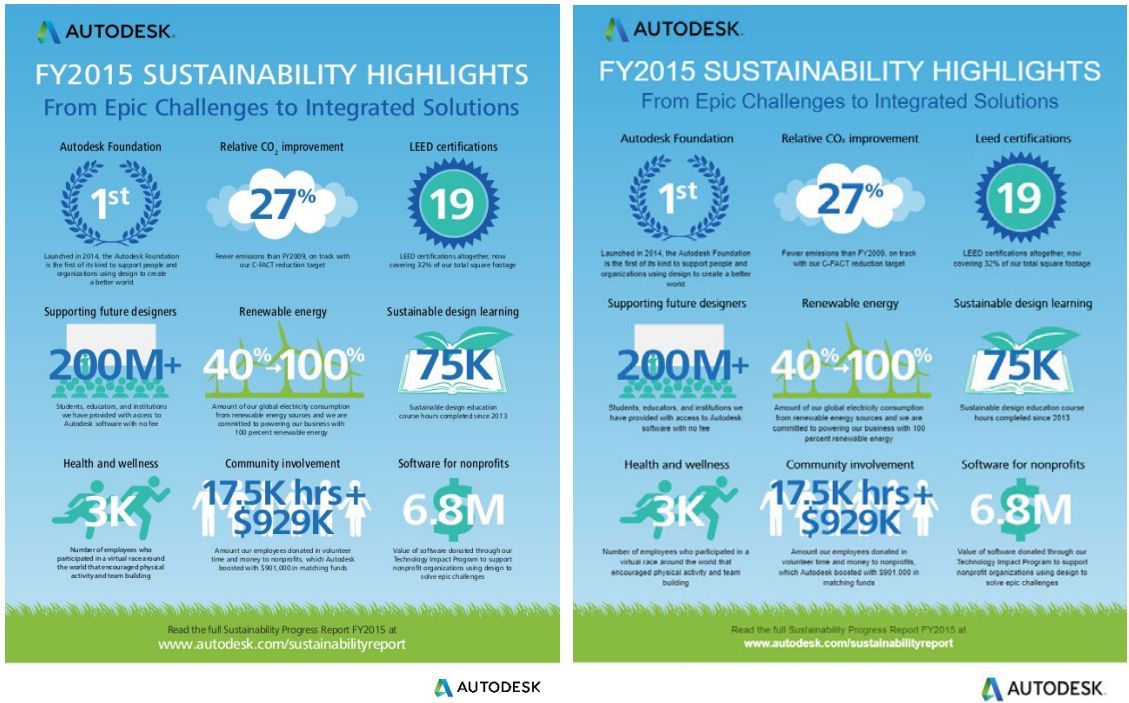

**Fig. 72.** Original infographic by Autodesk (left) and its corresponding generated infographic (right), using the specification in Table 10

**Table 10.** Specification of the infografic shown at the right of Fig. 72

```
bgpattern: "https://i.imgur.com/03juVGq.png"
bgsize: 800x1036
head:
  bgcolor: off
  title:
    font: 40px sans-serif
    value: "FY2015 SUSTAINABILITY HIGHLIGHTS"
    color: ffffff
    position: 400x110
    maxwidth: 800
    align: center
  subtitle:
    font: 30px sans-serif
    value: "From Epic Challenges to Integrated Solutions"
    color: 1557a2
    position: 400x150
    maxwidth: 800
    align: center
image1:
  size: 200x35
  position: 46x18
  src: "https://upload.wikimedia.org/wikipedia/commons/thumb/b/b5/Autodesk_Logo.svg/1200px-Autodesk_Logo.svg.png"
image2:
  size: 150x130
  position: 79x216
  src: "https://i.imgur.com/Ln2Z0F1.png"
```

```
image3:
  size: 222x107
  position: 289x228
  src: "https://i.imgur.com/FwS1wZm.png"
image4:
  size: 138x138
  position: 578x211
  src: "https://i.imgur.com/1wtn1s6.png"
image5:
  size: 193x112
  position: 63x463
  src: "https://i.imgur.com/lRwGKXo.png"
image6:
  size: 237x118
  position: 283x459
  src: "https://i.imgur.com/SEfIqus.png"
image7:
  size: 175x104
  position: 562x466
  src: "https://i.imgur.com/vtIHDlV.png"
image8:
  size: 176x96
  position: 65x685
  src: "https://i.imgur.com/jzcalRT.png"
image9:
  size: 264x94
  position: 267x686
  src: "https://i.imgur.com/t1YnCPa.png"
image10:
  size: 149x98
  position: 572x686
  src: "https://i.imgur.com/X1noK4q.png"
image11:
  size: 180x35
  position: 590x979
  src: "https://upload.wikimedia.org/wikipedia/commons/thumb/b/b5/Autodesk_Logo.svg/1200px-Autodesk_Logo.svg.png"
foot: off
titletext1:
  font: bold 18px FrutigerNextLT
  position: 150x200
  color: black
  maxwidth: 200
  value: "Autodesk Foundation"
  align: center
titletext2:
  font: bold 18px FrutigerNextLT
  position: 400x200
  color: black
  maxwidth: 200
  value: "Relative CO‚ÇÇ improvement"
  align: center
titletext3:
  font: bold 18px FrutigerNextLT
  position: 650x200
  color: black
  maxwidth: 200
  value: "Leed certifications"
  align: center
titletext4:
  font: bold 18px FrutigerNextLT
  position: 150x440
  color: black
  maxwidth: 200
  value: "Supporting future designers"
  align: center
titletext5:
  font: bold 18px FrutigerNextLT
  position: 400x440
  color: black
  maxwidth: 200
  value: "Renewable energy"
  align: center
titletext6:
  font: bold 18px FrutigerNextLT
  position: 650x440
```

```
  color: black
  maxwidth: 200
  value: "Sustainable design learning"
  align: center
titletext7:
  font: bold 18px FrutigerNextLT
  position: 150x675
  color: black
  maxwidth: 200
  value: "Health and wellness"
  align: center
titletext8:
  font: bold 18px FrutigerNextLT
  position: 400x675
  color: black
  maxwidth: 200
  value: "Community involvement"
  align: center
titletext9:
  font: bold 18px FrutigerNextLT
  position: 650x675
  color: black
  maxwidth: 200
  value: "Software for nonprofits"
  align: center
text1:
  font: 11px sans-serif
  position: 150x365
  value: "Launched in 2014, the Autodesk Foundation is the first of its kind to support people and organizations using design to create a better world"
  maxwidth: 222
  align: center
  lineheight: 15
text2:
  font: 11px sans-serif
  position: 400x365
  value: "Fewer emissions than FY2009, on track with our C-FACT reduction target"
  maxwidth: 215
  align: center
  lineheight: 15
text3:
  font: 11px sans-serif
  position: 650x365
  value: "LEED certifications altogether, now covering 32% of our total square footage"
  maxwidth: 215
  align: center
  lineheight: 15
text4:
  font: 11px sans-serif
  position: 150x590
  value: "Students, educators, and institutions we have provided with access to Autodesk software with no fee"
  maxwidth: 215
  align: center
  lineheight: 15
text5:
  font: 11px sans-serif
  position: 400x590
  value: "Amount of our global electricity consumption from renewable energy sources and we are committed to powering our business with 100 percent renewable energy"
  maxwidth: 225
  align: center
  lineheight: 15
text6:
  font: 11px sans-serif
  position: 650x590
  value: "Sustainable design education course hours completed since 2013"
  maxwidth: 200
  align: center
  lineheight: 15
text7:
  font: 11px sans-serif
  position: 150x800
  value: "Number of employees who participated in a virtual race around the world that encouraged physical activity and team building"
  maxwidth: 215
  align: center
```

```
    lineheight: 15
text8:
  font: 11px sans-serif
  position: 400x800
  value: "Amount our employees donated in volunteer time and money to nonprofits, which Autodesk boosted with $901,000 in matching funds"
  maxwidth: 215
  align: center
  lineheight: 15
text9:
  font: 11px sans-serif
  position: 650x800
  value: "Value of software donated through our Technology Impact Program to support nonprofit organizations using design to solve epic challenges"
  maxwidth: 215
  align: center
  lineheight: 15
text10:
  font: 14px sans-serif
  position: 400x915
  color: 3a4416
  value: "Read the full Sustainability Progress Report FY2015 at"
  maxwidth: 600
  align: center
text11:
  font: bold 16px sans-serif
  position: 400x935
  color: ffffff
  value: "www.autodesk.com/sustainabilityreport"
  maxwidth: 600
  align: center
```

## 5.3 CookCounty

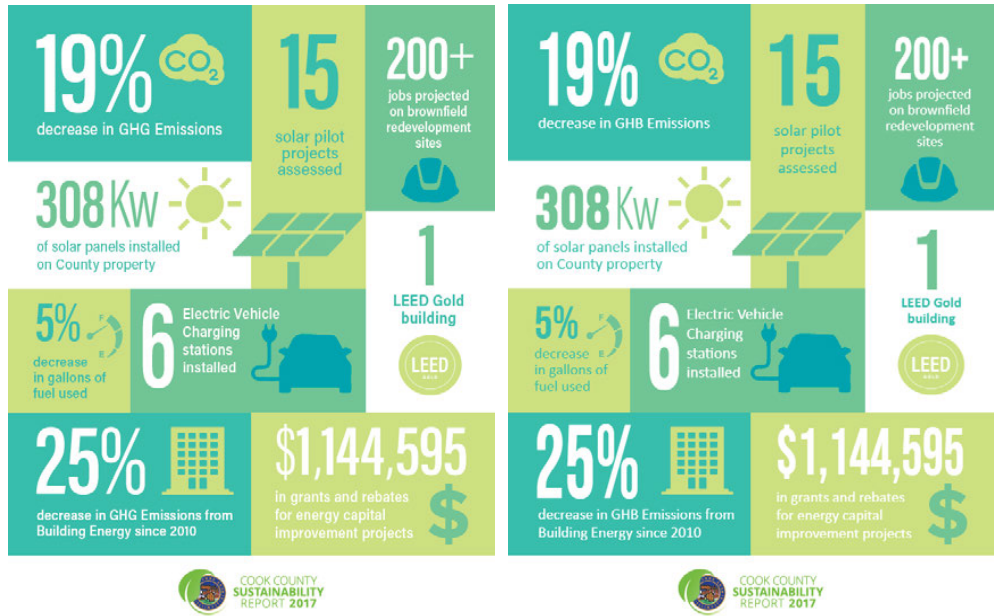

**Fig. 73.** Original infographic by CookCounty (left) and its corresponding generated infographic (right), using the specification in Table 11

**Table 11.** Specification of the infografic shown at the right of Fig. 73

```
bgcolor: ffffff
bgsize: 645x834
head: off
foot: off
box1:
  size: 322x209
  position: "0x0"
  bgcolor: 38beac
box2:
  size: 152x379
  position: 322x0
  bgcolor: cde081
box3:
  size: 171x275
  position: 474x0
  bgcolor: 71c596
box4:
  size: 162x166
  position: "0x379"
  bgcolor: cde081
box5:
  size: 312x166
  position: 162x379
  bgcolor: 71c596
box6:
  size: 322x190
  position: "0x545"
  bgcolor: 38beac
box7:
  size: 323x190
  position: 322x545
```

```
  bgcolor: cde081
image1:
  size: 89x71
  position: 199x37
  src: "https://i.imgur.com/vTpBEP4.png"
image2:
  size: 79x63
  position: 520x202
  src: "https://i.imgur.com/aG3ufHe.png"
image3:
  size: 257x162
  position: 213x217
  src: "https://i.imgur.com/chNEEUI.png"
image4:
  size: 50x60
  position: 103x411
  src: "https://i.imgur.com/WVNCbQT.png"
image5:
  size: 135x93
  position: 322x426
  src: "https://i.imgur.com/8rrZG3t.png"
image6:
  size: 92x84
  position: 516x439
  src: "https://i.imgur.com/1SN2w2W.png"
image7:
  size: 97x92
  position: 203x562
  src: "https://i.imgur.com/iwCQ1Zj.png"
image8:
  size: 55x82
  position: 556x639
  src: "https://i.imgur.com/xFjkdd5.png"
image9:
  size: 194x64
  position: 226x752
  src: "https://i.imgur.com/0GeGqKN.png"
titletext1:
  position: 35x135
  font: 140px Bebas Neue
  maxwidth: 140
  value: "19%"
  color: white
text1:
  position: 40x165
  font: 21px Calibri
  color: white
  value: "decrease in GHB Emissions"
titletext2:
  position: 35x300
  font: bold 80px Bebas Neue
  maxwidth: 308
  value: "308"
  color: 71c596
titletext3:
  position: 141x300
  font: 100 90px Calibri
  maxwidth: 60
  value: "Kw"
  color: 71c596
text2:
  position: 38x330
  font: 21px Calibri
  color: 38beac
  maxwidth: 220
  lineheight: 23
  value: "of solar panels installed on County property"
titletext4:
  position: 337x135
  font: 135px Bebas Neue
  maxwidth: 140
  value: "15"
  color: 38beac
text3:
  position: 400x175
  font: 22px Calibri
```

```
    color: 38beac
    align: center
    maxwidth: 140
    lineheight: 25
    value: "solar pilot projects assessed"
titletext5:
    position: 562x100
    font: 80px Bebas Neue
    maxwidth: 100
    value: "200+"
    align: center
    color: white
text4:
    position: 562x126
    font: 19px Calibri
    color: white
    align: center
    maxwidth: 130
    lineheight: 21
    value: "jobs projected on brownfield redevelopment sites"
titletext6:
    position: 557x381
    font: 115px Bebas Neue
    maxwidth: 80
    value: "1"
    align: center
    color: 71c596
text5:
    position: 564x404
    font: bold 19px Calibri
    color: 38beac
    align: center
    maxwidth: 100
    lineheight: 21
    value: "LEED Gold building"
titletext7:
    position: 35x445
    font: 60px Bebas Neue
    value: "5%"
    color: 38beac
text6:
    position: 35x472
    font: 20px Calibri
    color: 38beac
    maxwidth: 110
    lineheight: 22
    value: "decrease    in gallons of fuel used"
titletext8:
    position: 184x513
    font: 160px Bebas Neue
    value: "6"
    maxwidth: 46
    color: white
text7:
    position: 237x420
    font: 21px Calibri
    color: white
    maxwidth: 145
    lineheight: 26
    value: "Electric Vehicle Charging stations installed"
titletext9:
    position: 35x659
    font: 140px Bebas Neue
    maxwidth: 140
    value: "25%"
    color: white
text8:
    position: 40x687
    font: 20px Calibri
    color: white
    maxwidth: 280
    lineheight: 24
    value: "decrease in GHB Emissions from Building Energy since 2010"
titletext10:
    position: 356x628
    font: 90px Bebas Neue
```

```
    maxwidth: 250
    value: "$1,144,595"
    color: white
text9:
    position: 356x664
    font: 20px Calibri
    color: white
    maxwidth: 200
    lineheight: 24
    value: "in grants and rebates for energy capital improvement projects"
```

## 5.4 Crocs

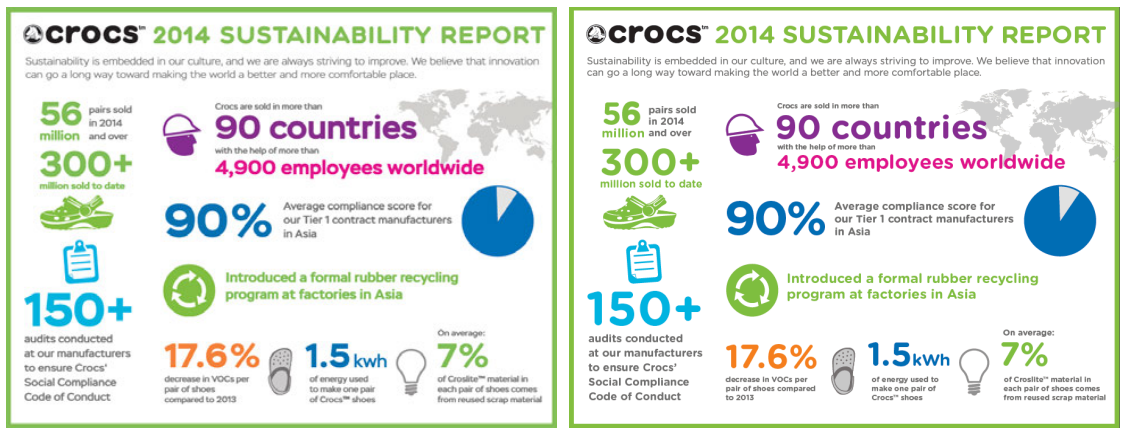

**Fig. 74.** Original infographic by Crocs (left) and its corresponding generated infographic (right), using the specification in Table 12

**Table 12.** Specification of the infografic shown at the right of Fig. 74

```
bgcolor: white
bgsize: 904x697
head: off
foot: off
box1:
    size: 8x697
    position: "0x0"
    bgcolor: 83bf56
box2:
    size: 8x697
    position: "896x0"
    bgcolor: 83bf56
box3:
    size: 896x8
    position: "0x0"
    bgcolor: 83bf56
box4:
    size: 904x8
    position: "0x689"
    bgcolor: 83bf56
image1:
    size: 201x32
    src: "https://i.imgur.com/XTciBlO.png"
    position: 28x29
image2:
    size: 122x148
    src: "https://i.imgur.com/Xf124Jq.png"
    position: 56x311
image3:
```

```
    size: 64x71
    src: "https://i.imgur.com/gYwhwXA.png"
    position: 257x176
image4:
    size: 86x87
    src: "https://i.imgur.com/hfGMT33.png"
    position: 258x425
image5:
    size: 261x279
    src: "https://i.imgur.com/Xy04KFa.png"
    position: 635x134
image6:
    size: 42x79
    src: "https://i.imgur.com/DWEvSaP.png"
    position: 429x564
image7:
    size: 50x79
    src: "https://i.imgur.com/t2nrr8f.png"
    position: 640x565
titletext1:
    position: 240x60
    font: 40px CentralW01-Bold
    value: "2014 SUSTAINABILITY REPORT"
    color: 83bf56
    maxwidth: 650
text1:
    position: 32x95
    value: "Sustainability is embedded in our culture, and we are always striving to improve. We believe that innovation can go a long way toward making the world a better and more comfortable place."
    color: 717271
    font: 17px Stem ☞ Light
    lineheight: 22
    maxwidth: 850
text2:
    color: 7fbe56
    position: 55x195
    value: "56"
    font: 55px Stem ☞ Heavy
text3:
    color: 7fbe56
    position: 55x220
    value: "million"
    font: 22px Stem ☞ Heavy
text4:
    color: 717271
    position: 136x175
    value: "pairs sold in 2014 and over"
    font: 16px Stem ☞ Regular
    maxwidth: 80
    lineheight: 22
text5:
    color: 7fbe56
    position: 55x280
    value: "300+"
    font: 60px Stem ☞ Heavy
text6:
    color: 7fbe56
    position: 55x301
    value: "million sold to date"
    font: 16px Stem ☞ Heavy
    lineheight: 22
text7:
    color: 02b4df
    position: 30x528
    value: "150+"
    font: 78px Stem ☞ Heavy
text8:
    color: 717271
    position: 30x555
    value: "audits conducted     at our manufacturers to ensure Crocs' Social Compliance Code of Conduct"
    font: 18px Stem ☞ Regular
    lineheight: 24
    maxwidth: 180
text9:
```

```
  color: 717271
  position: 343x169
  value: "Crocs are sold in more than"
  font: 14px Stem ☞ Regular
titletext2:
  color: 872d8c
  position: 343x218
  value: "90 countries"
  font: 56px Stem ☞ Heavy
text10:
  color: 717271
  position: 343x242
  value: "with the help of more than "
  font: 14px Stem ☞ Regular
titletext3:
  color: e4138b
  position: 343x277
  value: "4,900 employees worldwide"
  font: 34px Stem ☞ Heavy
  maxwidth: 500
titletext4:
  color: 0672b3
  position: 260x380
  value: "90"
  font: 78px Stem ☞ Heavy
titletext5:
  color: 0672b3
  position: 368x380
  value: "%"
  font: 78px Stem ☞ Regular
text11:
  color: 717271
  position: 455x335
  value: "Average compliance score for      our Tier 1 contract manufacturers in Asia"
  font: 18px Stem ☞ Regular
  lineheight: 23
  maxwidth: 285
text12:
  color: 7fbe56
  position: 360x455
  value: "Introduced a formal rubber recycling program at factories in Asia"
  font: 23px Stem ☞ Heavy
  lineheight: 28
  maxwidth: 413
titletext6:
  color: f5813a
  position: 260x596
  value: "17.6"
  font: 56px Stem ☞ Heavy
titletext7:
  color: f5813a
  position: 367x596
  value: "%"
  font: 56px Stem ☞ Regular
text13:
  color: 717271
  position: 260x620
  value: "decrease in VOCs per pair of shoes compared to 2013"
  font: 14px Stem ☞ Regular
  lineheight: 17
  maxwidth: 150
titletext8:
  color: 0672b3
  position: 491x596
  value: "1.5"
  font: 56px Stem ☞ Heavy
titletext9:
  color: 0672b3
  position: 570x596
  value: "kwh"
  font: 30px Stem ☞ Regular
text14:
  color: 717271
```

```
      position: 498x620
      value: "of energy used to make one pair of Crocs™ shoes"
      font: 14px Stem ☞ Regular
      lineheight: 17
      maxwidth: 115
    text15:
      color: 717271
      position: 708x544
      value: "On average:"
      font: 14px Stem ☞ Regular
      lineheight: 17
      maxwidth: 175
    titletext10:
      color: 7fbe56
      position: 708x596
      value: "7"
      font: 56px Stem ☞ Heavy
    titletext11:
      color: 7fbe56
      position: 742x596
      value: "%"
      font: 56px Stem ☞ Regular
    text16:
      color: 717271
      position: 708x620
      value: "of Croslite™ material in each pair of shoes comes from reused scrap material"
      font: 14px Stem ☞ Regular
      lineheight: 17
      maxwidth: 175
```

## 5.5 First Horizon

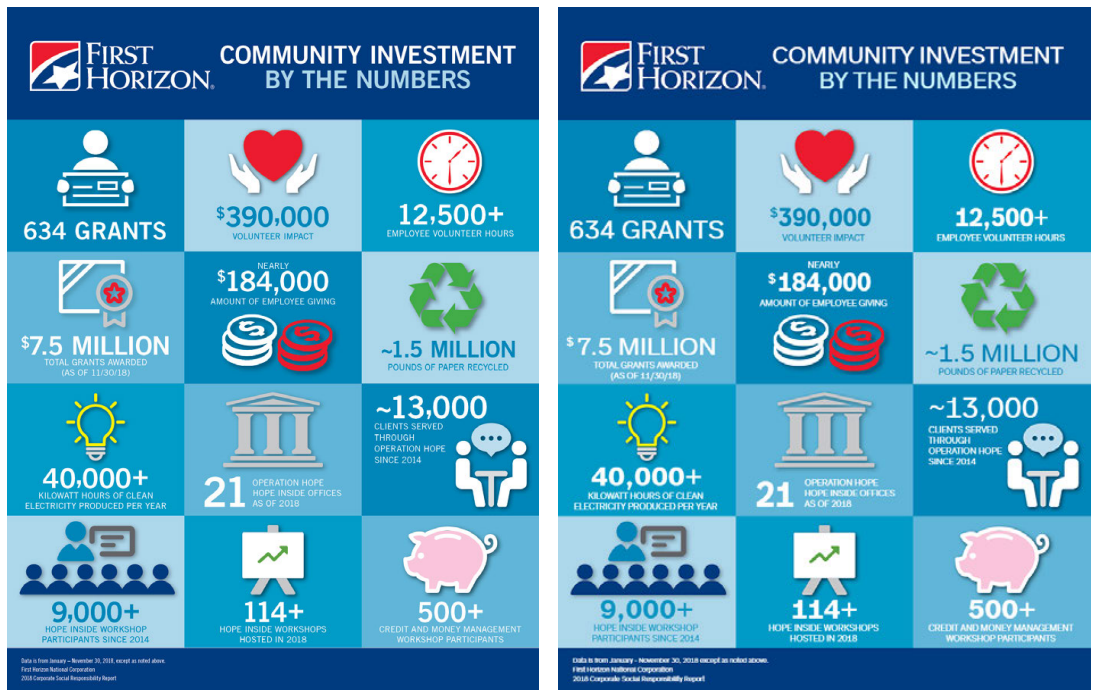

**Fig. 75.** Original infographic by First Horizon (left) and its corresponding generated infographic (right), using the specification in Table 13

**Table 13.** Specification of the infografic shown at the right of Fig. 75

```
bgcolor: 003b7f
bgsize: 800x1035
head:
  bgcolor: 003b7f
  size: 1x1
  position: 1x1
  title:
    align: center
    color: white
    font: 35px Studio Sans
    position: 540x85
    value: "COMMUNITY INVESTMENT"
  subtitle:
    align: center
    color: c3eef9
    position: 540x123
    font: 35px Studio Sans
    value: "BY THE NUMBERS"
foot: off
box1:
  bgcolor: 0086b9
  size: 267x200
  position: "0x171"
box2:
  bgcolor: acd7ea
  size: 266x200
  position: 267x171
box3:
  bgcolor: 009bc7
  size: 267x200
  position: "533x171"
box4:
  bgcolor: 60b6d9
  size: 267x200
  position: "0x371"
box5:
  bgcolor: 0086b9
  size: 266x200
  position: "267x371"
box6:
  bgcolor: acd7ea
  size: 267x200
  position: "533x371"
box7:
  bgcolor: 009bc7
  size: 267x200
  position: "0x571"
box8:
  bgcolor: 60b6d9
  size: 266x200
  position: "267x571"
box9:
  bgcolor: 0086b9
  size: 267x200
  position: "533x571"
box10:
  bgcolor: acd7ea
  size: 267x200
  position: "0x771"
box11:
  bgcolor: 009bc7
  size: 266x200
  position: "267x771"
box12:
  bgcolor: 60b6d9
  size: 267x200
  position: "533x771"
image1:
  size: 283x81
  position: 32x48
  src: "https://i.imgur.com/NCNlhkc.png"
image2:
  size: 132x132
```

```
    position: 71x182
    src: "https://i.imgur.com/u9WtN92.png"
image3:
    size: 148x109
    position: 328x180
    src: "https://i.imgur.com/wYixn7e.png"
image4:
    size: 110x106
    position: 612x182
    src: "https://i.imgur.com/27Rj5ZU.png"
image5:
    size: 130x116
    position: 70x378
    src: "https://i.imgur.com/mMVh52R.png"
image6:
    size: 182x96
    position: 318x462
    src: "https://i.imgur.com/mhLbss7.png"
image7:
    size: 128x120
    position: 599x382
    src: "https://i.imgur.com/yQuz0uK.png"
image8:
    size: 100x114
    position: 85x582
    src: "https://i.imgur.com/3Tlm9O5.png"
image9:
    size: 163x126
    position: 322x582
    src: "https://i.imgur.com/oOJ42jh.png"
image10:
    size: 127x139
    position: 667x626
    src: "https://i.imgur.com/pKiNdso.png"
image11:
    size: 257x122
    position: 5x775
    src: "https://i.imgur.com/8YlelSN.png"
image12:
    size: 117x123
    position: 338x776
    src: "https://i.imgur.com/EZSVC7V.png"
image13:
    size: 158x119
    position: 590x779
    src: "https://i.imgur.com/mER6luK.png"
titletext1:
      align: center
      color: white
      font: 39px Studio Sans
      position: 133x351
      value: "634 GRANTS"
titletext2:
      color: 0086b7
      font: 20px Studio Sans
      position: 318x318
      value: "$"
titletext3:
      align: center
      color: 0086b7
      font: bold 39px Studio Sans
      position: 400x332
      value: "390,000"
      maxwidth: 140
text1:
      align: center
      color: 0086b7
      font: 13px Studio Sans Light
      position: 400x354
      value: "VOLUNTEER IMPACT"
titletext4:
      align: center
      color: white
      font: bold 39px Studio Sans
      position: 666x332
      value: "12,500+"
```

```
        maxwidth: 140
text2:
    align: center
    color: white
    font: 13px Studio Sans Light
    position: 666x354
    value: "EMPLOYEE VOLUNTEER HOURS"
titletext5:
    align: center
    color: white
    font: 18px Studio Sans
    position: 18x513
    value: "$"
titletext6:
    align: center
    color: white
    font: 37px Studio Sans
    position: 133x527
    value: "7.5 MILLION"
text3:
    align: center
    color: white
    font: 13px Studio Sans Light
    position: 133x547
    value: "TOTAL GRANTS AWARDED (AS OF 11/30/18)"
    maxwidth: 170
    lineheight: 16
text4:
    align: center
    color: white
    font: 13px Studio Sans Light
    position: 400x395
    value: "NEARLY"
titletext7:
    color: white
    font: 20px Studio Sans
    position: 315x418
    value: "$"
titletext8:
    align: center
    color: white
    font: bold 39px Studio Sans
    position: 400x430
    value: "184,000"
    maxwidth: 140
text5:
    align: center
    color: white
    font: 13px Studio Sans Light
    position: 400x453
    value: "AMOUNT OF EMPLOYEE GIVING"
titletext9:
    align: center
    color: 0083b7
    font: 37px Studio Sans
    position: 666x537
    value: "~1.5 MILLION"
text6:
    align: center
    color: 0083b7
    font: 13px Studio Sans Light
    position: 666x557
    value: "POUNDS OF PAPER RECYCLED"
titletext10:
    align: center
    color: white
    font: bold 39px Studio Sans
    position: 133x725
    value: "40,000+"
text7:
    align: center
    color: white
    font: 13px Studio Sans Light
    position: 133x745
    value: "KILOWATT HOURS OF CLEAN ELECTRICITY PRODUCED PER YEAR"
    maxwidth: 220
```

```
    lineheight: 16
titletext11:
    color: white
    font: bold 55px Studio Sans
    position: 296x757
    value: "21"
    maxwidth: 60
text8:
    color: white
    font: 13px Studio Sans Light
    position: 370x725
    value: "OPERATION HOPE    HOPE INSIDE OFFICES AS OF 2018"
    maxwidth: 140
    lineheight: 16
titletext12:
    color: white
    font: 43px Studio Sans
    position: 556x622
    value: "~13,000"
text9:
    color: white
    font: 13px Studio Sans Light
    position: 556x645
    value: "CLIENTS SERVED THROUGH OPERATION HOPE SINCE 2014"
    maxwidth: 120
    lineheight: 16
titletext13:
    align: center
    color: 0d95c3
    font: bold 39px Studio Sans
    position: 133x926
    value: "9,000+"
text10:
    align: center
    color: 0d95c3
    font: 13px Studio Sans Light
    position: 133x945
    value: "HOPE INSIDE WORKSHOP PARTICIPANTS SINCE 2014"
    maxwidth: 172
    lineheight: 16
titletext14:
    align: center
    color: white
    font: bold 39px Studio Sans
    position: 400x926
    value: "114+"
text11:
    align: center
    color: white
    font: 13px Studio Sans Light
    position: 400x945
    value: "HOPE INSIDE WORKSHOPS HOSTED IN 2018"
    maxwidth: 170
    lineheight: 16
titletext15:
    align: center
    color: white
    font: bold 39px Studio Sans
    position: 666x926
    value: "500+"
text12:
    align: center
    color: white
    font: 13px Studio Sans Light
    position: 666x945
    value: "CREDIT AND MONEY MANAGEMENT WORKSHOP PARTICIPANTS"
    maxwidth: 225
    lineheight: 16
text13:
    color: white
    font: 10px Studio Sans Light
    position: 22x993
    value: "Data is from January - November 30, 2018 except as noted above."
    maxwidth: 800
    lineheight: 16
text14:
```

```
            color: white
            font: 10px Studio Sans Light
            position: 22x1007
            value: "First Horizon National Corporation"
            maxwidth: 800
            lineheight: 16
text15:
            color: white
            font: 10px Studio Sans Light
            position: 22x1021
            value: "2018 Corporate Social Responsibility Report"
            maxwidth: 800
            lineheight: 16
```

## 5.6 GSI

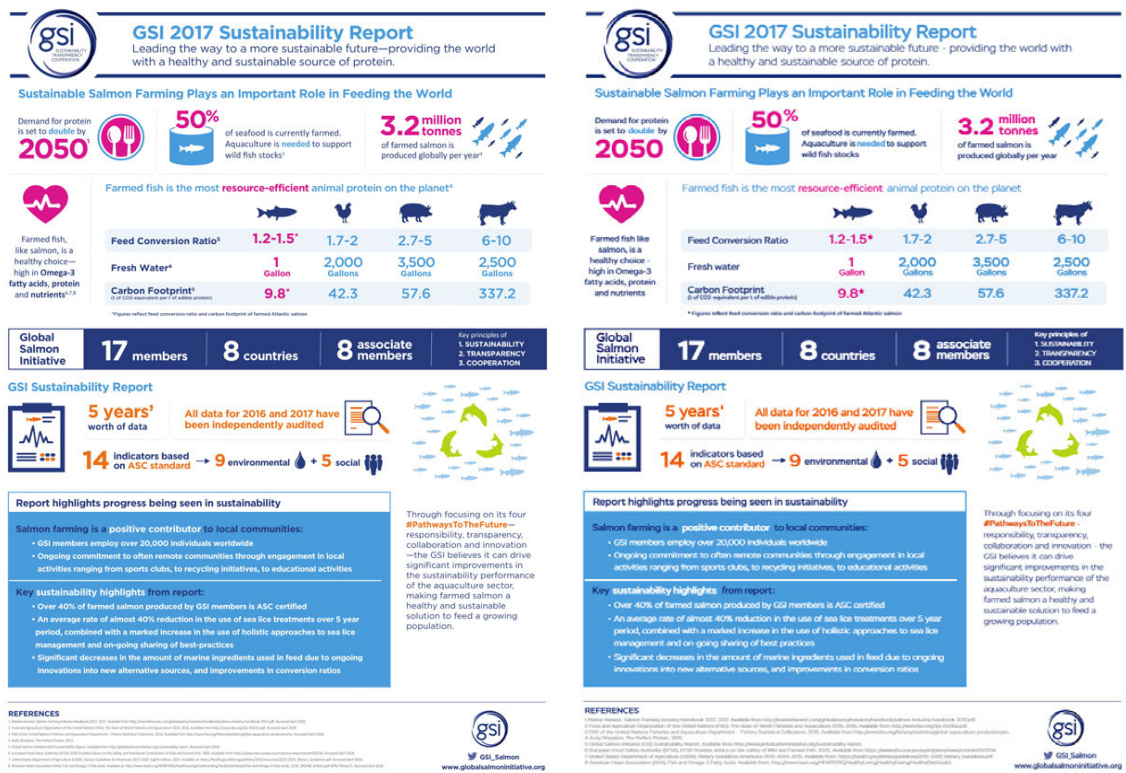

**Fig. 76.** Original infographic by GSI (left) and its corresponding generated infographic (right), using the specification in Table 14

**Table 14.** Specification of the infografic shown at the right of Fig. 76

```
bgcolor: ffffff
bgsize: 827x1169
foot: off
head:
  title:
    color: 4799da
```

```
    font: 28px Stem ☞
    position: 202x56
    value: "GSI 2017 Sustainability Report"
  subtitle: off
  bgimage: "https://i.imgur.com/scJFYEk.png"
  size: 785x104
  position: 22x13
box1:
  bgimage: "https://i.imgur.com/O2YhV3v.png"
  size: 769x291
  position: 20x161
box2:
  bgcolor: 273a7c
  size: 791x62
  position: 19x485
box3:
  bgcolor: white
  size: 109x58
  position: 21x487
box4:
  bgcolor: 6c7caa
  size: 2x50
  position: 309x491
box5:
  bgcolor: 6c7caa
  size: 2x50
  position: 476x491
box6:
  bgcolor: 6c7caa
  size: 2x50
  position: 639x491
box7:
  bgimage: "https://i.imgur.com/k2EXplo.png"
  size: 560x112
  position: 19x590
box8:
  bgcolor: 489cdc
  size: 562x290
  position: 18x726
box9:
  bgcolor: white
  size: 558x28
  position: 20x728
box10:
  bgcolor: white
  size: 546x2
  position: 20x857
box11:
  size: 789x2
  position: 18x1036
  bgimage: "https://i.imgur.com/suFegI3.png"
image1:
  size: 181x145
  position: 611x567
  src: "https://i.imgur.com/eROAsfj.png"
image2:
  size: 61x74
  position: 692x1049
  src: "https://i.imgur.com/2oUfi4Q.png"
image3:
  size: 7x7
  position: 53x799
  src: "https://i.imgur.com/lqTauUh.png"
image4:
  size: 7x7
  position: 53x818
  src: "https://i.imgur.com/lqTauUh.png"
image5:
  size: 7x7
  position: 53x892
  src: "https://i.imgur.com/lqTauUh.png"
image6:
  size: 7x7
  position: 53x912
  src: "https://i.imgur.com/lqTauUh.png"
image7  :
```

```
    size: 7x7
    position: 53x968
    src: "https://i.imgur.com/lqTauUh.png"
text1:
    font: 16px Stem ☞ Light
    color: 233678
    lineheight: 20
    position: 202x76
    maxwidth: 540
    value: "Leading the way to a more sustainable future – providing the world with a healthy and sustainable source of protein."
titletext1:
    font: 18px Stem ☞ Medium
    color: 4092d5
    position: 35x143
    maxwidth: 700
    value: "Sustainable Salmon Farming Plays an Important Role in Feeding the World"
text2:
    font: 12px Stem ☞
    color: 233678
    lineheight: 15
    position: 35x182
    maxwidth: 120
    value: "Demand for protein is set to                   by"
text3:
    font: 12px Stem ☞
    color: 4799da
    position: 84x197
    value: "double"
titletext2:
    font: 39px Stem ☞ Heavy
    color: de138b
    position: 35x233
    value: "2050"
titletext3:
    font: 35px Stem ☞ Heavy
    color: de138b
    position: 264x189
    value: "50"
titletext4:
    font: 25px Stem ☞ Medium
    color: de138b
    position: 311x179
    value: "%"
text4:
    font: 12px Stem ☞
    color: 233678
    lineheight: 16
    position: 338x200
    maxwidth: 195
    value: "of seafood is currently farmed. Aquaculture is                 to support wild fish stocks"
text5:
    font: 12px Stem ☞
    color: 4799da
    position: 419x216
    value: "needed"
titletext5:
    font: 35px Stem ☞ Heavy
    color: de138b
    position: 567x199
    value: "3.2"
text6:
    font: 18px Stem ☞ Heavy
    color: de138b
    lineheight: 17
    maxwidth: 65
    position: 627x182
    value: "million tonnes"
text7:
    font: 12px Stem ☞
    color: 233678
    lineheight: 16
    position: 567x218
```

```
    maxwidth: 160
    value: "of farmed salmon is produced globally per year"
text8:
    font: 12px Stem ☞
    align: center
    color: 233678
    lineheight: 16
    position: 73x357
    maxwidth: 110
    value: "Farmed fish like salmon, is a healthy choice - high in Omega-3 fatty acids, protein and nutrients"
text9:
    font: 14px Stem ☞
    color: 233678
    lineheight: 16
    position: 171x360
    value: "Feed Conversion Ratio"
text10:
    font: 14px Stem ☞
    color: 233678
    lineheight: 16
    position: 171x399
    value: "Fresh water"
text11:
    font: 14px Stem ☞
    color: 233678
    lineheight: 16
    position: 171x433
    value: "Carbon Footprint"
text12:
    font: 8px Stem ☞
    color: 233678
    lineheight: 16
    position: 171x442
    value: "(t of CO2-equivalent per t of edible protein)"
text13:
    font: 19px Stem ☞
    align: center
    color: da0081
    lineheight: 16
    position: 413x360
    value: "1.2-1.5*"
text14:
    font: 19px Stem ☞
    align: center
    color: da0081
    lineheight: 16
    position: 413x394
    value: "1"
text15:
    font: 13px Stem ☞
    align: center
    color: da0081
    lineheight: 16
    position: 413x407
    value: "Gallon"
text16:
    font: 19px Stem ☞
    align: center
    color: da0081
    lineheight: 16
    position: 413x440
    value: "9.8*"
text17:
    font: 19px Stem ☞
    align: center
    color: 4494d3
    lineheight: 16
    position: 510x360
    value: "1.7-2"
text18:
    font: 19px Stem ☞
    align: center
    color: 4494d3
```

```
    lineheight: 16
    position: 510x394
    value: "2,000"
text19:
    font: 13px Stem ☞
    align: center
    color: 4494d3
    lineheight: 16
    position: 510x407
    value: "Gallons"
text20:
    font: 19px Stem ☞
    align: center
    color: 4494d3
    lineheight: 16
    position: 510x440
    value: "42.3"
text21:
    font: 19px Stem ☞
    align: center
    color: 4494d3
    lineheight: 16
    position: 618x360
    value: "2.7-5"
text22:
    font: 19px Stem ☞
    align: center
    color: 4494d3
    lineheight: 16
    position: 618x394
    value: "3,500"
text23:
    font: 13px Stem ☞
    align: center
    color: 4494d3
    lineheight: 16
    position: 618x407
    value: "Gallons"
text24:
    font: 19px Stem ☞
    align: center
    color: 4494d3
    lineheight: 16
    position: 618x440
    value: "57.6"
text25:
    font: 19px Stem ☞
    align: center
    color: 4494d3
    lineheight: 16
    position: 736x360
    value: "6-10"
text26:
    font: 19px Stem ☞
    align: center
    color: 4494d3
    lineheight: 16
    position: 736x394
    value: "2,500"
text27:
    font: 13px Stem ☞
    align: center
    color: 4494d3
    lineheight: 16
    position: 736x407
    value: "Gallons"
text28:
    font: 19px Stem ☞
    align: center
    color: 4494d3
    lineheight: 16
    position: 736x440
    value: "337.2"
text29:
```

```
    font: 8px Stem ☞
    color: 233678
    lineheight: 16
    position: 171x466
    maxwidth: 700
    value: "* Figures reflect feed conversion ratio and carbon footprint of farmed Atlantic salmon"
text30:
    font: 18px Stem ☞
    color: 000a5c
    lineheight: 16
    maxwidth: 80
    position: 37x505
    value: "Global Salmon Initiative"
text31:
    font: 40px Stem ☞ Heavy
    color: white
    position: 157x532
    value: "17"
text32:
    font: 18px Stem ☞ Medium
    color: white
    position: 202x532
    value: "members"
text33:
    font: 40px Stem ☞ Heavy
    color: white
    position: 335x532
    value: "8"
text34:
    font: 18px Stem ☞ Medium
    color: white
    position: 367x532
    value: "countries"
text35:
    font: 40px Stem ☞ Heavy
    color: white
    position: 501x532
    value: "8"
text36:
    font: 18px Stem ☞ Medium
    color: white
    lineheight: 16
    maxwidth: 90
    position: 536x515
    value: "associate members"
text37:
    font: 10px Stem ☞ Light
    color: white
    position: 680x498
    value: "Key principles of"
text38:
    font: 10px Stem ☞ Light
    color: white
    position: 680x512
    value: "1. SUSTAINABILITY"
text39:
    font: 10px Stem ☞ Light
    color: white
    position: 680x526
    value: "2. TRANSPARENCY"
text40:
    font: 10px Stem ☞ Light
    color: white
    position: 680x540
    value: "3. COOPERATION"
titletext6:
    font: 18px Stem ☞ Medium
    color: 4092d5
    position: 20x577
    maxwidth: 700
    value: "GSI Sustainability Report"
text41:
    font: 24px Stem ☞ Medium
```

```
  color: ed5700
  position: 138x616
  value: "5 years'"
text42:
  font: 13px Stem ☞ Medium
  color: 14266f
  position: 138x634
  value: "worth of data"
text43:
  font: 16px Stem ☞ Medium
  color: ed5700
  lineheight: 20
  maxwidth: 240
  position: 270x615
  value: "All data for 2016 and 2017 have been independently audited"
text44:
  font: 33px Stem ☞ Medium
  color: ed5700
  position: 131x691
  value: "14"
text45:
  font: 14px Stem ☞ Medium
  color: 14266f
  lineheight: 15
  maxwidth: 120
  position: 175x676
  value: "indicators based on"
text46:
  font: 14px Stem ☞ Medium
  color: ed5700
  position: 195x691
  value: "ASC standard"
text47:
  font: 26px Stem ☞ Medium
  color: ed5700
  position: 320x690
  value: "9"
text48:
  font: 14px Stem ☞ Medium
  color: 14266f
  position: 342x687
  value: "environmental"
text49:
  font: 26px Stem ☞ Medium
  color: ed5700
  position: 479x690
  value: "5"
text50:
  font: 14px Stem ☞ Medium
  color: 14266f
  position: 500x687
  value: "social"
titletext7:
  font: 15px Stem ☞ Medium
  color: 14266f
  position: 32x747
  maxwidth: 700
  value: "Report highlights progress being seen in sustainability"
titletext8:
  font: 14px Stem ☞ Medium
  color: 14266f
  position: 31x785
  maxwidth: 700
  value: "Salmon farming is a                                      to local communities:"
titletext9:
  font: 14px Stem ☞ Medium
  color: white
  position: 163x785
  value: "positive contributor"
text51:
  font: 12px Stem ☞ Light
  color: white
  maxwidth: 480
```

```
  position: 63x806
  value: "GSI members employ over 20,000 individuals worldwide"
text52:
  font: 12px Stem ☞ Light
  color: white
  lineheight: 18
  maxwidth: 480
  position: 63x825
  value: "Ongoing commitment to often remote communities through engagement in local activities ranging from sports clubs, to recycling initiatives, to educational activities"
titletext10:
  font: 14px Stem ☞ Medium
  color: 14266f
  position: 31x878
  maxwidth: 700
  value: "Key                                              from report:"
titletext11:
  font: 14px Stem ☞ Medium
  color: white
  position: 61x878
  maxwidth: 700
  value: "sustainability highlights"
text53:
  font: 12px Stem ☞ Light
  color: white
  lineheight: 18
  maxwidth: 480
  position: 63x899
  value: "Over 40% of farmed salmon produced by GSI members is ASC certified"
text54:
  font: 12px Stem ☞ Light
  color: white
  lineheight: 18
  maxwidth: 490
  position: 63x919
  value: "An average rate of almost 40% reduction in the use of sea lice treatments over 5 year period, combined with a marked increase in the use of hollistic approaches to sea lice management and on-going sharing of best practices"
text55:
  font: 12px Stem ☞ Light
  color: white
  lineheight: 18
  maxwidth: 490
  position: 63x976
  value: "Significant decreases in the amount of marine ingredients used in feed due to ongoing innovations into new alternative sources, and improvements in conversion ratios"
text56:
  font: 12px Stem ☞ Light
  color: 14266f
  lineheight: 16
  maxwidth: 195
  position: 605x763
  value: "Through focusing on its four                                                 – responsibility, transparency, collaboration and innovation – the GSI believes it can drive significant improvements in the sustainability performance of the aquaculture sector, making farmed salmon a healthy and sustainable solution to feed a growing population."
text57:
  font: 12px Stem ☞
  color: da4500
  position: 605x778
  value: "#PathwaysToTheFuture"
titletext12:
  font: 12px Stem ☞ Medium
  color: 14266f
  position: 20x1055
  value: "REFERENCES"
text58:
  font: 7px Stem ☞ Light
  color: a7abae
  position: 20x1067
  maxwidth: 1000
  value: "1 Marine Harvest. Salmon Farming Industry Handbook 2017. 2017. Available from http://marineharvest.com/globalassets/investors/handbook/salmon-industry-handbook-2017.pdf."
text59:
```

```
    font: 7px Stem ☞ Light
    color: a7abae
    position: 20x1076
    maxwidth: 1000
    value: "2 Food and Agriculture Organization of the United Nations (FAO). The state of World Fisheries and Aquaculture 2016. 2016. Available from http://www.fao.org/3/a-i5555e.pdf."
text60:
    font: 7px Stem ☞ Light
    color: a7abae
    position: 20x1085
    maxwidth: 1000
    value: "3 FAO of the United Nations Fisheries and Aquaculture Department – Fishery Statistical Collections. 2016. Available from http://www.fao.org/fishery/statistics/global-aquaculture-production/en."
text61:
    font: 7px Stem ☞ Light
    color: a7abae
    position: 20x1094
    maxwidth: 1000
    value: "4 Andy Sharpless. The Perfect Protein. 2015."
text62:
    font: 7px Stem ☞ Light
    color: a7abae
    position: 20x1103
    maxwidth: 1000
    value: "5 Global Salmon Initiative (GSI) Sustainability Report. Available from http://www.globalsalmoninitiative.org/sustainability-report."
text63:
    font: 7px Stem ☞ Light
    color: a7abae
    position: 20x1112
    maxwidth: 1000
    value: "6 European Food Safety Authority (EFSA). EFSA Provides Advice on the Safety of Wild and Farmed Fish. 2005. Available from https://www.efsa.europa.eu/en/press/news/contam050704."
text64:
    font: 7px Stem ☞ Light
    color: a7abae
    position: 20x1121
    maxwidth: 1000
    value: "7 United States Department of Agriculture (USDA). Dietary Guidelines Americans 2015-2020. 2015. Available from: https://health.gov/dietaryguidelines/2015-2020_Dietary_Guidelines.pdf."
text65:
    font: 7px Stem ☞ Light
    color: a7abae
    position: 20x1130
    maxwidth: 1000
    value: "8 American Heart Association (AHA). Fish and Omega-3 Fatty Acids. Available from: http://www.heart.org/HEARTORG/HealthyLiving/HealthyEating/HealthyDietGoals1."
text66:
    font: 11px Stem ☞
    color: 273a7c
    position: 711x1121
    value: "GSI_Salmon"
text67:
    font: 11px Stem ☞
    color: 273a7c
    position: 650x1136
    value: "www.globalsalmoninitiative.org"
titletext13:
    font: 15px Stem ☞ Light
    color: 4092d5
    position: 163x283
    maxwidth: 700
    value: "Farmed fish is the most                          animal protein on the planet"
titletext14:
    font: 15px Stem ☞ Medium
    color: de138b
    position: 333x283
    maxwidth: 700
    value: "resource-efficient"
```

## 5.7 Lenovo

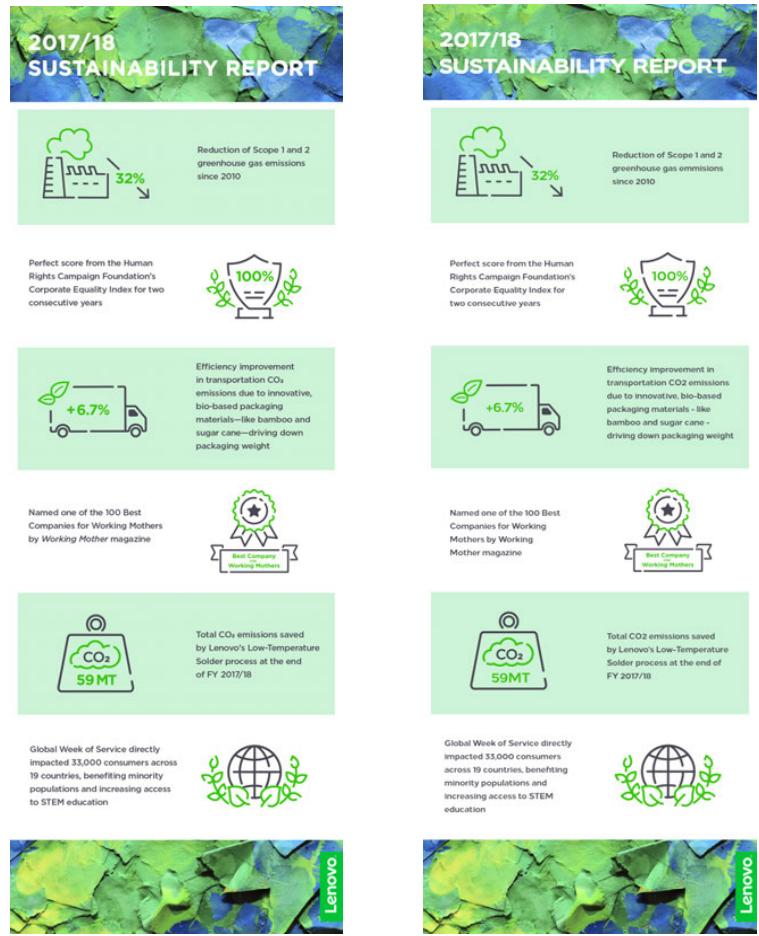

**Fig. 77.** Original infographic by Lenovo (left) and its corresponding generated infographic (right), using the specification in Table 15

**Table 15.** Specification of the infografic shown at the right of Fig. 77

```
bgcolor: ffffff
bgsize: 625x1755
box1:
  bgcolor: ccf3d7
  position: 15x195
  size: 595x218
box2:
  bgcolor: ccf3d7
  position: 15x645
  size: 595x232
box3:
  bgcolor: ccf3d7
  position: 15x1109
  size: 595x231
foot:
```

```
    bgimage: 'https://i.imgur.com/WZ08A9o.png'
    size: 625x178
    text: 'off'
head:
  bgimage: 'https://i.imgur.com/tTDygTH.png'
  size: 625x181
  subtitle:
    color: ffffff
    font: 37px BiondiSansW00-Bold
    position: 30x130
    value: "SUSTAINABILITY REPORT"
    maxwidth: 1100
  title:
    color: ffffff
    font: 37px BiondiSansW00-Bold
    position: 30x80
    value: "2017/18"
image1:
  position: 59x229
  size: 203x135
  src: 'https://i.imgur.com/MLKHyOi.png'
image2:
  position: 366x467
  size: 182x124
  src: 'https://i.imgur.com/LTySR3G.png'
image3:
  position: 50x708
  size: 208x108
  src: 'https://i.imgur.com/PaF56B0.png'
image4:
  position: 376x912
  size: 166x161
  src: 'https://i.imgur.com/muS4Gfm.png'
image5:
  position: 87x1144
  size: 147x152
  src: 'https://i.imgur.com/PJDNMQo.png'
image6:
  position: 357x1396
  size: 203x120
  src: 'https://i.imgur.com/KSSgmOf.png'
text1:
  color: 474651
  font:  bold 15px Loew
  maxwidth: 220
  position: 354x290
  value: "Reduction of Scope 1 and 2 greenhouse gas emmisions since 2010"
text10:
  color: 3bc511
  font: bold 25px Loew
  position: 128x1280
  value: 59MT
text2:
  color: 474651
  font:  bold 15px Loew
  maxwidth: 240
  position: 50x495
  value: "Perfect score from the Human Rights Campaign Foundation's Corporate Equality Index for two consecutive years"
text3:
  color: 474651
  font:  bold 15px Loew
  maxwidth: 250
  position: 344x695
  value: "Efficiency improvement in transportation CO2 emissions due to innovative, bio-based packaging materials - like bamboo and sugar cane - driving down packaging weight"
text4:
  color: 474651
  font:  bold 15px Loew
  maxwidth: 220
  position: 50x965
  value: "Named one of the 100 Best Companies for Working Mothers by Working Mother magazine"
text5:
  color: 474651
  font:  bold 15px Loew
  maxwidth: 240
```

```
    position: 344x1198
    value: "Total CO2 emissions saved     by Lenovo's Low-Temperature Solder process at the end of FY 2017/18"
text6:
  color: 474651
  font:  bold 15px Loew
  maxwidth: 260
  position: 40x1400
  value: "Global Week of Service directly impacted 33,000 consumers across 19 countries, benefiting minority populations and increasing access to STEM education"
text7:
  color: 3bc511
  font: bold 25px Loew
  position: 202x332
  value: "32%"
text8:
  color: 3bc511
  font: bold 25px Loew
  position: 428x522
  value: "100%"
text9:
  color: 3bc511
  font: bold 25px Loew
  position: 118x770
  value: "+6.7%"
```

## 5.8 The Home Depot

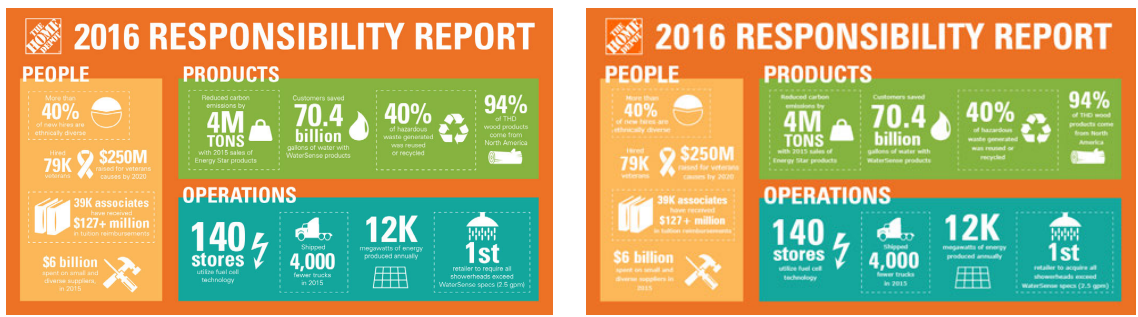

**Fig. 78.** Original infographic by The Home Depot (left) and its corresponding generated infographic (right), using the specification in Table 16

**Table 16.** Specification of the infografic shown at the right of Fig. 78

```
bgcolor: ec7124
bgsize: 1200x675
head:
  bgcolor: off
  title:
    value: "2016 RESPONSIBILITY REPORT"
    font: bold 79px Zurich Cn BT
    color: ffffff
    position: 150x90
    maxwidth: 1000
  subtitle: off
foot: off
image1:
  src: "https://corporate.homedepot.com/sites/default/files/styles/thumbnail/public/image_gallery/THD_logo.jpg"
  position: 43x22
  size: 80x80
image2:
  position: 57x178
  size: 267x442
  src: "https://i.imgur.com/H0DDTFD.png"
image3:
```

```
    position: 400x180
    size: 733x177
    src: "https://i.imgur.com/8AWXIlK.png"
image4:
    position: 538x448
    size: 613x178
    src: "https://i.imgur.com/DomjXF8.png"
box1:
    size: 312x490
    position: 31x155
    bgcolor: fdb859
box2:
    size: 790x221
    position: 378x155
    bgcolor: 85b440
box3:
    size: 790x228
    position: 378x417
    bgcolor: 14a79d
titletext1:
    font: bold 52px Zurich Cn BT
    color: ffffff
    position: 40x164
    value: "PEOPLE"
titletext2:
    font: bold 52px Zurich Cn BT
    color: ffffff
    position: 387x164
    value: "PRODUCTS"
titletext3:
    font: bold 52px Zurich Cn BT
    color: ffffff
    position: 387x430
    value: "OPERATIONS"
text1:
    font: 14px Verdana
    color: ffffff
    position: 125x203
    value: "More than"
    align: center
text2:
    font: bold 42px Zurich Cn BT
    color: ffffff
    position: 125x239
    value: "40%"
    align: center
text3:
    font: 14px Verdana
    color: ffffff
    position: 125x255
    maxwidth: 150
    value: "of new hires are ethnically diverse"
    align: center
    lineheight: 18
text4:
    font: 14px Verdana
    color: ffffff
    position: 110x319
    value: "Hired"
    align: center
text5:
    font: bold 42px Zurich Cn BT
    color: ffffff
    position: 110x356
    value: "79K"
    align: center
text6:
    font: 14px Verdana
    color: ffffff
    position: 110x373
    value: "veterans"
    align: center
text7:
    font: bold 42px Zurich Cn BT
    color: ffffff
    position: 270x335
```

```
    value: "$250M"
    align: center
text8:
    font: 14px Verdana
    color: ffffff
    position: 270x355
    value: "raised for veterans causes by 2020"
    align: center
    maxwidth: 150
    lineheight: 20
text9:
    font: bold 25px Zurich Cn BT
    color: ffffff
    position: 235x430
    value: "39K associates"
    align: center
text10:
    font: 14px Verdana
    color: ffffff
    position: 235x449
    value: "have received"
    align: center
    maxwidth: 150
    lineheight: 20
text11:
    font: bold 25px Zurich Cn BT
    color: ffffff
    position: 235x475
    value: "$127+ million"
    align: center
text12:
    font: 13px Verdana
    color: ffffff
    position: 235x494
    value: "in tuition reimbursements"
    align: center
    maxwidth: 200
    lineheight: 20
text13:
    font: bold 36px Zurich Cn BT
    color: ffffff
    position: 130x560
    value: "$6 billion"
    align: center
text14:
    font: 13px Verdana
    color: ffffff
    position: 130x580
    value: "spent on small and diverse suppliers in 2015"
    align: center
    maxwidth: 150
    lineheight: 20
text15:
    font: 13px Verdana
    color: ffffff
    position: 480x200
    value: "Reduced carbon emissions by"
    align: center
    maxwidth: 150
    lineheight: 20
text16:
    font: bold 64px Zurich Cn BT
    color: ffffff
    position: 480x272
    value: "4M"
    align: center
    maxwidth: 50
    lineheight: 34
text17:
    font: bold 38px Zurich Cn BT
    color: ffffff
    position: 480x305
    value: "TONS"
    align: center
    maxwidth: 50
    lineheight: 34
```

```
text18:
  font: 13px Verdana
  color: ffffff
  position: 480x320
  value: "with 2015 sales of Energy Star products"
  align: center
  maxwidth: 150
  lineheight: 20
text19:
  font: 13px Verdana
  color: ffffff
  position: 685x200
  value: "Customers saved"
  align: center
  maxwidth: 150
  lineheight: 20
text20:
  font: bold 72px Zurich Cn BT
  color: ffffff
  position: 690x263
  value: "70.4"
  align: center
  lineheight: 34
text21:
  font: bold 42px Zurich Cn BT
  color: ffffff
  position: 685x302
  value: "billion"
  align: center
  maxwidth: 40
  lineheight: 34
text22:
  font: 13px Verdana
  color: ffffff
  position: 685x320
  value: "gallons of water with WaterSense products"
  align: center
  maxwidth: 150
  lineheight: 20
text23:
  font: bold 62px Zurich Cn BT
  color: ffffff
  position: 890x250
  value: "40%"
  align: center
  maxwidth: 50
  lineheight: 34
text24:
  font: 13px Verdana
  color: ffffff
  position: 890x273
  value: "of hazardous waste generated was reused or recycled"
  align: center
  maxwidth: 130
  lineheight: 20
text25:
  font: bold 52px Zurich Cn BT
  color: ffffff
  position: 1105x220
  value: "94%"
  align: center
  maxwidth: 50
  lineheight: 34
text26:
  font: 13px Verdana
  color: ffffff
  position: 1105x240
  value: "of THD wood products come from North America"
  align: center
  maxwidth: 130
  lineheight: 20
text27:
  font: bold 80px Zurich Cn BT
  color: ffffff
  position: 472x520
  value: "140"
```

```
    align: center
    maxwidth: 35
    lineheight: 34
text28:
    font: bold 45px Zurich Cn BT
    color: ffffff
    position: 470x555
    value: "stores"
    align: center
    maxwidth: 50
    lineheight: 34
text29:
    font: 13px Verdana
    color: ffffff
    position: 470x577
    value: "utilize fuel cell technology"
    align: center
    maxwidth: 150
    lineheight: 20
text30:
    font: 13px Verdana
    color: ffffff
    position: 678x529
    value: "Shipped"
    align: center
    maxwidth: 150
    lineheight: 20
text31:
    font: bold 50px Zurich Cn BT
    color: ffffff
    position: 678x572
    value: "4,000"
    align: center
    maxwidth: 50
    lineheight: 34
text32:
    font: 13px Verdana
    color: ffffff
    position: 678x590
    value: "fewer trucks in 2015"
    align: center
    maxwidth: 100
    lineheight: 20
text33:
    font: bold 80px Zurich Cn BT
    color: ffffff
    position: 850x510
    value: "12K"
    align: center
    maxwidth: 50
    lineheight: 34
text34:
    font: 13px Verdana
    color: ffffff
    position: 850x530
    value: "megawatts of energy produced annually"
    align: center
    maxwidth: 150
    lineheight: 20
text35:
    font: bold 56px Zurich Cn BT
    color: ffffff
    position: 1048x555
    value: "1st"
    align: center
    maxwidth: 50
    lineheight: 34
text36:
    font: 13px Verdana
    color: ffffff
    position: 1048x575
    value: "retailer to acquire all showerheads exceed WaterSense specs (2.5 gpm)"
    align: center
    maxwidth: 200
    lineheight: 20
```

## 5.9 Trinseo

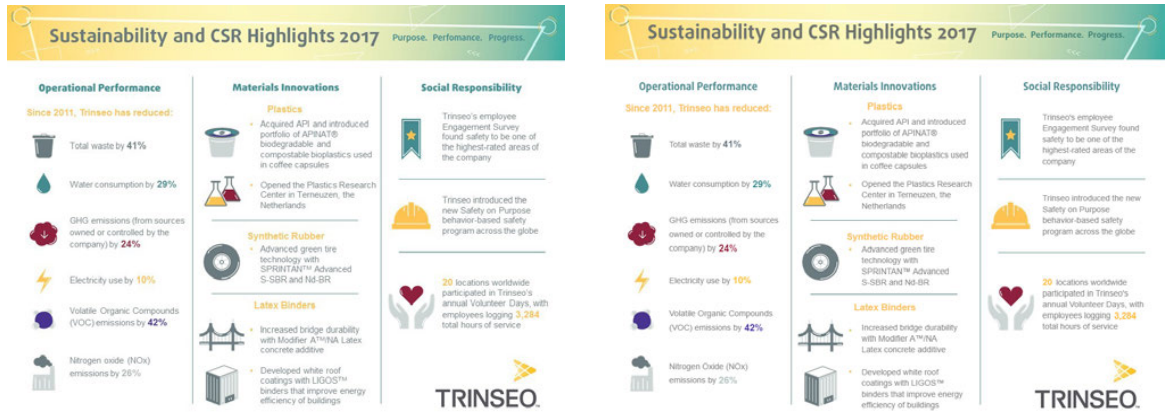

**Fig. 79.** Original infographic by Trinseo (left) and its corresponding generated infographic (right), using the specification in Table 17

**Table 17.** Specification of the infografic shown at the right of Fig. 79

```
bgimage: "https://i.imgur.com/0giWB5X.png"
bgsize: 960x720
head:
  size: 960x96
  bgimage: "https://i.imgur.com/7ZEKp4R.png"
  title:
    value: "Sustainability and CSR Highlights 2017"
    color: 6f746d
    position: 73x60
    font: bold 35px DaxlinePro
    maxwidth: 565
  subtitle:
    color: 3d7f73
    position: 663x58
    font: bold 15px DaxlinePro
    value: "Purpose.  Performance.  Progress."
foot: off
titletext1:
  font: bold 18px DaxlinePro
  value: "Operational Performance"
  position: 160x148
  align: center
  color: 4e898f
titletext2:
  font: bold 16px sans-serif
  value: "Since 2011, Trinseo has reduced:"
  position: 160x185
  align: center
  color: f8c671
titletext3:
  font: bold 18px DaxlinePro
  value: Materials Innovations
  position: 480x148
  align: center
  color: 4e898f
titletext4:
  font: bold 16px sans-serif
  value: "Plastics"
  position: 480x182
  align: center
  color: f8c671
titletext5:
```

```
    font: bold 16px sans-serif
    value: "Synthetic Rubber"
    position: 480x407
    align: center
    color: f8c671
titletext6:
    font: bold 16px sans-serif
    value: "Latex Binders"
    position: 480x529
    align: center
    color: f8c671
titletext7:
    font: bold 18px DaxlinePro
    value: "Social Responsibility"
    position: 800x148
    align: center
    color: 4e898f
text1:
    font: 14px sans-serif
    color: 828483
    value: "Total waste by"
    position: 110x247
text2:
    font: bold 16px sans-serif
    value: "41%"
    color: 6c7a83
    position: 202x247
text3:
    font: 14px sans-serif
    color: 828483
    value: "Water consumption by"
    position: 110x315
text4:
    font: bold 16px sans-serif
    value: "29%"
    color: 3f8d91
    position: 253x316
text5:
    font: 14px sans-serif
    color: 828483
    value: "GHG emissions (from sources owned or controlled by the company) by"
    position: 110x378
    maxwidth: 200
text6:
    font: bold 16px sans-serif
    value: "24%"
    color: 921f3c
    position: 195x427
text7:
    font: 14px sans-serif
    color: 828483
    value: "Electricity use by"
    position: 110x482
    maxwidth: 200
text8:
    font: bold 16px sans-serif
    value: "10%"
    color: fac84f
    position: 219x482
text9:
    font: 14px sans-serif
    color: 828483
    value: "Volatile Organic Compounds (VOC) emissions by"
    position: 110x539
    maxwidth: 200
text10:
    font: bold 16px sans-serif
    value: "42%"
    color: 48319b
    position: 239x564
text11:
    font: 14px sans-serif
    color: 828483
    value: "Nitrogen Oxide (NOx) emissions by"
    position: 110x630
    maxwidth: 200
```

```
text12:
  font: bold 16px sans-serif
  value: "26%"
  color: bdc9c8
  position: 195x655
text13:
  font: 14px sans-serif
  lineheight: 19
  value: "Acquired API and introduced portfolio of APINAT® biodegradable and compostable bioplastics used in coffee capsules"
  color: 828483
  position: 440x208
  maxwidth: 200
text14:
  font: 14px sans-serif
  lineheight: 19
  value: "Opened the Plastics Research Center in Terneuzen, the Netherlands"
  color: 828483
  position: 440x315
  maxwidth: 200
text15:
  font: 14px sans-serif
  lineheight: 19
  value: "Advanced green tire technology with SPRINTAN™ Advanced S-SBR and Nd-BR"
  color: 828483
  position: 440x428
  maxwidth: 170
text16:
  font: 14px sans-serif
  lineheight: 19
  value: "Increased bridge durability with Modifier A™/NA Latex concrete additive"
  color: 828483
  position: 440x563
  maxwidth: 170
text17:
  font: 14px sans-serif
  lineheight: 19
  value: "Developed white roof coatings with LIGOS™ binders that improve energy efficiency of buildings"
  color: 828483
  position: 440x640
  maxwidth: 190
text18:
  font: 14px sans-serif
  lineheight: 19
  value: "Trinseo's employee Engagement Survey found safety to be one of the highest-rated areas of the company"
  color: 828483
  position: 750x200
  maxwidth: 190
text19:
  font: 14px sans-serif
  lineheight: 19
  value: "Trinseo introduced the new Safety on Purpose behavior-based safety program across the globe"
  color: 828483
  position: 750x340
  maxwidth: 190
text20:
  font: 14px sans-serif
  lineheight: 19
  value: "      locations worldwide participated in Trinseo's annual Volunteer Days, with employees logging      total hours of service"
  color: 828483
  position: 750x485
  maxwidth: 190
text21:
  font: bold 15px sans-serif
  lineheight: 19
  value: "20"
  color: ffc253
  position: 750x485
  maxwidth: 190
text22:
  font: bold 15px sans-serif
  lineheight: 19
  value: "3,284"
  color: ffc253
  position: 873x543
```

```
  maxwidth: 190
image1:
  size: 40x43
  position: 44x222
  src: "https://i.imgur.com/PRGWIKE.png"
image2:
  size: 25x36
  position: 51x291
  src: "https://i.imgur.com/eW0JEDy.png"
image3:
  size: 48x42
  position: 39x376
  src: "https://i.imgur.com/abdIFQ4.png"
image4:
  size: 26x44
  position: 50x457
  src: "https://i.imgur.com/MMQYQ1g.png"
image5:
  size: 42x39
  position: 43x525
  src: "https://i.imgur.com/e1VUtT8.png"
image6:
  size: 42x62
  position: 42x607
  src: "https://i.imgur.com/yghRV47.png"
image7:
  size: 60x52
  position: 340x212
  src: "https://i.imgur.com/wKJO9YZ.png"
image8:
  size: 64x53
  position: 338x296
  src: "https://i.imgur.com/GDKRFWA.png"
image9:
  size: 63x63
  position: 338x414
  src: "https://i.imgur.com/ZhPRdFl.png"
image10:
  size: 77x56
  position: 331x544
  src: "https://i.imgur.com/laYrvcw.png"
image11:
  size: 56x71
  position: 342x624
  src: "https://i.imgur.com/er6QXdS.png"
image12:
  size: 35x72
  position: 678x196
  src: "https://i.imgur.com/QATJJDa.png"
image13:
  size: 68x48
  position: 662x340
  src: "https://i.imgur.com/bBv6W2i.png"
image14:
  size: 80x78
  position: 656x485
  src: "https://i.imgur.com/pKPvXnn.png"
image15:
  size: 180x90
  position: 737x608
  src: "https://i.imgur.com/FyuctZR.png"
```

## 5.10 Vanderbilt University

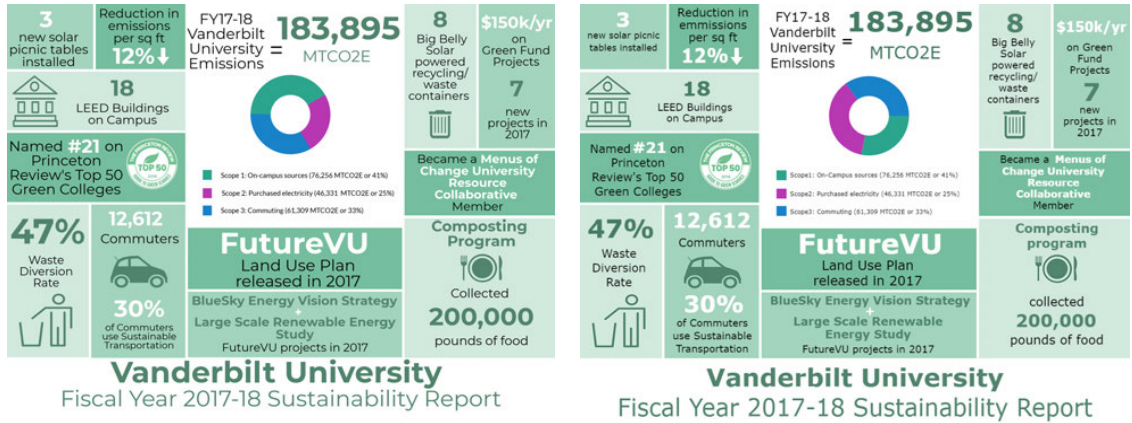

**Fig. 80.** Original infographic by Vanderbilt University (left) and its corresponding generated infographic (right), using the specification in Table 18

**Table 18.** Specification of the infografic shown at the right of Fig. 80

```
bgsize: 1100x850
head:
  position: '0x713'
  size: 1100x137
  bgcolor: ffffff
  title:
    position: 550x770
    value: "Vanderbilt University"
    maxwidth: 900
    align: center
    color: 51846e
    font: bold 50px Verdana
  subtitle:
    position: 550x830
    value: "Fiscal Year 2017-18 Sustainability Report"
    maxwidth: 950
    align: center
    color: 51846e
    font: 50px Verdana
foot: off
bgcolor: white
box1:
  size: 173x131
  position: 0x0
  bgcolor: a3d4be
box2:
  size: 180x131
  position: 178x0
  bgcolor: 75be9e
box3:
  size: 358x119
  position: '0x137'
  bgcolor: d1e9de
box4:
  size: 358x140
  position: '0x260'
  bgcolor: 75be9e
box5:
  size: 164x308
  position: '0x405'
  bgcolor: d1e9de
box6:
```

```
    size: 189x308
    position: 169x405
    bgcolor: a3d4be
box7:
    size: 430x124
    position: 363x451
    bgcolor: 75be9e
box8:
    size: 430x133
    position: 363x580
    bgcolor: a3d4be
box9:
    size: 144x291
    position: 798x0
    bgcolor: d1e9de
box10:
    size: 153x291
    position: 947x0
    bgcolor: a3d4be
box11:
    size: 302x131
    position: 798x296
    bgcolor: 75be9e
box12:
    size: 302x280
    position: 798x433
    bgcolor: d1e9de
box13:
    size: 20x20
    position: 393x335
    bgcolor: 2ca58d
box14:
    size: 20x20
    position: 393x372
    bgcolor: b737b2
box15:
    size: 20x20
    position: 393x407
    bgcolor: 1982c9
titletext1:
    font: bold 40px Verdana
    color: ffffff
    position: 86x46
    align: center
    maxwidth: 145
    value: "3"
titletext2:
    font: bold 40px Verdana
    color: ffffff
    position: 250x118
    align: center
    maxwidth: 95
    value: "12%"
titletext3:
    font: bold 40px Verdana
    color: 477c65
    position: 233x185
    align: center
    value: "18"
titletext4:
    font: bold 50px Verdana
    color: 477c65
    position: 82x470
    align: center
    value: "47%"
titletext5:
    font: bold 40px Verdana
    color: ffffff
    position: 263x450
    align: center
    value: "12,612"
titletext6:
    font: bold 45px Verdana
    color: ffffff
    position: 268x620
    align: center
```

```
    value: "30%"
titletext7:
  font: bold 55px Verdana
  color: 477c65
  position: 660x55
  align: center
  value: "183,895"
titletext8:
  font: bold 55px Verdana
  color: ffffff
  position: 580x502
  align: center
  value: "FutureVU"
titletext9:
  font: bold 23px Verdana
  color: 4e826c
  position: 580x605
  align: center
  value: "BlueSky Energy Vision Strategy"
  maxwidth: 400
titletext10:
  font: bold 23px Verdana
  color: ffffff
  position: 580x626
  align: center
  value: "+"
titletext11:
  font: bold 50px Verdana
  color: 51846e
  position: 870x56
  align: center
  maxwidth: 145
  value: "8"
titletext12:
  font: bold 30px Verdana
  color: ffffff
  position: 1023x56
  align: center
  maxwidth: 140
  value: "$150k/yr"
titletext13:
  font: bold 50px Verdana
  color: 51846e
  position: 1023x195
  align: center
  maxwidth: 140
  value: "7"
text1:
  font: 18px Verdana
  color: 000000
  maxwidth: 165
  align: center
  position: 90x69
  value: "new solar picnic tables installed"
  lineheight: 25
text2:
  font: 23px Verdana
  color: 000000
  maxwidth: 165
  align: center
  position: 270x25
  value: "Reduction in emmissions per sq ft"
  lineheight: 25
text3:
  font: 20px Verdana
  color: 000000
  maxwidth: 160
  align: center
  position: 233x215
  value: "LEED Buildings on Campus"
  lineheight: 22
text4:
  font: 23px Verdana
  color: 000000
  maxwidth: 145
  position: 20x300
```

```
    value: "Named"
text5:
  font: bold 30px Verdana
  color: ffffff
  maxwidth: 145
  position: 105x300
  value: "#21"
text6:
  font: 23px Verdana
  color: 000000
  maxwidth: 145
  position: 180x300
  value: "on"
text7:
  font: 23px Verdana
  color: 000000
  maxwidth: 200
  align: center
  position: 120x328
  value: "Princeton Review's Top 50 Green Colleges"
  lineheight: 29
text8:
  font: 23px Verdana
  color: 000000
  maxwidth: 100
  align: center
  position: 82x510
  value: "Waste Diversion Rate"
  lineheight: 29
text9:
  font: 23px Verdana
  color: 000000
  maxwidth: 100
  align: center
  position: 268x488
  value: "Commuters"
  lineheight: 29
text10:
  font: 20px Verdana
  color: 000000
  maxwidth: 180
  align: center
  position: 268x649
  value: "of Commuters use Sustainable Transportation"
  lineheight: 26
text11:
  font: 26px Verdana
  color: 000000
  maxwidth: 180
  align: center
  position: 450x35
  value: "FY17-18 Vanderbilt University Emissions"
  lineheight: 31
text12:
  font: 32px Verdana
  color: 000000
  maxwidth: 180
  align: center
  position: 540x85
  value: "="
  lineheight: 29
text13:
  font: 34px Verdana
  color: 407860
  maxwidth: 180
  align: center
  position: 656x110
  value: "MTCO2E"
  lineheight: 31
text14:
  font: 12px Verdana
  color: 0
  maxwidth: 400
  position: 420x349
  value: "Scope1: On-Campus sources (76,256 MTCO2E or 41%)"
  lineheight: 31
```

```
text15:
  font: 12px Verdana
  color: 0
  maxwidth: 400
  position: 420x387
  value: "Scope2: Purchased electricity (46,331 MTCO2E or 25%)"
  lineheight: 31
text16:
  font: 12px Verdana
  color: 0
  maxwidth: 400
  position: 420x423
  value: "Scope3: Commuting (61,309 MTCO2E or 33%)"
  lineheight: 31
text17:
  font: 26px Verdana
  color: 000000
  maxwidth: 250
  align: center
  position: 580x535
  value: "Land Use Plan released in 2017"
  lineheight: 31
text18:
  font: bold 23px Verdana
  color: 4e826c
  position: 580x647
  align: center
  value: "Large Scale Renewable Energy Study"
  maxwidth: 400
  lineheight: 26
text19:
  font: 20px Verdana
  color: 0
  position: 580x702
  align: center
  value: "FutureVU projects in 2017"
  maxwidth: 400
  lineheight: 26
text20:
  font: 20px Verdana
  color: 0
  position: 870x84
  align: center
  value: "Big Belly Solar powered recycling/ waste containers"
  maxwidth: 150
  lineheight: 23
text21:
  font: 20px Verdana
  color: 0
  position: 1023x94
  align: center
  value: "on Green Fund Projects"
  maxwidth: 120
  lineheight: 23
text22:
  font: 20px Verdana
  color: 0
  position: 1023x220
  align: center
  value: "new projects in 2017"
  maxwidth: 120
  lineheight: 23
text23:
  font: 20px Verdana
  color: 0
  position: 840x320
  value: "Became a"
  maxwidth: 120
  lineheight: 23
text24:
  font: bold 20px Verdana
  color: ffffff
  position: 950x320
  value: "Menus of"
  maxwidth: 120
  lineheight: 23
```

```
text25:
  font: bold 20px Verdana
  color: ffffff
  position: 950x344
  align: center
  value: "Change University Resource Collaborative"
  maxwidth: 250
  lineheight: 23
text26:
  font: 20px Verdana
  color: 0
  position: 950x413
  align: center
  value: "Member"
  maxwidth: 250
  lineheight: 23
text27:
  font: bold 25px Verdana
  color: 51846e
  position: 958x460
  align: center
  maxwidth: 140
  lineheight: 30
  value: "Composting program"
text28:
  font: 25px Verdana
  color: 0
  position: 958x608
  align: center
  maxwidth: 140
  lineheight: 30
  value: "Collected"
text29:
  font: bold 35px Verdana
  color: 51846e
  position: 958x650
  align: center
  maxwidth: 140
  lineheight: 30
  value: "200,000"
text30:
  font: 25px Verdana
  color: 0
  position: 958x685
  align: center
  maxwidth: 200
  lineheight: 30
  value: "pounds of food"
image1:
  position: 300x86
  size: 27x38
  src: "https://i.imgur.com/j66QQWf.png"
image2:
  position: 9x141
  size: 95x109
  src: "https://i.imgur.com/ylG9moU.png"
image3:
  position: 236x270
  size: 109x106
  src: "https://i.imgur.com/Mb9M0Uh.png"
image4:
  position: 18x582
  size: 118x110
  src: "https://i.imgur.com/roGotLf.png"
image5:
  position: 206x504
  size: 118x72
  src: "https://i.imgur.com/hzby3M2.png"
image6:
  position: 846x213
  size: 47x60
  src: "https://i.imgur.com/7Am24hF.png"
image7:
  position: 907x501
  size: 94x64
  src: "https://i.imgur.com/u64yStm.png"
```

```
piechart1:
  colors: 2ca58d,b737b2,1982c9
  position: 488x135
  type: donut
  size: 80
  padding: 10
  title: "Female/Male ratio 2018"
  bgcolor: ffffff
  showpercentage: off
  showtitle: off
  showlegend: off
  data:
    "Scope1": "76,25"
    "Scope2": "46,33"
    "Scope3": "61,31"
```

# 6 User stories for the DSL and the interpreter

In this section we present the list of requirements, expressed as user stories [12]. In Table 19, the columns contain the following information:

- **ID**. An identifier for the user story under with the format US*i*, where *i* is just an incremental number.
- **Statement**. The actual user story, expressed according to the Connextra template, which is widely adopted in industry [13].
- **Pri**. The priority we assigned to this user story, using the MoSCoW method [14]. The cells contain one of these values:
    - **M**. It means that the requirement *must* be implemented for the proof of concept.
    - **S**. It means that the requirement *should* be implemented for the proof of concept.
    - **C**. It means that the requirement *could* be implemented for the proof of concept.
    - **W**. It means that the requirement *won't* be implemented for the proof of concept.
- **Imp**. Whether the requirement has been implemented in the first proof of concept (1), or not (0).

**Table 19.** List of requirements

| ID | Statement | Pri | Imp |
| --- | --- | --- | --- |
| US1 | As a user, I want to create an infographic summarizing the social and environmental accounting report, so I can easily share the results of the social and environmental accounting report as an image with my stakeholders. | M | 1 |
| US2 | As a user, I want to upload my own infographic specification, so that I can tweak the infographic based on my needs. | M | 1 |
| US3 | As a user, I want to select which infographic specification I will use, so that I can adhere to the design of my organisation or network. | M | 1 |
| US4 | As a user, I want to remove the infographic specification, so that the infographic specification is deleted from the database. | M | 1 |
| US5 | As a network user, I want to upload an infographic specification, so that users of organisations in my network can use this infographic specification. | M | 1 |
| US6 | As an organisation user, I want to upload an infographic specification, so that users in my organisation can use this infographic specification. | M | 1 |
| US7 | As a user, I want to define the background of the infographic by selecting a color hex code, so I can manipulate the color of the background. | M | 1 |
| US8 | As a user, I want to add text to the infographic, so that the infographic is populated with text. | M | 1 |
| US9 | As a user, I want to change the size of the canvas of the infographic, so that I can tweak the infographic based on my needs. | M | 1 |

| ID | Statement | Pri | Imp |
|---|---|---|---|
| US10 | As a user, I want to create pie charts on my infographic, so I can display data in a visual manner. | M | 1 |
| US11 | As a user, I want to create bar charts on my infographic, so I can display data in a visual manner. | M | 1 |
| US12 | As a user, I want to export the infographic as a PNG image, so I can use this to share with my stakeholders. | M | 1 |
| US13 | As a user, I want to use the values of indicators (direct and indirect) to populate my charts, so that my charts contain valuable data. | M | 1 |
| US14 | As a user, I want to use the values of indicators (direct and indirect) to populate parts of my text, so that my text contains valuable data. | M | 1 |
| US15 | As a user, I want to be able to use a head, body and foot section, so that my infographic is divided into sections. | M | 1 |
| US16 | As a user, I want to be able to create boxes, so that my infographic is divided into configurable sections. | M | 1 |
| US17 | As a user, I want to add a type basic element, so that my infographic is styled the same way as reports are style in OpenESEA. | M | 1 |
| US18 | As a user, I want to create picturegraphs on my infographic, so I can display data in a visual manner. | S | 1 |
| US19 | As a user, I want to define the background of the infographic by selecting color swatches, so I can manipulate the color of the background. | S | 0 |
| US20 | As a user, I want to define the background of the infographic by selecting an image from a database, so I can manipulate the looks of the background. | C | 0 |
| US21 | As a user, I want to define the background of the infographic by uploading an image, so I can manipulate the looks of the background. | C | 1 |
| US22 | As a user, I want to define the background of the infographic by selecting a pattern, so I can manipulate the looks of the background. | C | 1 |
| US23 | As a user, I want to be able to select an image from a database and use this image on the infographic, so that I can manipulate the looks of the infographic. | C | 0 |
| US24 | As a user, I want to be able to upload an image and use this image on the infographic, so that I can manipulate the looks of the infographic. | C | 1 |
| US25 | As a user, I want to be able to select icons and shapes and use these on the infographic, so that I can manipulate the looks of the infographic. | C | 0 |
| US26 | As a user, I want to create tables on my infographic, so I can display data in a visual manner. | C | 0 |
| US27 | As a user, I want to create grouped bar charts on my infographic, so I can display data in a visual manner. | C | 0 |
| US28 | As a user, I want to create stacked bar charts on my infographic, so I can display data in a visual manner. | C | 0 |
| US29 | As a user, I want to export the infographic as a PDF file, so I can use this to share with my stakeholders. | C | 0 |
| US30 | As a user, I want to export the infographic as a JPG image, so I can use this to share with my stakeholders. | C | 0 |
| US31 | As a user, I want to create scatterplots on my infographic, so I can display data in a visual manner. | W | 0 |
| US32 | As a user, I want to create line charts on my infographic, so I can display data in a visual manner. | W | 0 |
| US33 | As a user, I want to create Gantt charts on my infographic, so I can display data in a visual manner. | W | 0 |
| US34 | As a user, I want to create area charts on my infographic, so I can display data in a visual manner. | W | 0 |
| US35 | As a user, I want to create bubble charts on my infographic, so I can display data in a visual manner. | W | 0 |
| US36 | As a user, I want to create map charts on my infographic, so I can display data in a visual manner. | W | 0 |

In the following, we present the pre-requirements specification traceability. Table 20 and Table 21 trace the user stories to the following sources:

- **O**. The project manager and product owner of the openESEA tool. User stories having a 1 in this cell correspond to needs expressed by these stakeholders. Since these are the authors Sergio España and Vijanti Ramautar, respectively, these requirements can be considered as having been formulated by the research team.
- **CM** refers to the conceptual model presented in Fig. 1.
- **Ci** (that is, columns C1 to C22) refer to the infographic component types presented in Table 6.
- **Fi** (that is, F1 to F28) refer to the generic features of infographic design tools, presented in Table 8.

**Table 20.** Traceability matrix, part 1

| ID | O | CM | C1 | C2 | C3 | C4 | C5 | C6 | C7 | C8 | C9 | C10 | C11 | C12 | C13 | C14 | C15 | C16 | C17 | C18 | C19 | C20 | C21 | C22 |
|---|---|---|---|---|---|---|---|---|---|---|---|---|---|---|---|---|---|---|---|---|---|---|---|---|
| US1 | 1 | | | | | | | | | | | | | | | | | | | | | | | |
| US2 | 1 | | | | | | | | | | | | | | | | | | | | | | | |
| US3 | 1 | | | | | | | | | | | | | | | | | | | | | | | |
| US4 | 1 | | | | | | | | | | | | | | | | | | | | | | | |
| US5 | 1 | | | | | | | | | | | | | | | | | | | | | | | |
| US6 | 1 | | | | | | | | | | | | | | | | | | | | | | | |
| US7 | | | | | | | | | | | | | | | | | | | | | | | | |
| US8 | | 1 | | 1 | 1 | | | | | 1 | 1 | | | | | | | | | | 1 | 1 | | 1 |
| US9 | | | | | | | | | | | | | | | | | | | | | | | | |
| US10 | | | | | | | | | | | | | 1 | | | | | | | | | | | |
| US11 | | | | | | | | | | | | | | | 1 | | | | | | | | | |
| US12 | | | | | | | | | | | | | | | | | | | | | | | | |
| US13 | | 1 | | | | | | | | | | | | | | | | | | | | | | |
| US14 | | 1 | | | | | | | | | | | | | | | | | | | | | | |
| US15 | | 1 | 1 | | | | 1 | | | | | | | | | | | | | 1 | | | | |
| US16 | | 1 | | | | | | 1 | | | | | | | | | | | | | | | | |
| US17 | 1 | | | | | | | | | | | | | | | | | | | | | | | |
| US18 | | | | | | | | | | | | | | | | 1 | | | | | | | | |
| US19 | | | | | | | | | | | | | | | | | | | | | | | | |
| US20 | | | | | | | | | | | | | | | | | | | | | | | | |
| US21 | | | | | | | | | | | | | | | | | | | | | | | | |
| US22 | | | | | | | | | | | | | | | | | | | | | | | | |
| US23 | | | | | | | | | | | | | | | | | | | | | | | | |
| US24 | | | | | | | | | | | | | | | | | | | | | | | | |
| US25 | | | | | | | | | | | | | | | | | | | | | | | | |
| US26 | | | | | | | | | | | | | 1 | | | | | | | | | | | |
| US27 | | | | | | | | | | | | | | | 1 | | | | | | | | | |
| US28 | | | | | | | | | | | | | | | | | 1 | | | | | | | |
| US29 | | | | | | | | | | | | | | | | | | | | | | | | |
| US30 | | | | | | | | | | | | | | | | | | | | | | | | |
| US31 | | | | | | | | | | | | | | | | | | | | | | | | |
| US32 | | | | | | | | | | | | | | | | | | | 1 | | | | | |
| US33 | | | | | | | | | | | | | | | | | | | | | | | | |
| US34 | | | | | | | | | | | | | | | | | | | | | | | | |
| US35 | | | | | | | | | | | | | | | | | | | | | | | | |
| US36 | | | | | | | | | | | | | | | | | | | | 1 | | | | |

**Table 21.** Traceability matrix, part 2

| ID | F1 | F2 | F3 | F4 | F5 | F6 | F7 | F8 | F9 | F10 | F11 | F12 | F13 | F14 | F15 | F16 | F17 | F18 | F19 | F20 | F21 | F22 | F23 | F24 | F25 | F26 | F27 | F28 |
|---|---|---|---|---|---|---|---|---|---|---|---|---|---|---|---|---|---|---|---|---|---|---|---|---|---|---|---|---|
| US1 | | | | | | | | | | | | | | | | | | | | | | | | | | | | |
| US2 | | | | | | | | | | | | | | | | | | | | | | | | | | | | |
| US3 | | | | | | | | | | | | | | | | | | | | | | | | | | | | |
| US4 | | | | | | | | | | | | | | | | | | | | | | | | | | | | |
| US5 | | | | | | | | | | | | | | | | | | | | | | | | | | | | |
| US6 | | | | | | | | | | | | | | | | | | | | | | | | | | | | |

| ID | F1 | F2 | F3 | F4 | F5 | F6 | F7 | F8 | F9 | F10 | F11 | F12 | F13 | F14 | F15 | F16 | F17 | F18 | F19 | F20 | F21 | F22 | F23 | F24 | F25 | F26 | F27 | F28 |
|---|---|---|---|---|---|---|---|---|---|---|---|---|---|---|---|---|---|---|---|---|---|---|---|---|---|---|---|---|
| US7 | 1 | | | | | | | | | | | | | | | | | | | | | | | | | | | |
| US8 | | | | | | | 1 | 1 | 1 | | | | | | | | | | | | | | | | | | | |
| US9 | | | | | | | | | | 1 | | | | | | | | | | | | | | | | | | |
| US10 | | | | | | | | | | | | | | | | 1 | | | | | | | | | | | | |
| US11 | | | | | | | | | | | | | | | | | | 1 | | | | | | | | | | |
| US12 | | | | | | | | | | | | | | | | | | | | | | | | | | | 1 | |
| US13 | | | | | | | | | | | | | | | | | | | | | | | | | | | | |
| US14 | | | | | | | | | | | | | | | | | | | | | | | | | | | | |
| US15 | | | | | | | | | | | | | | | | | | | | | | | | | | | | |
| US16 | | | | | | | | | | | | | | | | | | | | | | | | | | | | |
| US17 | | | | | | | | | | | | | | | | | | | | | | | | | | | | |
| US18 | | | | | | | | | | | | | | | 1 | | | | | | | | | | | | | |
| US19 | | 1 | | | | | | | | | | | | | | | | | | | | | | | | | | |
| US20 | | | 1 | | | | | | | | | | | | | | | | | | | | | | | | | |
| US21 | | | | 1 | | | | | | | | | | | | | | | | | | | | | | | | |
| US22 | | | | | 1 | | | | | | | | | | | | | | | | | | | | | | | |
| US23 | | | | | | | | | | | | 1 | | | | | | | | | | | | | | | | |
| US24 | | | | | | | | | | | | | 1 | | | | | | | | | | | | | | | |
| US25 | | | | | | | | | | | | | | 1 | | | | | | | | | | | | | | |
| US26 | | | | | | | | | | | | | | | | 1 | | | | | | | | | | | | |
| US27 | | | | | | | | | | | | | | | | | | | 1 | | | | | | | | | |
| US28 | | | | | | | | | | | | | | | | | | | | 1 | | | | | | | | |
| US29 | | | | | | | | | | | | | | | | | | | | | | | | | | 1 | | |
| US30 | | | | | | | | | | | | | | | | | | | | | | | | | | | | 1 |
| US31 | | | | | | | | | | | | | | | | | | 1 | | | | | | | | | | |
| US32 | | | | | | | | | | | | | | | | | | | | | 1 | | | | | | | |
| US33 | | | | | | | | | | | | | | | | | | | | | | | 1 | | | | | |
| US34 | | | | | | | | | | | | | | | | | | | | | | | | 1 | | | | |
| US35 | | | | | | | | | | | | | | | | | | | | | | | | | 1 | | | |
| US36 | | | | | | | | | | | | | | | | | | | | | | | | | | 1 | | |

# 7 Comparative experiment

## 7.1 Example of the survey

**Fig. 81.** Survey example (page 1 of 32)

3. **What is the highest degree or level of school you have completed? \***
   *Mark only one oval.*

   ◯ Less than a high school diploma
   ◯ High school degree or equivalent
   ◯ Bachelor's degree
   ◯ Master's degree
   ◯ Doctorate

4. **What is your version number? \***
   *Mark only one oval.*

   ◯ 1
   ◯ 2    *After the last question in this section, skip to "Explanation infographic ratings."*

5. **What is your e-mailaddress? (not mandatory)**

   _______________________________
   _______________________________
   _______________________________
   _______________________________

## Explanation infographic ratings

You will be rating the infographics on ten different terms. The first five regarding "classical aesthetics" refers to the historical notion of aesthetics, which entails orderliness, symmetry, proportion.

Aesthetic design refers to the visual attractiveness of a product.
Symmetrical design refers to beauty due to balanced proportions.
Pleasant design refers to an enjoyable feeling due to the design.
Organized design refers to a design arranged or structured in a systematic way.
Clean design refers to a design that is precise and consistent.

The last five regarding "expressive aesthetics" referring to the creativity and originality of the designer, including his/her ability to go beyond conventions in the enrichment of his/her creation.

Creative design refers to the inventivity of the design.
Sophisticated design refers to elegant, well-thought out and carefully planned design.
Original design refers to the uniqueness of the design.
Using special effects refers to the use of illusions or visual tricks to enhance the design.
Fascinating design refers to a design that is extremely interesting.

## Infographic #1

Right-click the infographic and select 'Open image in new tab' for a larger version of the infographic.

A description for the different terms used in the rating of infographics may be found here (http://tiny.cc/infographics).

**Fig. 82.** Survey example (page 2 of 32)

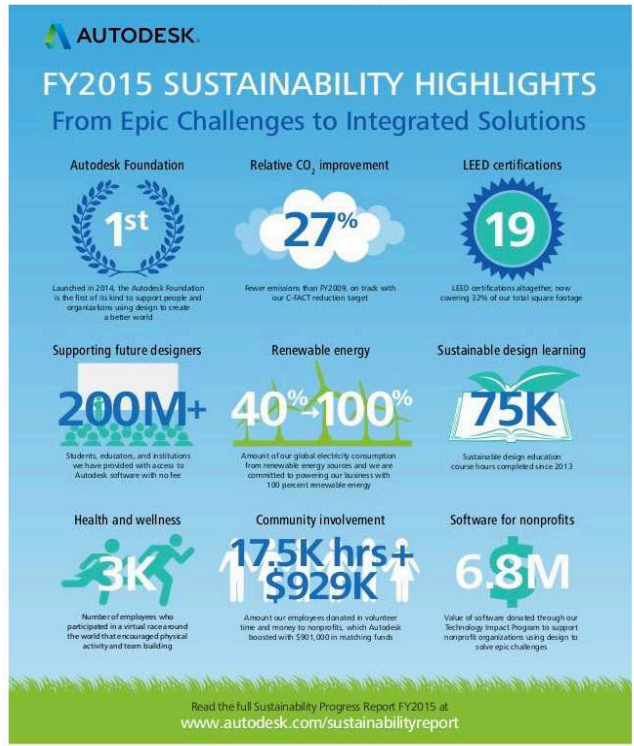

**Fig. 83.** Survey example (page 3 of 32)

**7. Expressive aesthetics *** 
Mark only one oval per row.

|  | Disagree 1 | 2 | 3 | 4 | 5 Agree |
|---|---|---|---|---|---|
| Creative design | ○ | ○ | ○ | ○ | ○ |
| Sophisticated design | ○ | ○ | ○ | ○ | ○ |
| Original design | ○ | ○ | ○ | ○ | ○ |
| Using special effects | ○ | ○ | ○ | ○ | ○ |
| Fascinating design | ○ | ○ | ○ | ○ | ○ |

**8. Do you think the infographic is created manually by a designer (using image-editing software) or whether it is generated using an automated software tool? *** 
Mark only one oval.

○ Manually created by a designer

○ Generated using an automated tool

## Infographic #2
Right-click the infographic and select 'Open image in new tab' for a larger version of the infographic.

A description for the different terms used in the rating of infographics may be found here (http://tiny.cc/infographics).

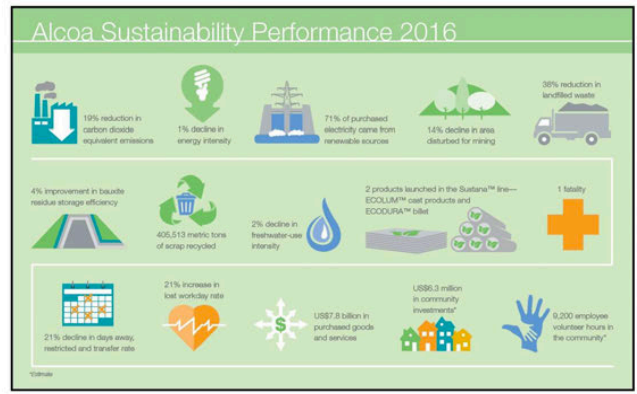

**9. Classical aesthetics *** 
Mark only one oval per row.

|  | Disagree 1 | 2 | 3 | 4 | 5 Agree |
|---|---|---|---|---|---|
| Aesthetic design | ○ | ○ | ○ | ○ | ○ |
| Symmetrical design | ○ | ○ | ○ | ○ | ○ |
| Pleasant design | ○ | ○ | ○ | ○ | ○ |
| Organized design | ○ | ○ | ○ | ○ | ○ |
| Clean design | ○ | ○ | ○ | ○ | ○ |

**Fig. 84.** Survey example (page 4 of 32)

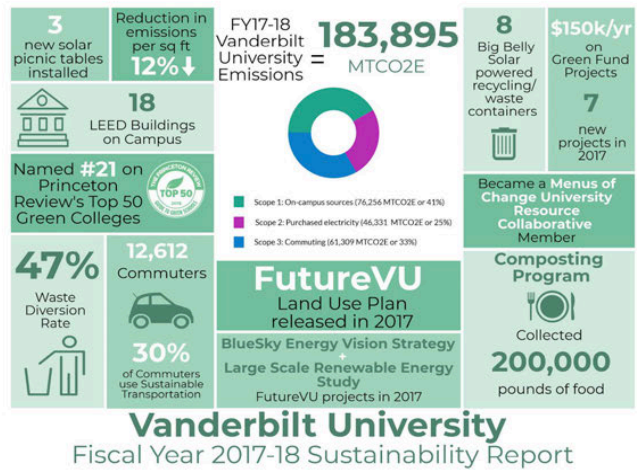

**Fig. 85.** Survey example (page 5 of 32)

**12. Classical aesthetics** *

Mark only one oval per row.

|  | Disagree 1 | 2 | 3 | 4 | 5 Agree |
|---|---|---|---|---|---|
| Aesthetic design | ◯ | ◯ | ◯ | ◯ | ◯ |
| Symmetrical design | ◯ | ◯ | ◯ | ◯ | ◯ |
| Pleasant design | ◯ | ◯ | ◯ | ◯ | ◯ |
| Organized design | ◯ | ◯ | ◯ | ◯ | ◯ |
| Clean design | ◯ | ◯ | ◯ | ◯ | ◯ |

**13. Expressive aesthetics** *

Mark only one oval per row.

|  | Disagree 1 | 2 | 3 | 4 | 5 Agree |
|---|---|---|---|---|---|
| Creative design | ◯ | ◯ | ◯ | ◯ | ◯ |
| Sophisticated design | ◯ | ◯ | ◯ | ◯ | ◯ |
| Original design | ◯ | ◯ | ◯ | ◯ | ◯ |
| Using special effects | ◯ | ◯ | ◯ | ◯ | ◯ |
| Fascinating design | ◯ | ◯ | ◯ | ◯ | ◯ |

**14. Do you think the infographic is created manually by a designer (using image-editing software) or whether it is generated using an automated software tool?** *

Mark only one oval.

◯ Manually created by a designer

◯ Generated using an automated tool

## Infographic #4

Right-click the infographic and select 'Open image in new tab' for a larger version of the infographic.

A description for the different terms used in the rating of infographics may be found here (http://tiny.cc/infographics).

**Fig. 86.** Survey example (page 6 of 32)

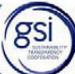

**Fig. 87.** Survey example (page 7 of 32)

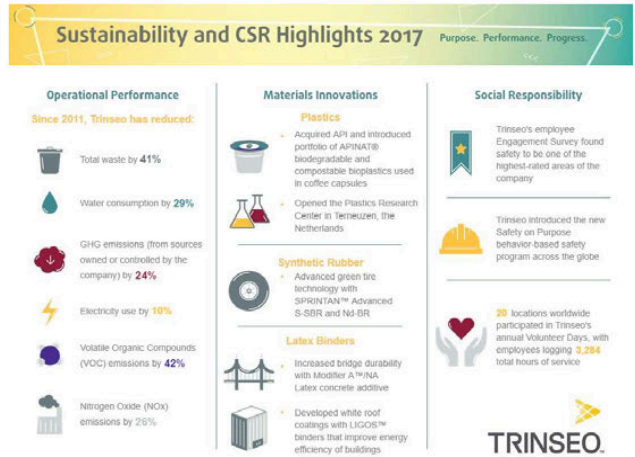

**Fig. 88.** Survey example (page 8 of 32)

**18. Classical aesthetics \***

Mark only one oval per row.

|                    | Disagree 1 | 2 | 3 | 4 | 5 Agree |
|--------------------|------------|---|---|---|---------|
| Aesthetic design   | ○          | ○ | ○ | ○ | ○       |
| Symmetrical design | ○          | ○ | ○ | ○ | ○       |
| Pleasant design    | ○          | ○ | ○ | ○ | ○       |
| Organized design   | ○          | ○ | ○ | ○ | ○       |
| Clean design       | ○          | ○ | ○ | ○ | ○       |

**19. Expressive aesthetics \***

Mark only one oval per row.

|                       | Disagree 1 | 2 | 3 | 4 | 5 Agree |
|-----------------------|------------|---|---|---|---------|
| Creative design       | ○          | ○ | ○ | ○ | ○       |
| Sophisticated design  | ○          | ○ | ○ | ○ | ○       |
| Original design       | ○          | ○ | ○ | ○ | ○       |
| Using special effects | ○          | ○ | ○ | ○ | ○       |
| Fascinating design    | ○          | ○ | ○ | ○ | ○       |

**20. Do you think the infographic is created manually by a designer (using image-editing software) or whether it is generated using an automated software tool? \***

Mark only one oval.

○ Manually created by a designer

○ Generated using an automated tool

## Infographic #6

Right-click the infographic and select 'Open image in new tab' for a larger version of the infographic.

A description for the different terms used in the rating of infographics may be found here (http://tiny.cc/infographics).

**Fig. 89.** Survey example (page 9 of 32)

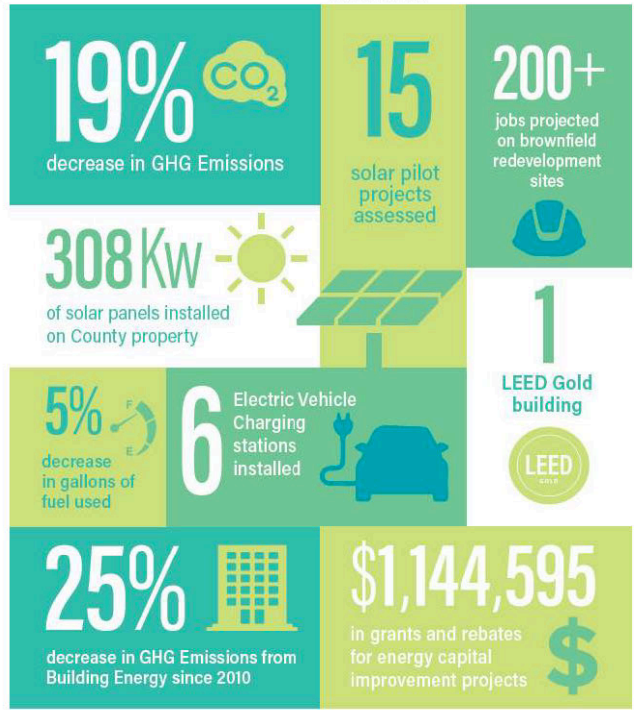
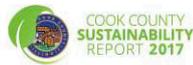

**Fig. 90.** Survey example (page 10 of 32)

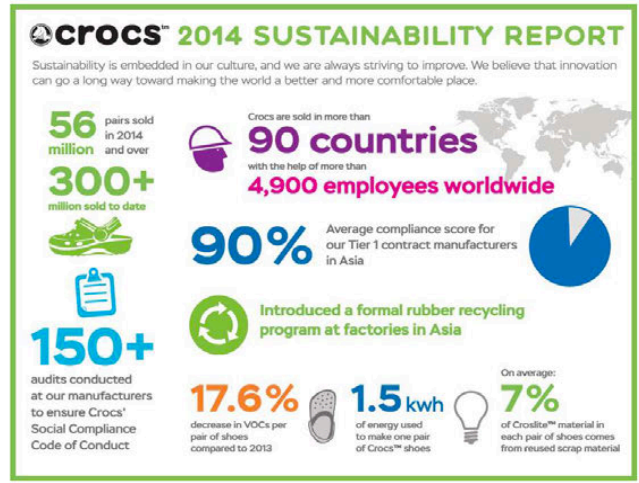

**Fig. 91.** Survey example (page 11 of 32)

### 24. Classical aesthetics *
Mark only one oval per row.

|  | Disagree 1 | 2 | 3 | 4 | 5 Agree |
|---|---|---|---|---|---|
| Aesthetic design | ○ | ○ | ○ | ○ | ○ |
| Symmetrical design | ○ | ○ | ○ | ○ | ○ |
| Pleasant design | ○ | ○ | ○ | ○ | ○ |
| Organized design | ○ | ○ | ○ | ○ | ○ |
| Clean design | ○ | ○ | ○ | ○ | ○ |

### 25. Expressive aesthetics *
Mark only one oval per row.

|  | Disagree 1 | 2 | 3 | 4 | 5 Agree |
|---|---|---|---|---|---|
| Creative design | ○ | ○ | ○ | ○ | ○ |
| Sophisticated design | ○ | ○ | ○ | ○ | ○ |
| Original design | ○ | ○ | ○ | ○ | ○ |
| Using special effects | ○ | ○ | ○ | ○ | ○ |
| Fascinating design | ○ | ○ | ○ | ○ | ○ |

### 26. Do you think the infographic is created manually by a designer (using image-editing software) or whether it is generated using an automated software tool? *
Mark only one oval.

○ Manually created by a designer

○ Generated using an automated tool

## Infographic #8
Right-click the infographic and select 'Open image in new tab' for a larger version of the infographic.

A description for the different terms used in the rating of infographics may be found here (http://tiny.cc/infographics).

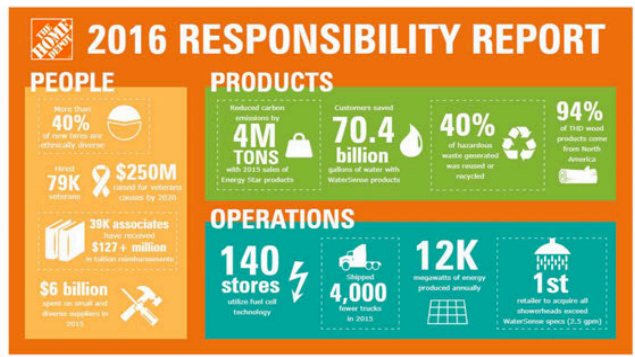

**Fig. 92.** Survey example (page 12 of 32)

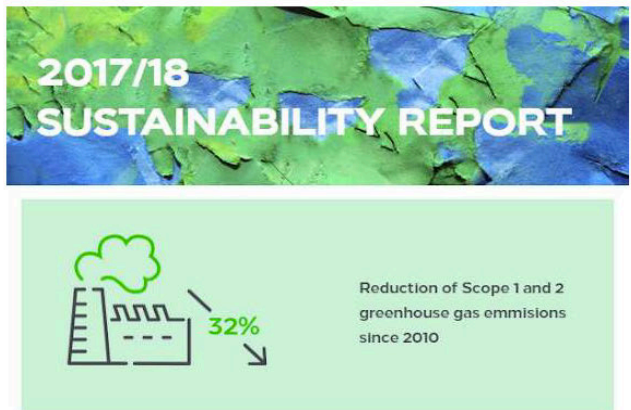

**Fig. 93.** Survey example (page 13 of 32)

Perfect score from the Human Rights Campaign Foundation's Corporate Equality Index for two consecutive years

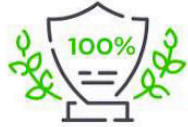

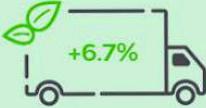

Efficiency improvement in transportation CO2 emissions due to innovative, bio-based packaging materials – like bamboo and sugar cane – driving down packaging weight

Named one of the 100 Best Companies for Working Mothers by Working Mother magazine

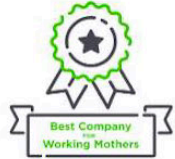

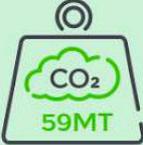

Total CO2 emissions saved by Lenovo's Low-Temperature Solder process at the end of FY 2017/18

Global Week of Service directly impacted 33,000 consumers across 19 countries, benefiting minority populations and increasing access to STEM education

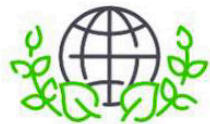

**Fig. 94.** Survey example (page 14 of 32)

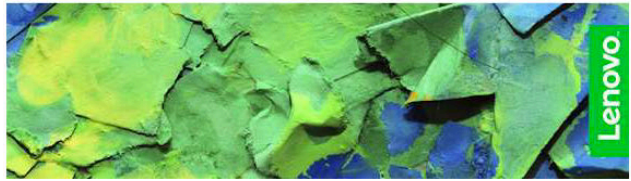

**Fig. 95.** Survey example (page 15 of 32)

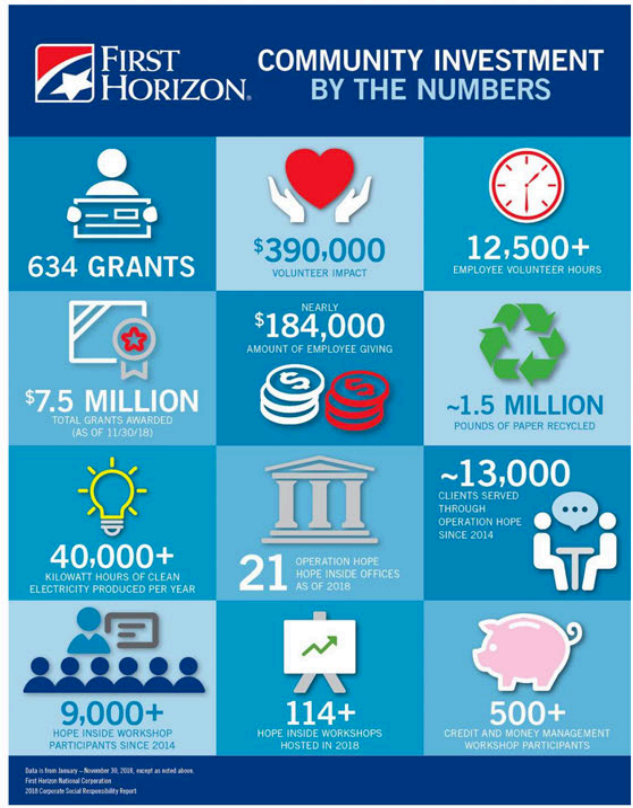

**Fig. 96.** Survey example (page 16 of 32)

**34. Expressive aesthetics** *
Mark only one oval per row.

|  | Disagree 1 | 2 | 3 | 4 | 5 Agree |
|---|---|---|---|---|---|
| Creative design | ○ | ○ | ○ | ○ | ○ |
| Sophisticated design | ○ | ○ | ○ | ○ | ○ |
| Original design | ○ | ○ | ○ | ○ | ○ |
| Using special effects | ○ | ○ | ○ | ○ | ○ |
| Fascinating design | ○ | ○ | ○ | ○ | ○ |

**35. Do you think the infographic is created manually by a designer (using image-editing software) or whether it is generated using an automated software tool?** *
Mark only one oval.

○ Manually created by a designer

○ Generated using an automated tool

*Stop filling out this form.*

## Explanation infographic ratings

You will be rating the infographics on ten different terms. The first five regarding "classical aesthetics" refers to the historical notion of aesthetics, which entails orderliness, symmetry, proportion.

Aesthetic design refers to the visual attractiveness of a product.
Symmetrical design refers to beauty due to balanced proportions.
Pleasant design refers to an enjoyable feeling due to the design.
Organized design refers to a design arranged or structured in a systematic way.
Clean design refers to a design that is precise and consistent.

The last five regarding "expressive aesthetics" referring to the creativity and originality of the designer, including his/her ability to go beyond conventions in the enrichment of his/her creation.

Creative design refers to the inventivity of the design.
Sophisticated design refers to elegant, well-thought out and carefully planned design.
Original design refers to the uniqueness of the design.
Using special effects refers to the use of illusions or visual tricks to enhance the design.
Fascinating design refers to a design that is extremely interesting.

*Skip to question 36.*

## Infographic #1

Right-click the infographic and select 'Open image in new tab' for a larger version of the infographic.

A description for the different terms used in the rating of infographics may be found here (http://tiny.cc/infographics).

**Fig. 97.** Survey example (page 17 of 32)

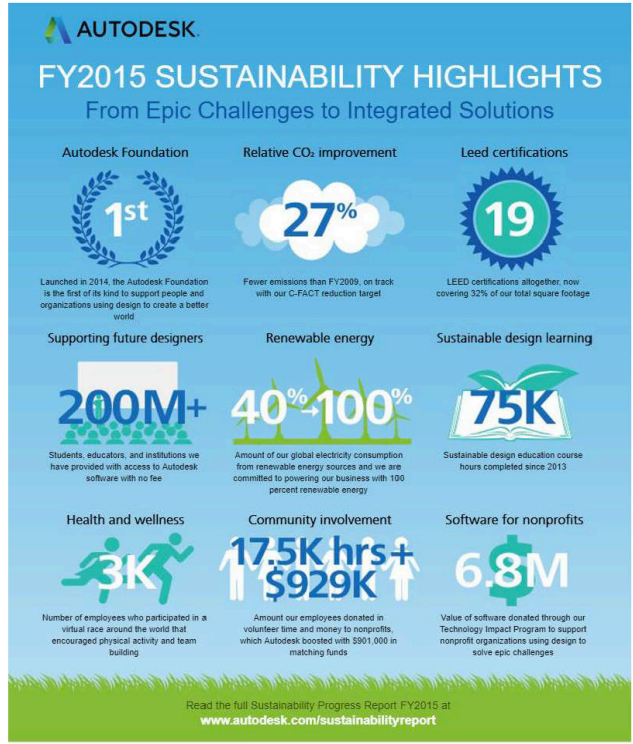
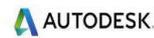

**Fig. 98.** Survey example (page 18 of 32)

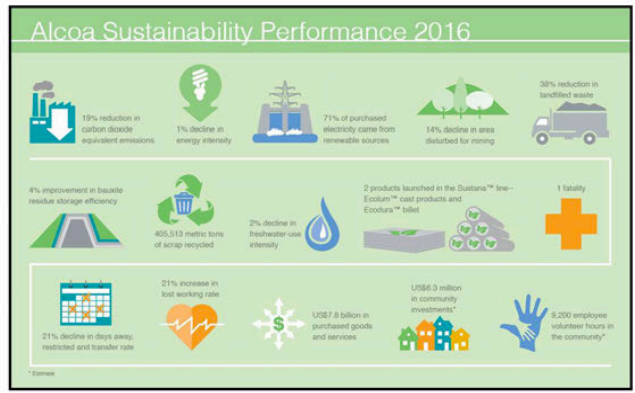

**Fig. 99.** Survey example (page 19 of 32)

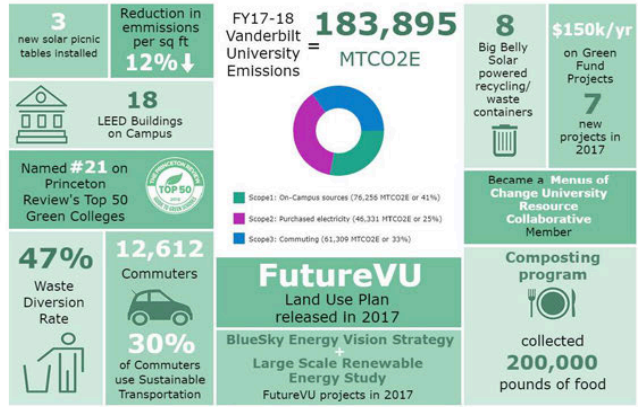

**Fig. 100.** Survey example (page 20 of 32)

**42. Classical aesthetics** *

Mark only one oval per row.

|  | Disagree 1 | 2 | 3 | 4 | 5 Agree |
|---|---|---|---|---|---|
| Aesthetic design |  |  |  |  |  |
| Symmetrical design |  |  |  |  |  |
| Pleasant design |  |  |  |  |  |
| Organized design |  |  |  |  |  |
| Clean design |  |  |  |  |  |

**43. Expressive aesthetics** *

Mark only one oval per row.

|  | Disagree 1 | 2 | 3 | 4 | 5 Agree |
|---|---|---|---|---|---|
| Creative design |  |  |  |  |  |
| Sophisticated design |  |  |  |  |  |
| Original design |  |  |  |  |  |
| Using special effects |  |  |  |  |  |
| Fascinating design |  |  |  |  |  |

**44. Do you think the infographic is created manually by a designer (using image-editing software) or whether it is generated using an automated software tool?** *

Mark only one oval.

◯ Manually created by a designer

◯ Generated using an automated tool

## Infographic #4

Right-click the infographic and select 'Open image in new tab' for a larger version of the infographic.

A description for the different terms used in the rating of infographics may be found here (http://tiny.cc/infographics).

**Fig. 101.** Survey example (page 21 of 32)

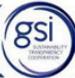

**Fig. 102.** Survey example (page 22 of 32)

**46. Expressive aesthetics** *

Mark only one oval per row.

|  | Disagree 1 | 2 | 3 | 4 | 5 Agree |
|---|---|---|---|---|---|
| Creative design | ○ | ○ | ○ | ○ | ○ |
| Sophisticated design | ○ | ○ | ○ | ○ | ○ |
| Original design | ○ | ○ | ○ | ○ | ○ |
| Using special effects | ○ | ○ | ○ | ○ | ○ |
| Fascinating design | ○ | ○ | ○ | ○ | ○ |

**47. Do you think the infographic is created manually by a designer (using image-editing software) or whether it is generated using an automated software tool?** *

Mark only one oval.

○ Manually created by a designer

○ Generated using an automated tool

## Infographic #5

Right-click the infographic and select 'Open image in new tab' for a larger version of the infographic.

A description for the different terms used in the rating of infographics may be found here (http://tiny.cc/infographics).

### Sustainability and CSR Highlights 2017 — Purpose. Performance. Progress.

**Operational Performance**

Since 2011, Trinseo has reduced:
- Total waste by **41%**
- Water consumption by **29%**
- GHG emissions (from sources owned or controlled by the company) by **24%**
- Electricity use by **10%**
- Volatile Organic Compounds (VOC) emissions by **42%**
- Nitrogen oxide (NOx) emissions by **26%**

**Materials Innovations**

Plastics
- Acquired API and introduced portfolio of APINAT® biodegradable and compostable bioplastics used in coffee capsules
- Opened the Plastics Research Center in Terneuzen, the Netherlands

Synthetic Rubber
- Advanced green tire technology with SPRINTAN™ Advanced S-SBR and Nd-BR

Latex Binders
- Increased bridge durability with Modifier A™/NA Latex concrete additive
- Developed white roof coatings with LIGOS™ binders that improve energy efficiency of buildings

**Social Responsibility**

- Trinseo's employee Engagement Survey found safety to be one of the highest-rated areas of the company
- Trinseo introduced the new Safety on Purpose behavior-based safety program across the globe
- 26 locations worldwide participated in Trinseo's annual Volunteer Days, with employees logging 3,284 total hours of service

**TRINSEO**

**Fig. 103.** Survey example (page 23 of 32)

**48. Classical aesthetics** *

Mark only one oval per row.

|  | Disagree 1 | 2 | 3 | 4 | 5 Agree |
|---|---|---|---|---|---|
| Aesthetic design | ◯ | ◯ | ◯ | ◯ | ◯ |
| Symmetrical design | ◯ | ◯ | ◯ | ◯ | ◯ |
| Pleasant design | ◯ | ◯ | ◯ | ◯ | ◯ |
| Organized design | ◯ | ◯ | ◯ | ◯ | ◯ |
| Clean design | ◯ | ◯ | ◯ | ◯ | ◯ |

**49. Expressive aesthetics** *

Mark only one oval per row.

|  | Disagree 1 | 2 | 3 | 4 | 5 Agree |
|---|---|---|---|---|---|
| Creative design | ◯ | ◯ | ◯ | ◯ | ◯ |
| Sophisticated design | ◯ | ◯ | ◯ | ◯ | ◯ |
| Original design | ◯ | ◯ | ◯ | ◯ | ◯ |
| Using special effects | ◯ | ◯ | ◯ | ◯ | ◯ |
| Fascinating design | ◯ | ◯ | ◯ | ◯ | ◯ |

**50. Do you think the infographic is created manually by a designer (using image-editing software) or whether it is generated using an automated software tool?** *

Mark only one oval.

◯ Manually created by a designer

◯ Generated using an automated tool

### Infographic #6

Right-click the infographic and select 'Open image in new tab' for a larger version of the infographic.

A description for the different terms used in the rating of infographics may be found here (http://tiny.cc/infographics).

**Fig. 104.** Survey example (page 24 of 32)

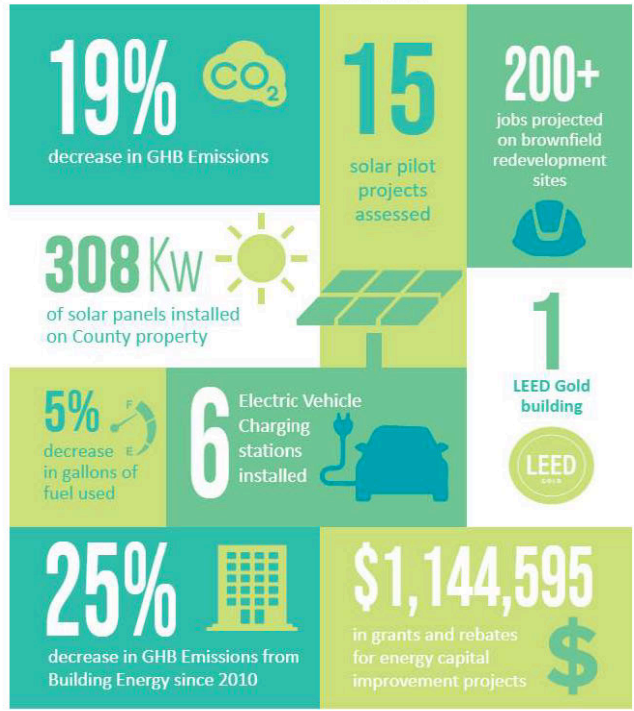

**Fig. 105.** Survey example (page 25 of 32)

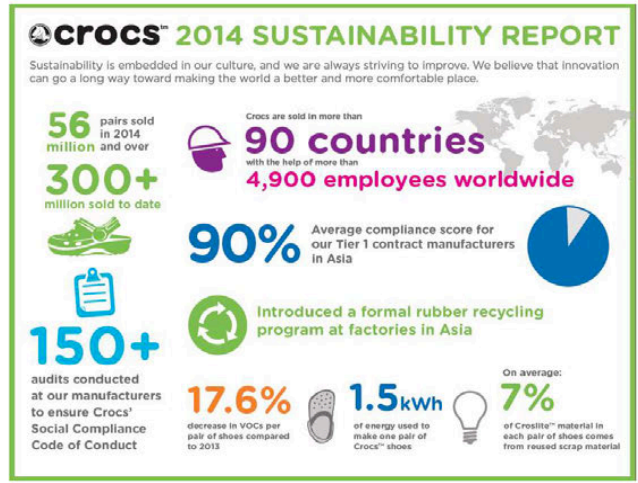

**Fig. 106.** Survey example (page 26 of 32)

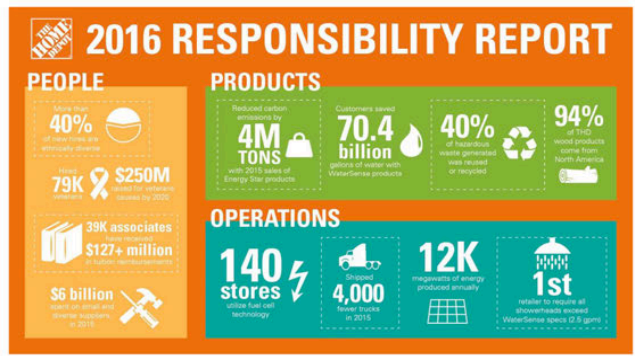

**Fig. 107.** Survey example (page 27 of 32)

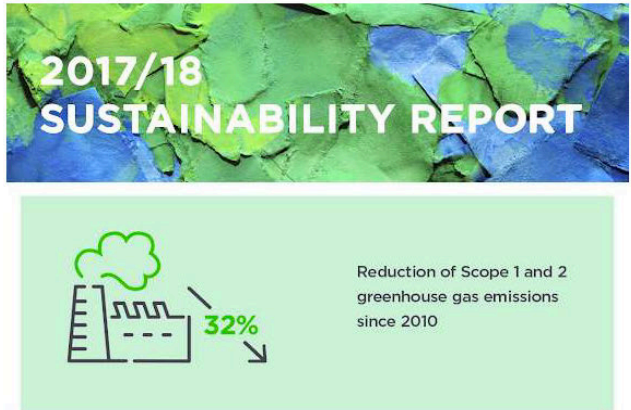

**Fig. 108.** Survey example (page 28 of 32)

Perfect score from the Human Rights Campaign Foundation's Corporate Equality Index for two consecutive years

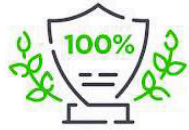

Efficiency improvement in transportation $CO_2$ emissions due to innovative, bio-based packaging materials—like bamboo and sugar cane—driving down packaging weight

+6.7%

Named one of the 100 Best Companies for Working Mothers by *Working Mother* magazine

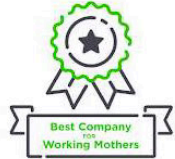

Total $CO_2$ emissions saved by Lenovo's Low-Temperature Solder process at the end of FY 2017/18

59 MT

Global Week of Service directly impacted 33,000 consumers across 19 countries, benefiting minority populations and increasing access to STEM education

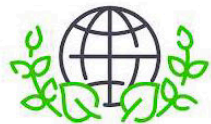

**Fig. 109.** Survey example (page 29 of 32)

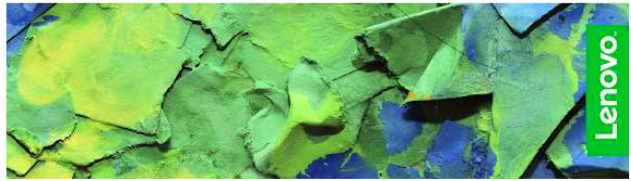

**Fig. 110.** Survey example (page 30 of 32)

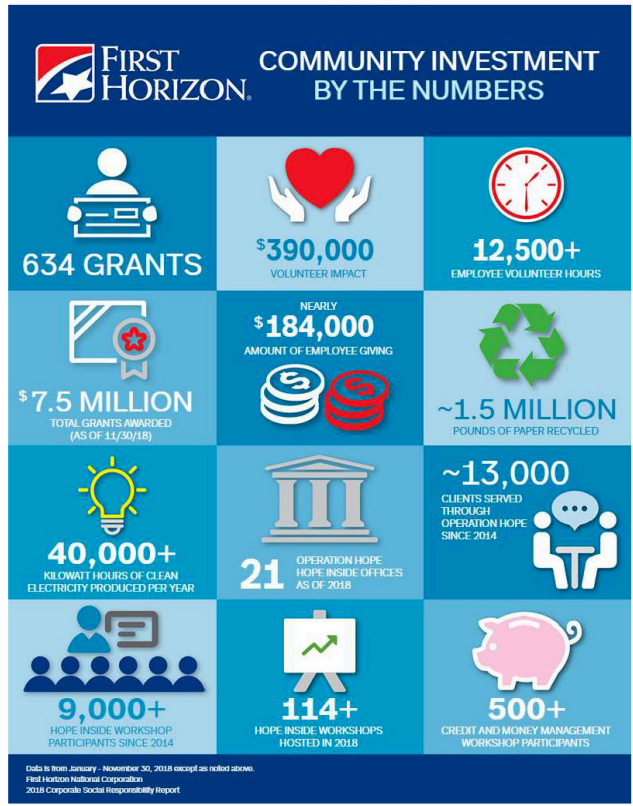

**Fig. 111.** Survey example (page 31 of 32)

**Fig. 112.** Survey example (page 32 of 32)

### 7.2 Raw data

This section provides the raw data of the experiment in Table 22 to Table 32. Then Table 33 to Table 38 present some calculations over the raw data to facilitate undersanding the results. Table 35 and Table 38 contain the two datasets analysed with SPSS.

**Table 22.** Demographic data of the experiment participants and group (or survey variant) they have received

| Participant | Timestamp | Age | Gender | Education | Group |
|---|---|---|---|---|---|
| p1 | 1/7/2020 14:38:23 | 25-34 | Male | BSc | 1 |
| p2 | 1/8/2020 11:16:58 | 25-34 | Female | PhD | 1 |
| p3 | 1/8/2020 11:50:28 | 25-34 | Male | BSc | 1 |
| p4 | 1/8/2020 12:45:57 | 25-34 | Female | MSc | 1 |
| p5 | 1/8/2020 13:32:13 | 25-34 | Female | MSc | 1 |
| p6 | 1/8/2020 13:39:46 | 18-24 | Male | High | 1 |
| p7 | 1/8/2020 13:55:23 | 25-34 | Male | MSc | 1 |
| p8 | 1/8/2020 14:32:57 | 25-34 | Male | MSc | 1 |
| p9 | 1/8/2020 15:14:15 | 25-34 | Female | MSc | 1 |
| p10 | 1/8/2020 17:21:25 | 25-34 | Male | MSc | 1 |
| p11 | 1/8/2020 19:42:56 | 25-34 | Male | BSc | 1 |
| p12 | 1/8/2020 20:38:10 | 25-34 | Male | High | 1 |
| p13 | 1/8/2020 21:24:08 | 25-34 | Female | MSc | 1 |
| p14 | 1/8/2020 22:10:12 | 25-34 | Male | BSc | 1 |
| p15 | 1/10/2020 15:02:03 | 25-34 | Female | MSc | 1 |
| p16 | 1/11/2020 14:41:29 | 25-34 | Male | MSc | 1 |
| p17 | 1/12/2020 11:43:16 | 35-44 | Male | MSc | 1 |
| p18 | 1/16/2020 17:39:29 | 25-34 | Male | High | 1 |
| p19 | 1/16/2020 17:54:42 | 18-24 | Female | BSc | 1 |
| p20 | 1/21/2020 13:36:55 | 18-24 | Female | High | 1 |
| p21 | 1/8/2020 10:32:46 | 35-44 | Female | MSc | 2 |
| p22 | 1/8/2020 12:43:44 | 18-24 | Male | High | 2 |
| p23 | 1/8/2020 13:58:23 | 25-34 | Female | MSc | 2 |
| p24 | 1/8/2020 15:28:12 | 18-24 | Male | High | 2 |
| p25 | 1/8/2020 18:52:14 | 25-34 | Female | BSc | 2 |
| p26 | 1/8/2020 19:32:01 | 18-24 | Female | High | 2 |
| p27 | 1/8/2020 19:36:36 | 18-24 | Male | BSc | 2 |
| p28 | 1/9/2020 2:11:30 | 25-34 | Male | BSc | 2 |
| p29 | 1/9/2020 14:53:03 | 25-34 | Male | BSc | 2 |
| p30 | 1/9/2020 17:16:58 | 25-34 | Male | BSc | 2 |
| p31 | 1/9/2020 20:04:45 | 25-34 | Male | MSc | 2 |
| p32 | 1/9/2020 20:51:01 | 25-34 | Female | MSc | 2 |
| p33 | 1/9/2020 23:15:16 | 18-24 | Male | MSc | 2 |
| p34 | 1/11/2020 15:15:49 | 25-34 | Male | MSc | 2 |
| p35 | 1/11/2020 18:25:03 | 25-34 | Male | MSc | 2 |
| p36 | 1/14/2020 20:30:28 | 25-34 | Male | MSc | 2 |
| p37 | 1/15/2020 8:38:22 | 18-24 | Female | MSc | 2 |
| p38 | 1/16/2020 12:59:29 | 25-34 | Male | BSc | 2 |
| p39 | 1/16/2020 13:08:27 | 25-34 | Male | BSc | 2 |
| p40 | 1/21/2020 11:54:41 | 18-24 | Female | High | 2 |

**Table 23.** Raw data of the experiment (for the first infographic in the surveys)

| Participant | VA1 | VA2 | VA3 | VA4 | VA5 | VA6 | VA7 | VA8 | VA9 | VA10 | Type |
|---|---|---|---|---|---|---|---|---|---|---|---|
| p1 | 4 | 4 | 2 | 3 | 3 | 2 | 3 | 3 | 4 | 2 | G |
| p2 | 4 | 3 | 4 | 4 | 4 | 2 | 2 | 2 | 2 | 1 | G |
| p3 | 4 | 4 | 3 | 4 | 4 | 2 | 2 | 2 | 1 | 1 | G |
| p4 | 3 | 2 | 4 | 2 | 3 | 2 | 3 | 2 | 3 | 3 | G |

| Participant | VA1 | VA2 | VA3 | VA4 | VA5 | VA6 | VA7 | VA8 | VA9 | VA10 | Type |
|---|---|---|---|---|---|---|---|---|---|---|---|
| p5 | 3 | 3 | 2 | 4 | 2 | 3 | 2 | 2 | 3 | 3 | G |
| p6 | 4 | 5 | 4 | 4 | 5 | 4 | 4 | 5 | 3 | 3 | O |
| p7 | 4 | 5 | 5 | 4 | 4 | 2 | 2 | 1 | 2 | 2 | G |
| p8 | 5 | 5 | 4 | 3 | 3 | 4 | 3 | 3 | 3 | 4 | G |
| p9 | 4 | 5 | 4 | 4 | 4 | 2 | 4 | 2 | 2 | 2 | G |
| p10 | 1 | 2 | 2 | 2 | 2 | 2 | 2 | 2 | 2 | 1 | O |
| p11 | 4 | 3 | 4 | 3 | 4 | 4 | 3 | 3 | 4 | 2 | O |
| p12 | 4 | 3 | 3 | 2 | 4 | 4 | 4 | 1 | 2 | 2 | G |
| p13 | 2 | 1 | 1 | 2 | 1 | 1 | 1 | 1 | 1 | 1 | G |
| p14 | 3 | 3 | 2 | 2 | 2 | 2 | 2 | 2 | 2 | 2 | G |
| p15 | 2 | 3 | 3 | 2 | 3 | 3 | 3 | 2 | 3 | 2 | G |
| p16 | 3 | 4 | 3 | 4 | 2 | 1 | 2 | 1 | 2 | 1 | G |
| p17 | 4 | 5 | 2 | 4 | 4 | 3 | 4 | 2 | 1 | 2 | G |
| p18 | 4 | 3 | 2 | 2 | 2 | 4 | 4 | 2 | 4 | 3 | O |
| p19 | 2 | 3 | 2 | 2 | 1 | 3 | 3 | 1 | 1 | 1 | O |
| p20 | 2 | 1 | 2 | 1 | 2 | 3 | 2 | 2 | 2 | 2 | G |
| p21 | 2 | 3 | 1 | 1 | 2 | 2 | 2 | 1 | 1 | 1 | G |
| p22 | 4 | 4 | 4 | 4 | 4 | 2 | 3 | 1 | 1 | 1 | G |
| p23 | 2 | 3 | 1 | 3 | 2 | 1 | 2 | 1 | 3 | 1 | G |
| p24 | 4 | 3 | 4 | 3 | 2 | 4 | 3 | 4 | 3 | 2 | O |
| p25 | 3 | 2 | 2 | 2 | 2 | 1 | 3 | 1 | 2 | 1 | O |
| p26 | 3 | 5 | 3 | 2 | 2 | 2 | 3 | 2 | 3 | 3 | G |
| p27 | 4 | 3 | 4 | 5 | 3 | 1 | 4 | 1 | 1 | 1 | G |
| p28 | 3 | 4 | 2 | 4 | 2 | 2 | 3 | 2 | 3 | 2 | G |
| p29 | 1 | 5 | 1 | 5 | 3 | 3 | 1 | 1 | 1 | 1 | G |
| p30 | 2 | 2 | 1 | 3 | 2 | 2 | 2 | 2 | 2 | 1 | G |
| p31 | 3 | 5 | 4 | 4 | 3 | 2 | 3 | 3 | 3 | 2 | G |
| p32 | 3 | 3 | 2 | 3 | 2 | 1 | 2 | 1 | 1 | 1 | O |
| p33 | 3 | 3 | 3 | 3 | 1 | 1 | 2 | 2 | 2 | 2 | G |
| p34 | 2 | 1 | 2 | 3 | 3 | 1 | 2 | 1 | 1 | 1 | G |
| p35 | 4 | 4 | 4 | 3 | 4 | 3 | 3 | 2 | 3 | 3 | O |
| p36 | 4 | 4 | 3 | 4 | 4 | 2 | 2 | 2 | 3 | 3 | G |
| p37 | 4 | 5 | 4 | 4 | 4 | 5 | 4 | 3 | 2 | 2 | O |
| p38 | 1 | 3 | 1 | 1 | 1 | 1 | 1 | 1 | 1 | 1 | G |
| p39 | 4 | 4 | 2 | 3 | 1 | 1 | 2 | 1 | 3 | 2 | G |
| p40 | 3 | 4 | 3 | 4 | 3 | 2 | 2 | 2 | 3 | 3 | G |

**Table 24.** Raw data of the experiment (for the second infographic in the surveys)

| Participant | VA1 | VA2 | VA3 | VA4 | VA5 | VA6 | VA7 | VA8 | VA9 | VA10 | Type |
|---|---|---|---|---|---|---|---|---|---|---|---|
| p1 | 4 | 3 | 4 | 4 | 4 | 4 | 4 | 5 | 3 | 3 | O |
| p2 | 3 | 3 | 2 | 3 | 2 | 2 | 2 | 3 | 2 | 3 | G |
| p3 | 4 | 4 | 3 | 3 | 4 | 2 | 2 | 1 | 2 | 2 | O |
| p4 | 3 | 3 | 3 | 2 | 2 | 2 | 3 | 3 | 2 | 2 | G |
| p5 | 4 | 4 | 4 | 4 | 4 | 3 | 3 | 4 | 3 | 3 | G |
| p6 | 4 | 3 | 4 | 5 | 5 | 5 | 4 | 4 | 3 | 5 | O |
| p7 | 2 | 4 | 2 | 4 | 2 | 2 | 3 | 2 | 1 | 1 | G |
| p8 | 4 | 4 | 4 | 2 | 3 | 4 | 5 | 3 | 5 | 3 | O |
| p9 | 3 | 3 | 4 | 4 | 4 | 3 | 4 | 3 | 2 | 2 | O |

| Participant | VA1 | VA2 | VA3 | VA4 | VA5 | VA6 | VA7 | VA8 | VA9 | VA10 | Type |
|---|---|---|---|---|---|---|---|---|---|---|---|
| p10 | 3 | 2 | 3 | 2 | 3 | 3 | 3 | 3 | 3 | 3 | O |
| p11 | 4 | 4 | 4 | 3 | 4 | 4 | 3 | 4 | 4 | 4 | O |
| p12 | 2 | 4 | 3 | 4 | 3 | 4 | 3 | 3 | 1 | 3 | G |
| p13 | 1 | 1 | 2 | 2 | 2 | 1 | 2 | 1 | 1 | 1 | O |
| p14 | 3 | 3 | 2 | 2 | 2 | 2 | 2 | 2 | 2 | 2 | G |
| p15 | 4 | 4 | 5 | 4 | 4 | 3 | 4 | 3 | 3 | 4 | G |
| p16 | 4 | 4 | 4 | 4 | 4 | 4 | 4 | 2 | 2 | 3 | O |
| p17 | 3 | 4 | 1 | 4 | 1 | 4 | 3 | 1 | 1 | 2 | O |
| p18 | 4 | 4 | 3 | 4 | 3 | 4 | 3 | 4 | 2 | 4 | G |
| p19 | 2 | 1 | 2 | 1 | 1 | 2 | 2 | 3 | 1 | 3 | O |
| p20 | 3 | 2 | 4 | 3 | 3 | 2 | 2 | 3 | 3 | 3 | G |
| p21 | 3 | 2 | 3 | 2 | 2 | 3 | 2 | 2 | 2 | 2 | O |
| p22 | 3 | 2 | 2 | 1 | 1 | 2 | 2 | 2 | 1 | 1 | O |
| p23 | 2 | 2 | 3 | 2 | 2 | 2 | 2 | 1 | 1 | 2 | O |
| p24 | 5 | 4 | 5 | 5 | 4 | 3 | 4 | 4 | 4 | 3 | G |
| p25 | 4 | 4 | 5 | 5 | 5 | 3 | 4 | 3 | 3 | 3 | G |
| p26 | 2 | 3 | 2 | 4 | 4 | 4 | 3 | 3 | 2 | 2 | O |
| p27 | 5 | 2 | 4 | 4 | 5 | 1 | 3 | 2 | 1 | 2 | O |
| p28 | 4 | 3 | 4 | 4 | 3 | 3 | 4 | 3 | 2 | 3 | O |
| p29 | 3 | 1 | 3 | 1 | 3 | 3 | 1 | 1 | 1 | 1 | O |
| p30 | 3 | 3 | 3 | 3 | 3 | 2 | 3 | 2 | 3 | 2 | G |
| p31 | 4 | 4 | 4 | 4 | 4 | 4 | 3 | 4 | 3 | 3 | O |
| p32 | 3 | 2 | 3 | 3 | 2 | 3 | 2 | 3 | 2 | 3 | O |
| p33 | 3 | 3 | 4 | 4 | 3 | 2 | 2 | 2 | 1 | 1 | G |
| p34 | 2 | 3 | 2 | 3 | 3 | 1 | 2 | 2 | 1 | 1 | G |
| p35 | 2 | 3 | 3 | 3 | 2 | 3 | 3 | 2 | 3 | 2 | O |
| p36 | 3 | 4 | 3 | 3 | 4 | 3 | 3 | 4 | 3 | 3 | G |
| p37 | 2 | 2 | 2 | 2 | 1 | 3 | 1 | 1 | 1 | 1 | O |
| p38 | 1 | 3 | 1 | 1 | 1 | 1 | 1 | 1 | 1 | 1 | G |
| p39 | 3 | 4 | 3 | 3 | 2 | 2 | 2 | 3 | 3 | 2 | G |
| p40 | 2 | 4 | 3 | 3 | 4 | 4 | 3 | 4 | 3 | 4 | G |

**Table 25.** Raw data of the experiment (for the third infographic in the surveys)

| Participant | VA1 | VA2 | VA3 | VA4 | VA5 | VA6 | VA7 | VA8 | VA9 | VA10 | Type |
|---|---|---|---|---|---|---|---|---|---|---|---|
| p1 | 4 | 4 | 4 | 4 | 4 | 3 | 2 | 3 | 2 | 3 | G |
| p2 | 3 | 2 | 3 | 3 | 3 | 2 | 3 | 2 | 2 | 2 | G |
| p3 | 1 | 1 | 1 | 1 | 1 | 1 | 2 | 2 | 1 | 1 | G |
| p4 | 3 | 3 | 3 | 3 | 3 | 2 | 3 | 3 | 4 | 3 | G |
| p5 | 2 | 2 | 2 | 1 | 1 | 2 | 1 | 2 | 2 | 1 | O |
| p6 | 3 | 4 | 4 | 5 | 5 | 3 | 3 | 2 | 3 | 4 | G |
| p7 | 3 | 1 | 3 | 2 | 2 | 4 | 4 | 4 | 4 | 2 | O |
| p8 | 4 | 3 | 4 | 3 | 4 | 5 | 5 | 5 | 5 | 4 | O |
| p9 | 2 | 2 | 3 | 3 | 2 | 3 | 2 | 3 | 3 | 3 | G |
| p10 | 1 | 1 | 1 | 1 | 1 | 2 | 1 | 2 | 1 | 1 | G |
| p11 | 2 | 2 | 1 | 2 | 2 | 4 | 3 | 3 | 3 | 3 | G |
| p12 | 5 | 4 | 5 | 4 | 3 | 4 | 4 | 3 | 1 | 3 | G |
| p13 | 1 | 1 | 1 | 1 | 1 | 1 | 1 | 1 | 1 | 1 | G |
| p14 | 1 | 2 | 1 | 1 | 1 | 1 | 1 | 1 | 1 | 1 | O |

| Participant | VA1 | VA2 | VA3 | VA4 | VA5 | VA6 | VA7 | VA8 | VA9 | VA10 | Type |
|---|---|---|---|---|---|---|---|---|---|---|---|
| p15 | 3 | 2 | 2 | 2 | 3 | 3 | 2 | 3 | 3 | 3 | G |
| p16 | 2 | 2 | 2 | 2 | 3 | 2 | 2 | 2 | 2 | 2 | G |
| p17 | 3 | 3 | 3 | 4 | 4 | 4 | 4 | 2 | 3 | 1 | O |
| p18 | 2 | 1 | 1 | 1 | 1 | 3 | 2 | 2 | 2 | 2 | G |
| p19 | 3 | 2 | 3 | 3 | 2 | 2 | 3 | 3 | 2 | 2 | G |
| p20 | 2 | 3 | 2 | 2 | 3 | 3 | 3 | 2 | 2 | 3 | G |
| p21 | 2 | 2 | 2 | 2 | 2 | 2 | 2 | 2 | 2 | 2 | G |
| p22 | 4 | 3 | 3 | 4 | 3 | 3 | 3 | 2 | 1 | 1 | G |
| p23 | 2 | 3 | 2 | 3 | 2 | 3 | 2 | 2 | 3 | 2 | O |
| p24 | 3 | 5 | 2 | 1 | 2 | 4 | 4 | 4 | 3 | 2 | G |
| p25 | 1 | 1 | 1 | 3 | 2 | 1 | 1 | 3 | 3 | 2 | O |
| p26 | 4 | 1 | 3 | 1 | 1 | 4 | 2 | 4 | 4 | 4 | O |
| p27 | 3 | 2 | 4 | 4 | 3 | 1 | 2 | 1 | 1 | 1 | G |
| p28 | 2 | 2 | 3 | 3 | 3 | 3 | 2 | 4 | 4 | 3 | G |
| p29 | 3 | 3 | 3 | 5 | 5 | 3 | 3 | 3 | 1 | 3 | G |
| p30 | 2 | 3 | 3 | 3 | 2 | 2 | 2 | 2 | 1 | 2 | O |
| p31 | 2 | 5 | 3 | 4 | 3 | 3 | 3 | 3 | 2 | 2 | G |
| p32 | 1 | 1 | 1 | 3 | 3 | 2 | 2 | 1 | 1 | 2 | G |
| p33 | 2 | 2 | 1 | 1 | 1 | 1 | 1 | 1 | 1 | 1 | G |
| p34 | 3 | 1 | 2 | 2 | 1 | 3 | 2 | 3 | 2 | 3 | O |
| p35 | 3 | 4 | 3 | 4 | 4 | 2 | 3 | 2 | 2 | 2 | G |
| p36 | 3 | 4 | 3 | 3 | 3 | 2 | 4 | 2 | 3 | 3 | G |
| p37 | 2 | 3 | 2 | 1 | 3 | 2 | 2 | 1 | 1 | 1 | O |
| p38 | 1 | 1 | 3 | 2 | 2 | 1 | 1 | 1 | 1 | 1 | G |
| p39 | 2 | 3 | 1 | 4 | 1 | 1 | 1 | 1 | 1 | 1 | G |
| p40 | 2 | 3 | 2 | 2 | 3 | 3 | 2 | 3 | 3 | 2 | G |

**Table 26.** Raw data of the experiment (for the fourth infographic in the surveys)

| Participant | VA1 | VA2 | VA3 | VA4 | VA5 | VA6 | VA7 | VA8 | VA9 | VA10 | Type |
|---|---|---|---|---|---|---|---|---|---|---|---|
| p1 | 4 | 3 | 4 | 5 | 5 | 4 | 4 | 2 | 4 | 4 | O |
| p2 | 4 | 3 | 4 | 4 | 4 | 4 | 4 | 4 | 4 | 3 | O |
| p3 | 2 | 3 | 3 | 1 | 2 | 3 | 2 | 2 | 2 | 2 | G |
| p4 | 4 | 3 | 3 | 4 | 3 | 3 | 3 | 3 | 4 | 3 | G |
| p5 | 2 | 2 | 3 | 3 | 3 | 4 | 3 | 4 | 4 | 3 | O |
| p6 | 4 | 3 | 4 | 4 | 5 | 4 | 3 | 4 | 4 | 5 | O |
| p7 | 5 | 2 | 5 | 5 | 5 | 3 | 4 | 4 | 5 | 4 | O |
| p8 | 5 | 4 | 4 | 3 | 4 | 5 | 4 | 5 | 5 | 5 | O |
| p9 | 4 | 4 | 4 | 4 | 4 | 4 | 4 | 4 | 3 | 3 | O |
| p10 | 2 | 2 | 1 | 1 | 1 | 2 | 1 | 2 | 2 | 1 | O |
| p11 | 2 | 3 | 4 | 4 | 3 | 3 | 3 | 4 | 3 | 2 | G |
| p12 | 4 | 4 | 3 | 3 | 2 | 4 | 4 | 5 | 2 | 2 | O |
| p13 | 2 | 1 | 2 | 2 | 2 | 2 | 2 | 2 | 1 | 1 | O |
| p14 | 3 | 3 | 4 | 3 | 4 | 4 | 3 | 3 | 3 | 3 | G |
| p15 | 2 | 2 | 2 | 2 | 3 | 3 | 2 | 3 | 2 | 2 | G |
| p16 | 1 | 2 | 1 | 2 | 1 | 1 | 1 | 1 | 1 | 1 | O |
| p17 | 4 | 2 | 3 | 4 | 2 | 4 | 4 | 4 | 3 | 3 | G |
| p18 | 2 | 2 | 1 | 1 | 1 | 2 | 4 | 2 | 3 | 1 | O |
| p19 | 3 | 1 | 3 | 2 | 1 | 3 | 4 | 3 | 2 | 2 | O |

| Participant | VA1 | VA2 | VA3 | VA4 | VA5 | VA6 | VA7 | VA8 | VA9 | VA10 | Type |
|---|---|---|---|---|---|---|---|---|---|---|---|
| p20 | 3 | 4 | 4 | 2 | 3 | 2 | 4 | 3 | 2 | 3 | O |
| p21 | 2 | 1 | 1 | 1 | 1 | 2 | 2 | 2 | 1 | 1 | O |
| p22 | 2 | 3 | 2 | 1 | 1 | 2 | 2 | 3 | 1 | 1 | O |
| p23 | 2 | 1 | 2 | 2 | 2 | 1 | 2 | 1 | 1 | 1 | G |
| p24 | 2 | 2 | 3 | 4 | 3 | 2 | 3 | 4 | 3 | 2 | O |
| p25 | 3 | 1 | 2 | 2 | 3 | 3 | 2 | 2 | 3 | 2 | O |
| p26 | 3 | 1 | 2 | 1 | 2 | 3 | 2 | 2 | 3 | 3 | O |
| p27 | 3 | 3 | 5 | 4 | 4 | 4 | 3 | 3 | 4 | 4 | O |
| p28 | 4 | 2 | 2 | 2 | 3 | 3 | 2 | 3 | 3 | 3 | O |
| p29 | 3 | 1 | 3 | 1 | 1 | 3 | 1 | 5 | 1 | 1 | O |
| p30 | 3 | 3 | 3 | 2 | 2 | 3 | 3 | 2 | 3 | 3 | O |
| p31 | 4 | 3 | 4 | 4 | 4 | 5 | 4 | 5 | 3 | 4 | O |
| p32 | 1 | 1 | 2 | 2 | 2 | 2 | 2 | 3 | 2 | 2 | O |
| p33 | 3 | 3 | 3 | 4 | 3 | 3 | 4 | 3 | 3 | 3 | O |
| p34 | 2 | 1 | 2 | 2 | 2 | 3 | 2 | 3 | 3 | 3 | O |
| p35 | 5 | 3 | 4 | 4 | 4 | 3 | 4 | 4 | 4 | 4 | O |
| p36 | 4 | 3 | 4 | 4 | 3 | 4 | 3 | 4 | 3 | 3 | G |
| p37 | 3 | 2 | 3 | 2 | 4 | 2 | 2 | 2 | 2 | 2 | G |
| p38 | 2 | 1 | 2 | 1 | 1 | 1 | 1 | 1 | 1 | 1 | O |
| p39 | 3 | 3 | 4 | 4 | 3 | 3 | 3 | 3 | 4 | 3 | O |
| p40 | 4 | 2 | 4 | 4 | 4 | 4 | 4 | 4 | 3 | 4 | G |

**Table 27.** Raw data of the experiment (for the fifth infographic in the surveys)

| Participant | VA1 | VA2 | VA3 | VA4 | VA5 | VA6 | VA7 | VA8 | VA9 | VA10 | Type |
|---|---|---|---|---|---|---|---|---|---|---|---|
| p1 | 4 | 4 | 4 | 4 | 5 | 3 | 3 | 3 | 2 | 2 | G |
| p2 | 4 | 4 | 4 | 4 | 4 | 2 | 3 | 3 | 3 | 3 | G |
| p3 | 4 | 3 | 3 | 5 | 5 | 3 | 3 | 2 | 2 | 2 | G |
| p4 | 4 | 4 | 4 | 4 | 4 | 4 | 3 | 3 | 3 | 4 | G |
| p5 | 4 | 4 | 3 | 3 | 4 | 4 | 3 | 3 | 2 | 3 | O |
| p6 | 4 | 5 | 4 | 3 | 5 | 4 | 3 | 4 | 5 | 5 | O |
| p7 | 5 | 5 | 5 | 5 | 5 | 2 | 4 | 2 | 5 | 4 | O |
| p8 | 4 | 5 | 5 | 4 | 5 | 3 | 3 | 3 | 3 | 4 | G |
| p9 | 3 | 4 | 4 | 4 | 4 | 3 | 4 | 3 | 2 | 2 | G |
| p10 | 3 | 3 | 4 | 3 | 3 | 2 | 2 | 1 | 2 | 3 | G |
| p11 | 3 | 4 | 4 | 5 | 5 | 3 | 4 | 3 | 4 | 4 | O |
| p12 | 4 | 4 | 5 | 4 | 5 | 3 | 4 | 3 | 2 | 3 | O |
| p13 | 2 | 1 | 2 | 2 | 2 | 2 | 2 | 2 | 2 | 2 | O |
| p14 | 2 | 3 | 2 | 2 | 3 | 3 | 3 | 2 | 3 | 2 | O |
| p15 | 3 | 4 | 4 | 4 | 3 | 3 | 3 | 3 | 3 | 3 | G |
| p16 | 4 | 4 | 4 | 4 | 4 | 2 | 3 | 2 | 1 | 1 | O |
| p17 | 4 | 5 | 4 | 4 | 2 | 3 | 4 | 3 | 4 | 2 | O |
| p18 | 4 | 4 | 4 | 4 | 5 | 3 | 4 | 3 | 3 | 3 | G |
| p19 | 2 | 4 | 2 | 2 | 2 | 1 | 2 | 2 | 1 | 2 | O |
| p20 | 1 | 2 | 1 | 2 | 2 | 1 | 2 | 1 | 1 | 2 | G |
| p21 | 2 | 2 | 2 | 3 | 3 | 2 | 2 | 2 | 2 | 2 | O |
| p22 | 4 | 4 | 4 | 4 | 3 | 3 | 4 | 4 | 1 | 1 | O |
| p23 | 3 | 3 | 3 | 3 | 3 | 3 | 3 | 2 | 3 | 2 | O |
| p24 | 3 | 4 | 4 | 4 | 3 | 2 | 3 | 3 | 3 | 2 | G |

| Participant | VA1 | VA2 | VA3 | VA4 | VA5 | VA6 | VA7 | VA8 | VA9 | VA10 | Type |
|---|---|---|---|---|---|---|---|---|---|---|---|
| p25 | 4 | 4 | 4 | 4 | 4 | 4 | 3 | 4 | 3 | 4 | G |
| p26 | 3 | 3 | 4 | 4 | 5 | 3 | 2 | 2 | 2 | 3 | G |
| p27 | 2 | 3 | 3 | 4 | 2 | 2 | 2 | 1 | 1 | 1 | G |
| p28 | 4 | 4 | 4 | 4 | 4 | 5 | 4 | 3 | 2 | 3 | O |
| p29 | 1 | 3 | 3 | 3 | 3 | 1 | 1 | 1 | 1 | 3 | O |
| p30 | 3 | 2 | 3 | 3 | 3 | 2 | 3 | 2 | 3 | 2 | O |
| p31 | 2 | 4 | 2 | 4 | 3 | 2 | 3 | 3 | 3 | 2 | G |
| p32 | 2 | 3 | 3 | 3 | 3 | 2 | 3 | 2 | 2 | 2 | O |
| p33 | 3 | 3 | 3 | 4 | 4 | 2 | 2 | 2 | 3 | 3 | G |
| p34 | 2 | 3 | 3 | 3 | 3 | 2 | 3 | 2 | 1 | 2 | G |
| p35 | 2 | 3 | 2 | 2 | 2 | 2 | 2 | 3 | 1 | 2 | O |
| p36 | 4 | 3 | 4 | 4 | 3 | 2 | 3 | 2 | 3 | 2 | G |
| p37 | 3 | 2 | 3 | 2 | 3 | 3 | 3 | 2 | 2 | 1 | G |
| p38 | 1 | 3 | 1 | 1 | 1 | 1 | 1 | 1 | 1 | 1 | O |
| p39 | 3 | 4 | 4 | 4 | 4 | 3 | 3 | 3 | 4 | 3 | G |
| p40 | 4 | 5 | 4 | 5 | 5 | 4 | 4 | 5 | 3 | 5 | G |

**Table 28.** Raw data of the experiment (for the sixth infographic in the surveys)

| Participant | VA1 | VA2 | VA3 | VA4 | VA5 | VA6 | VA7 | VA8 | VA9 | VA10 | Type |
|---|---|---|---|---|---|---|---|---|---|---|---|
| p1 | 3 | 5 | 4 | 4 | 4 | 1 | 2 | 3 | 1 | 1 | G |
| p2 | 4 | 4 | 4 | 4 | 4 | 3 | 3 | 2 | 2 | 2 | G |
| p3 | 3 | 3 | 2 | 2 | 2 | 4 | 4 | 4 | 2 | 2 | G |
| p4 | 3 | 4 | 4 | 4 | 3 | 4 | 3 | 4 | 3 | 3 | G |
| p5 | 2 | 3 | 4 | 4 | 4 | 1 | 1 | 2 | 1 | 2 | G |
| p6 | 4 | 3 | 3 | 4 | 5 | 4 | 5 | 3 | 3 | 4 | O |
| p7 | 4 | 2 | 4 | 5 | 5 | 4 | 4 | 4 | 5 | 3 | G |
| p8 | 4 | 4 | 4 | 5 | 5 | 3 | 3 | 2 | 3 | 3 | G |
| p9 | 4 | 3 | 4 | 4 | 4 | 4 | 3 | 3 | 3 | 2 | O |
| p10 | 1 | 1 | 1 | 1 | 2 | 1 | 1 | 1 | 1 | 1 | G |
| p11 | 4 | 4 | 5 | 4 | 4 | 3 | 4 | 3 | 2 | 3 | G |
| p12 | 5 | 5 | 5 | 5 | 5 | 5 | 5 | 2 | 4 | 4 | O |
| p13 | 1 | 1 | 1 | 2 | 2 | 1 | 2 | 1 | 1 | 1 | O |
| p14 | 3 | 3 | 3 | 4 | 4 | 3 | 3 | 3 | 3 | 3 | G |
| p15 | 4 | 3 | 3 | 4 | 4 | 4 | 4 | 3 | 3 | 4 | G |
| p16 | 5 | 4 | 5 | 5 | 5 | 4 | 5 | 3 | 2 | 3 | O |
| p17 | 2 | 2 | 3 | 5 | 3 | 4 | 4 | 3 | 1 | 2 | O |
| p18 | 4 | 4 | 5 | 4 | 4 | 2 | 3 | 3 | 4 | 4 | G |
| p19 | 5 | 4 | 4 | 2 | 3 | 3 | 4 | 2 | 1 | 3 | G |
| p20 | 2 | 4 | 3 | 4 | 3 | 3 | 3 | 4 | 2 | 3 | O |
| p21 | 3 | 2 | 3 | 3 | 3 | 3 | 3 | 2 | 1 | 2 | O |
| p22 | 3 | 2 | 3 | 4 | 4 | 2 | 2 | 2 | 1 | 1 | O |
| p23 | 2 | 3 | 3 | 3 | 3 | 3 | 3 | 3 | 3 | 3 | O |
| p24 | 4 | 5 | 5 | 4 | 4 | 5 | 4 | 4 | 4 | 5 | G |
| p25 | 4 | 3 | 4 | 4 | 4 | 4 | 2 | 4 | 4 | 3 | G |
| p26 | 4 | 2 | 4 | 5 | 5 | 5 | 2 | 3 | 4 | 4 | G |
| p27 | 4 | 4 | 5 | 4 | 3 | 4 | 3 | 4 | 1 | 4 | G |
| p28 | 5 | 4 | 5 | 4 | 4 | 4 | 4 | 4 | 3 | 4 | O |
| p29 | 5 | 3 | 5 | 5 | 5 | 1 | 1 | 1 | 1 | 3 | G |

| Participant | VA1 | VA2 | VA3 | VA4 | VA5 | VA6 | VA7 | VA8 | VA9 | VA10 | Type |
|---|---|---|---|---|---|---|---|---|---|---|---|
| p30 | 4 | 4 | 4 | 4 | 4 | 2 | 4 | 3 | 3 | 4 | O |
| p31 | 1 | 4 | 2 | 3 | 4 | 2 | 2 | 2 | 2 | 2 | G |
| p32 | 3 | 2 | 4 | 3 | 3 | 3 | 3 | 2 | 2 | 2 | G |
| p33 | 4 | 4 | 4 | 3 | 3 | 2 | 2 | 1 | 1 | 1 | G |
| p34 | 1 | 1 | 1 | 2 | 2 | 3 | 1 | 3 | 1 | 4 | O |
| p35 | 5 | 4 | 4 | 4 | 4 | 2 | 3 | 2 | 1 | 2 | G |
| p36 | 2 | 2 | 3 | 3 | 4 | 2 | 2 | 2 | 2 | 3 | G |
| p37 | 3 | 4 | 4 | 4 | 4 | 2 | 3 | 2 | 4 | 2 | G |
| p38 | 2 | 1 | 2 | 1 | 1 | 1 | 1 | 1 | 1 | 1 | O |
| p39 | 2 | 3 | 1 | 2 | 2 | 1 | 2 | 1 | 2 | 1 | G |
| p40 | 4 | 4 | 3 | 4 | 3 | 4 | 2 | 5 | 4 | 3 | O |

**Table 29.** Raw data of the experiment (for the seventh infographic in the surveys)

| Participant | VA1 | VA2 | VA3 | VA4 | VA5 | VA6 | VA7 | VA8 | VA9 | VA10 | Type |
|---|---|---|---|---|---|---|---|---|---|---|---|
| p1 | 4 | 3 | 4 | 2 | 2 | 3 | 3 | 2 | 2 | 2 | O |
| p2 | 2 | 1 | 2 | 2 | 2 | 2 | 2 | 2 | 2 | 1 | O |
| p3 | 3 | 2 | 2 | 2 | 2 | 2 | 2 | 2 | 3 | 2 | G |
| p4 | 4 | 3 | 4 | 4 | 4 | 4 | 3 | 4 | 4 | 4 | G |
| p5 | 4 | 3 | 4 | 3 | 3 | 3 | 3 | 3 | 2 | 3 | O |
| p6 | 5 | 4 | 4 | 5 | 4 | 4 | 5 | 5 | 4 | 4 | O |
| p7 | 3 | 2 | 2 | 4 | 4 | 3 | 4 | 3 | 3 | 2 | G |
| p8 | 2 | 3 | 3 | 2 | 3 | 4 | 3 | 4 | 4 | 3 | O |
| p9 | 4 | 3 | 4 | 4 | 4 | 3 | 3 | 3 | 2 | 2 | O |
| p10 | 2 | 1 | 1 | 1 | 2 | 1 | 1 | 1 | 2 | 2 | O |
| p11 | 2 | 2 | 2 | 2 | 1 | 2 | 3 | 2 | 2 | 3 | O |
| p12 | 4 | 3 | 2 | 2 | 3 | 4 | 4 | 4 | 1 | 3 | O |
| p13 | 2 | 2 | 2 | 2 | 2 | 2 | 2 | 2 | 2 | 2 | G |
| p14 | 3 | 2 | 2 | 2 | 3 | 2 | 2 | 2 | 1 | 2 | O |
| p15 | 3 | 2 | 3 | 3 | 3 | 2 | 3 | 3 | 3 | 3 | G |
| p16 | 4 | 2 | 3 | 3 | 3 | 3 | 3 | 2 | 2 | 2 | O |
| p17 | 4 | 2 | 4 | 4 | 3 | 4 | 4 | 2 | 4 | 3 | G |
| p18 | 3 | 2 | 3 | 3 | 3 | 2 | 3 | 2 | 2 | 2 | G |
| p19 | 3 | 1 | 2 | 1 | 1 | 3 | 1 | 4 | 1 | 2 | G |
| p20 | 1 | 1 | 2 | 2 | 2 | 1 | 1 | 2 | 2 | 2 | G |
| p21 | 2 | 2 | 3 | 2 | 2 | 2 | 2 | 2 | 2 | 2 | G |
| p22 | 2 | 2 | 2 | 3 | 2 | 2 | 2 | 1 | 1 | 1 | G |
| p23 | 1 | 1 | 1 | 1 | 1 | 1 | 1 | 1 | 1 | 1 | G |
| p24 | 4 | 3 | 3 | 2 | 2 | 4 | 3 | 3 | 3 | 4 | O |
| p25 | 2 | 3 | 3 | 4 | 4 | 3 | 2 | 4 | 3 | 3 | O |
| p26 | 1 | 1 | 1 | 1 | 3 | 1 | 1 | 1 | 1 | 1 | O |
| p27 | 4 | 2 | 3 | 2 | 5 | 3 | 4 | 2 | 2 | 3 | G |
| p28 | 3 | 2 | 3 | 3 | 2 | 2 | 3 | 3 | 3 | 3 | G |
| p29 | 1 | 1 | 1 | 1 | 3 | 1 | 1 | 1 | 1 | 1 | O |
| p30 | 1 | 2 | 1 | 2 | 2 | 1 | 1 | 1 | 2 | 1 | G |
| p31 | 4 | 3 | 4 | 4 | 4 | 3 | 4 | 4 | 2 | 3 | G |
| p32 | 2 | 2 | 2 | 2 | 2 | 2 | 2 | 2 | 2 | 2 | G |
| p33 | 3 | 3 | 4 | 3 | 3 | 2 | 2 | 2 | 1 | 2 | G |
| p34 | 2 | 1 | 3 | 3 | 2 | 1 | 2 | 1 | 2 | 3 | O |

| Participant | VA1 | VA2 | VA3 | VA4 | VA5 | VA6 | VA7 | VA8 | VA9 | VA10 | Type |
|---|---|---|---|---|---|---|---|---|---|---|---|
| p35 | 3 | 3 | 4 | 4 | 3 | 3 | 3 | 3 | 3 | 2 | O |
| p36 | 3 | 4 | 3 | 4 | 3 | 3 | 4 | 3 | 3 | 3 | G |
| p37 | 2 | 1 | 2 | 1 | 1 | 1 | 1 | 1 | 2 | 1 | O |
| p38 | 1 | 1 | 1 | 1 | 1 | 1 | 1 | 1 | 1 | 1 | O |
| p39 | 1 | 2 | 2 | 1 | 3 | 2 | 2 | 3 | 3 | 2 | O |
| p40 | 5 | 4 | 5 | 4 | 4 | 5 | 4 | 5 | 5 | 5 | O |

**Table 30.** Raw data of the experiment (for the eighth infographic in the surveys)

| Participant | VA1 | VA2 | VA3 | VA4 | VA5 | VA6 | VA7 | VA8 | VA9 | VA10 | Type |
|---|---|---|---|---|---|---|---|---|---|---|---|
| p1 | 4 | 4 | 4 | 5 | 3 | 3 | 3 | 2 | 2 | 2 | G |
| p2 | 3 | 4 | 3 | 4 | 3 | 2 | 3 | 3 | 3 | 3 | G |
| p3 | 3 | 4 | 2 | 2 | 3 | 2 | 2 | 1 | 2 | 2 | G |
| p4 | 4 | 4 | 4 | 4 | 4 | 4 | 3 | 3 | 4 | 4 | G |
| p5 | 4 | 4 | 4 | 4 | 4 | 3 | 2 | 2 | 2 | 2 | G |
| p6 | 4 | 5 | 3 | 4 | 4 | 4 | 5 | 4 | 3 | 5 | G |
| p7 | 3 | 5 | 3 | 4 | 5 | 2 | 4 | 3 | 5 | 2 | G |
| p8 | 4 | 5 | 4 | 4 | 4 | 3 | 3 | 3 | 4 | 3 | G |
| p9 | 3 | 4 | 4 | 4 | 4 | 3 | 3 | 3 | 2 | 2 | G |
| p10 | 3 | 2 | 2 | 3 | 2 | 2 | 2 | 3 | 2 | 2 | G |
| p11 | 4 | 2 | 3 | 4 | 3 | 4 | 3 | 4 | 2 | 3 | G |
| p12 | 5 | 5 | 5 | 5 | 3 | 4 | 4 | 3 | 3 | 4 | O |
| p13 | 1 | 2 | 1 | 2 | 2 | 1 | 2 | 1 | 1 | 1 | G |
| p14 | 4 | 4 | 3 | 3 | 4 | 3 | 3 | 4 | 3 | 3 | O |
| p15 | 4 | 3 | 4 | 3 | 3 | 4 | 3 | 4 | 3 | 4 | G |
| p16 | 4 | 4 | 4 | 5 | 4 | 4 | 4 | 2 | 2 | 3 | G |
| p17 | 3 | 5 | 3 | 5 | 4 | 3 | 5 | 4 | 2 | 1 | G |
| p18 | 4 | 4 | 4 | 5 | 4 | 4 | 3 | 3 | 3 | 3 | G |
| p19 | 4 | 3 | 2 | 4 | 3 | 2 | 4 | 4 | 1 | 4 | O |
| p20 | 2 | 2 | 4 | 4 | 3 | 2 | 4 | 2 | 3 | 3 | G |
| p21 | 2 | 2 | 2 | 2 | 2 | 2 | 2 | 2 | 2 | 2 | G |
| p22 | 3 | 3 | 3 | 3 | 3 | 2 | 2 | 1 | 1 | 1 | G |
| p23 | 2 | 3 | 2 | 3 | 2 | 2 | 1 | 1 | 1 | 2 | G |
| p24 | 4 | 4 | 2 | 4 | 2 | 3 | 3 | 3 | 4 | 3 | G |
| p25 | 4 | 3 | 4 | 5 | 5 | 4 | 4 | 4 | 4 | 3 | O |
| p26 | 3 | 3 | 4 | 4 | 3 | 4 | 3 | 2 | 4 | 3 | G |
| p27 | 4 | 4 | 3 | 3 | 4 | 2 | 4 | 1 | 1 | 2 | G |
| p28 | 4 | 5 | 4 | 4 | 4 | 4 | 4 | 5 | 3 | 4 | G |
| p29 | 3 | 5 | 3 | 3 | 3 | 3 | 1 | 3 | 1 | 3 | G |
| p30 | 3 | 4 | 3 | 3 | 3 | 3 | 3 | 3 | 4 | 3 | G |
| p31 | 4 | 4 | 3 | 3 | 3 | 3 | 3 | 3 | 2 | 2 | G |
| p32 | 3 | 3 | 4 | 4 | 3 | 1 | 3 | 2 | 2 | 2 | O |
| p33 | 4 | 3 | 4 | 4 | 4 | 2 | 1 | 1 | 2 | 2 | G |
| p34 | 2 | 2 | 3 | 4 | 4 | 3 | 3 | 3 | 1 | 2 | G |
| p35 | 4 | 4 | 4 | 5 | 5 | 1 | 4 | 2 | 2 | 3 | G |
| p36 | 2 | 3 | 2 | 2 | 3 | 2 | 2 | 2 | 2 | 3 | G |
| p37 | 4 | 3 | 4 | 4 | 5 | 2 | 3 | 2 | 3 | 2 | G |
| p38 | 4 | 2 | 3 | 3 | 2 | 3 | 2 | 3 | 2 | 3 | O |
| p39 | 4 | 4 | 3 | 5 | 3 | 3 | 3 | 4 | 4 | 3 | O |

| Participant | VA1 | VA2 | VA3 | VA4 | VA5 | VA6 | VA7 | VA8 | VA9 | VA10 | Type |
|---|---|---|---|---|---|---|---|---|---|---|---|
| p40 | 3 | 2 | 4 | 3 | 3 | 4 | 3 | 2 | 4 | 3 | G |

**Table 31.** Raw data of the experiment (for the nineth infographic in the surveys)

| Participant | VA1 | VA2 | VA3 | VA4 | VA5 | VA6 | VA7 | VA8 | VA9 | VA10 | Type |
|---|---|---|---|---|---|---|---|---|---|---|---|
| p1 | 2 | 4 | 3 | 5 | 5 | 1 | 1 | 1 | 1 | 1 | O |
| p2 | 4 | 4 | 4 | 5 | 5 | 3 | 3 | 3 | 3 | 3 | G |
| p3 | 2 | 4 | 3 | 4 | 4 | 2 | 2 | 1 | 1 | 1 | O |
| p4 | 4 | 4 | 4 | 5 | 4 | 3 | 3 | 3 | 3 | 3 | O |
| p5 | 4 | 4 | 4 | 4 | 4 | 3 | 4 | 3 | 3 | 3 | O |
| p6 | 3 | 3 | 4 | 4 | 3 | 3 | 4 | 3 | 3 | 4 | G |
| p7 | 5 | 5 | 5 | 5 | 5 | 2 | 5 | 3 | 5 | 4 | O |
| p8 | 5 | 5 | 5 | 5 | 5 | 4 | 5 | 3 | 5 | 4 | G |
| p9 | 3 | 4 | 3 | 4 | 4 | 2 | 3 | 2 | 2 | 2 | G |
| p10 | 2 | 3 | 3 | 3 | 3 | 1 | 2 | 2 | 2 | 2 | G |
| p11 | 3 | 5 | 2 | 4 | 3 | 3 | 3 | 4 | 3 | 3 | G |
| p12 | 4 | 5 | 4 | 5 | 5 | 3 | 3 | 3 | 2 | 4 | G |
| p13 | 1 | 2 | 3 | 3 | 5 | 1 | 5 | 3 | 1 | 3 | G |
| p14 | 3 | 4 | 4 | 4 | 4 | 3 | 4 | 4 | 4 | 4 | O |
| p15 | 5 | 5 | 5 | 4 | 4 | 4 | 3 | 4 | 3 | 4 | G |
| p16 | 2 | 4 | 4 | 4 | 4 | 2 | 2 | 1 | 1 | 1 | G |
| p17 | 3 | 4 | 5 | 3 | 4 | 4 | 4 | 2 | 4 | 3 | O |
| p18 | 5 | 3 | 4 | 4 | 4 | 2 | 1 | 3 | 4 | 3 | G |
| p19 | 4 | 4 | 4 | 5 | 4 | 2 | 4 | 4 | 1 | 4 | O |
| p20 | 4 | 5 | 3 | 4 | 4 | 3 | 5 | 3 | 4 | 4 | O |
| p21 | 3 | 2 | 3 | 3 | 3 | 3 | 3 | 2 | 1 | 1 | O |
| p22 | 4 | 4 | 4 | 3 | 3 | 2 | 4 | 2 | 1 | 1 | G |
| p23 | 2 | 3 | 2 | 3 | 2 | 1 | 3 | 2 | 2 | 2 | O |
| p24 | 4 | 4 | 4 | 5 | 5 | 2 | 3 | 4 | 3 | 4 | G |
| p25 | 3 | 5 | 3 | 4 | 4 | 2 | 3 | 1 | 2 | 2 | G |
| p26 | 4 | 4 | 5 | 5 | 5 | 4 | 4 | 5 | 3 | 4 | O |
| p27 | 4 | 5 | 4 | 4 | 5 | 4 | 4 | 4 | 3 | 4 | G |
| p28 | 4 | 5 | 4 | 4 | 4 | 4 | 4 | 2 | 3 | 3 | O |
| p29 | 3 | 5 | 3 | 5 | 3 | 1 | 1 | 1 | 1 | 3 | G |
| p30 | 4 | 4 | 4 | 4 | 4 | 3 | 4 | 3 | 2 | 3 | O |
| p31 | 4 | 4 | 2 | 4 | 3 | 3 | 2 | 3 | 1 | 2 | G |
| p32 | 2 | 3 | 4 | 3 | 3 | 2 | 2 | 2 | 2 | 2 | O |
| p33 | 5 | 4 | 5 | 5 | 5 | 2 | 2 | 3 | 1 | 2 | G |
| p34 | 4 | 3 | 5 | 5 | 5 | 3 | 4 | 4 | 2 | 4 | G |
| p35 | 4 | 3 | 3 | 3 | 3 | 2 | 2 | 3 | 3 | 4 | O |
| p36 | 4 | 4 | 4 | 4 | 5 | 3 | 3 | 4 | 4 | 4 | G |
| p37 | 4 | 4 | 4 | 4 | 4 | 3 | 4 | 2 | 2 | 2 | G |
| p38 | 3 | 2 | 3 | 3 | 3 | 3 | 3 | 2 | 3 | 3 | O |
| p39 | 3 | 4 | 4 | 4 | 3 | 1 | 2 | 3 | 3 | 1 | G |
| p40 | 4 | 4 | 3 | 4 | 4 | 4 | 3 | 4 | 4 | 4 | O |

**Table 32.** Raw data of the experiment (for the tenth infographic in the surveys)

| Participant | VA1 | VA2 | VA3 | VA4 | VA5 | VA6 | VA7 | VA8 | VA9 | VA10 | Type |
|---|---|---|---|---|---|---|---|---|---|---|---|
| p1 | 3 | 5 | 3 | 5 | 4 | 2 | 2 | 2 | 2 | 2 | G |
| p2 | 3 | 4 | 4 | 4 | 3 | 3 | 3 | 3 | 3 | 3 | G |
| p3 | 1 | 1 | 1 | 1 | 1 | 2 | 2 | 3 | 2 | 2 | G |
| p4 | 4 | 4 | 3 | 2 | 3 | 3 | 3 | 3 | 4 | 4 | G |
| p5 | 4 | 4 | 4 | 3 | 4 | 3 | 3 | 2 | 3 | 2 | G |
| p6 | 4 | 5 | 4 | 4 | 5 | 4 | 4 | 4 | 3 | 4 | O |
| p7 | 3 | 4 | 3 | 5 | 3 | 3 | 3 | 3 | 4 | 3 | G |
| p8 | 4 | 4 | 3 | 3 | 4 | 4 | 2 | 3 | 2 | 3 | G |
| p9 | 3 | 3 | 3 | 4 | 3 | 3 | 4 | 2 | 2 | 2 | G |
| p10 | 1 | 3 | 1 | 1 | 1 | 1 | 1 | 1 | 1 | 1 | G |
| p11 | 3 | 3 | 3 | 4 | 4 | 2 | 3 | 3 | 3 | 3 | G |
| p12 | 4 | 4 | 4 | 4 | 3 | 2 | 3 | 2 | 1 | 3 | G |
| p13 | 1 | 2 | 2 | 3 | 1 | 1 | 2 | 1 | 1 | 1 | G |
| p14 | 3 | 4 | 3 | 3 | 4 | 3 | 3 | 4 | 3 | 3 | O |
| p15 | 3 | 4 | 4 | 5 | 4 | 4 | 3 | 3 | 4 | 4 | G |
| p16 | 4 | 5 | 3 | 4 | 3 | 2 | 2 | 1 | 1 | 1 | G |
| p17 | 5 | 5 | 4 | 5 | 4 | 2 | 5 | 3 | 4 | 4 | G |
| p18 | 4 | 5 | 3 | 4 | 3 | 2 | 3 | 2 | 2 | 3 | G |
| p19 | 4 | 5 | 2 | 2 | 2 | 1 | 1 | 2 | 1 | 2 | G |
| p20 | 3 | 5 | 3 | 4 | 4 | 3 | 2 | 2 | 1 | 5 | G |
| p21 | 2 | 2 | 3 | 3 | 3 | 2 | 2 | 2 | 2 | 2 | O |
| p22 | 3 | 3 | 2 | 3 | 2 | 3 | 3 | 2 | 1 | 1 | G |
| p23 | 1 | 3 | 2 | 2 | 1 | 2 | 1 | 1 | 1 | 1 | G |
| p24 | 4 | 5 | 4 | 5 | 4 | 2 | 3 | 3 | 3 | 3 | G |
| p25 | 3 | 4 | 4 | 2 | 2 | 4 | 2 | 3 | 3 | 3 | O |
| p26 | 4 | 4 | 3 | 2 | 3 | 4 | 3 | 3 | 2 | 2 | G |
| p27 | 2 | 3 | 4 | 4 | 2 | 2 | 2 | 2 | 1 | 1 | G |
| p28 | 4 | 4 | 3 | 4 | 3 | 3 | 3 | 2 | 3 | 3 | G |
| p29 | 3 | 5 | 3 | 5 | 5 | 1 | 1 | 1 | 1 | 3 | G |
| p30 | 2 | 3 | 2 | 3 | 3 | 3 | 2 | 2 | 2 | 2 | G |
| p31 | 3 | 3 | 3 | 3 | 3 | 3 | 3 | 3 | 3 | 3 | G |
| p32 | 2 | 3 | 3 | 3 | 3 | 2 | 2 | 2 | 2 | 2 | G |
| p33 | 1 | 1 | 1 | 2 | 3 | 1 | 1 | 1 | 1 | 1 | G |
| p34 | 2 | 5 | 3 | 4 | 4 | 2 | 3 | 1 | 1 | 3 | G |
| p35 | 4 | 5 | 4 | 5 | 5 | 3 | 3 | 3 | 4 | 4 | O |
| p36 | 3 | 4 | 3 | 3 | 3 | 3 | 2 | 3 | 3 | 3 | G |
| p37 | 5 | 5 | 5 | 4 | 4 | 4 | 4 | 3 | 3 | 2 | G |
| p38 | 1 | 3 | 2 | 2 | 2 | 1 | 1 | 1 | 1 | 1 | G |
| p39 | 2 | 4 | 2 | 4 | 3 | 2 | 2 | 2 | 3 | 2 | G |
| p40 | 2 | 4 | 2 | 3 | 3 | 3 | 4 | 3 | 4 | 3 | G |

**Table 33.** Type of infographic presented in each position of the survey, depending on the group (or survey variant).

|  | #1 | #2 | #3 | #4 | #5 | #6 | #7 | #8 | #9 | #10 |
|---|---|---|---|---|---|---|---|---|---|---|
| **Group 1** | O | O | O | G | G | O | G | G | G | O |
| **Group 2** | G | G | G | O | O | G | O | O | O | G |

**Table 34.** Computation of the visual attractiveness score per infographic and subject. It is based on the scores for each individual visual attractiveness variable presented in Table 23 to Table 32

| Participant | #1 | #2 | #3 | #4 | #5 | #6 | #7 | #8 | #9 | #10 |
|---|---|---|---|---|---|---|---|---|---|---|
| p1 | 3 | 5 | 3 | 5 | 4 | 2 | 2 | 2 | 2 | 2 |
| p2 | 3 | 4 | 4 | 4 | 3 | 3 | 3 | 3 | 3 | 3 |
| p3 | 1 | 1 | 1 | 1 | 1 | 2 | 2 | 3 | 2 | 2 |
| p4 | 4 | 4 | 3 | 2 | 3 | 3 | 3 | 3 | 4 | 4 |
| p5 | 4 | 4 | 4 | 3 | 4 | 3 | 3 | 2 | 3 | 2 |
| p6 | 4 | 5 | 4 | 4 | 5 | 4 | 4 | 4 | 3 | 4 |
| p7 | 3 | 4 | 3 | 5 | 3 | 3 | 3 | 3 | 4 | 3 |
| p8 | 4 | 4 | 3 | 3 | 4 | 4 | 2 | 3 | 2 | 3 |
| p9 | 3 | 3 | 3 | 4 | 3 | 3 | 4 | 2 | 2 | 2 |
| p10 | 1 | 3 | 1 | 1 | 1 | 1 | 1 | 1 | 1 | 1 |
| p11 | 3 | 3 | 3 | 4 | 4 | 2 | 3 | 3 | 3 | 3 |
| p12 | 4 | 4 | 4 | 4 | 3 | 2 | 3 | 2 | 1 | 3 |
| p13 | 1 | 2 | 2 | 3 | 1 | 1 | 2 | 1 | 1 | 1 |
| p14 | 3 | 4 | 3 | 3 | 4 | 3 | 3 | 4 | 3 | 3 |
| p15 | 3 | 4 | 4 | 5 | 4 | 4 | 3 | 3 | 4 | 4 |
| p16 | 4 | 5 | 3 | 4 | 3 | 2 | 2 | 1 | 1 | 1 |
| p17 | 5 | 5 | 4 | 5 | 4 | 2 | 5 | 3 | 4 | 4 |
| p18 | 4 | 5 | 3 | 4 | 3 | 2 | 3 | 2 | 2 | 3 |
| p19 | 4 | 5 | 2 | 2 | 2 | 1 | 1 | 2 | 1 | 2 |
| p20 | 3 | 5 | 3 | 4 | 4 | 3 | 2 | 2 | 1 | 5 |
| p21 | 2 | 2 | 3 | 3 | 3 | 2 | 2 | 2 | 2 | 2 |
| p22 | 3 | 3 | 2 | 3 | 2 | 3 | 3 | 2 | 1 | 1 |
| p23 | 1 | 3 | 2 | 2 | 1 | 2 | 1 | 1 | 1 | 1 |
| p24 | 4 | 5 | 4 | 5 | 4 | 2 | 3 | 3 | 3 | 3 |
| p25 | 3 | 4 | 4 | 2 | 2 | 4 | 2 | 3 | 3 | 3 |
| p26 | 4 | 4 | 3 | 2 | 3 | 4 | 3 | 3 | 2 | 2 |
| p27 | 2 | 3 | 4 | 4 | 2 | 2 | 2 | 2 | 1 | 1 |
| p28 | 4 | 4 | 3 | 4 | 3 | 3 | 3 | 2 | 3 | 3 |
| p29 | 3 | 5 | 3 | 5 | 5 | 1 | 1 | 1 | 1 | 3 |
| p30 | 2 | 3 | 2 | 3 | 3 | 3 | 2 | 2 | 2 | 2 |
| p31 | 3 | 3 | 3 | 3 | 3 | 3 | 3 | 3 | 3 | 3 |
| p32 | 2 | 3 | 3 | 3 | 3 | 2 | 2 | 2 | 2 | 2 |
| p33 | 1 | 1 | 1 | 2 | 3 | 1 | 1 | 1 | 1 | 1 |
| p34 | 2 | 5 | 3 | 4 | 4 | 2 | 3 | 1 | 1 | 3 |
| p35 | 4 | 5 | 4 | 5 | 5 | 3 | 3 | 3 | 4 | 4 |
| p36 | 3 | 4 | 3 | 3 | 3 | 3 | 2 | 3 | 3 | 3 |
| p37 | 5 | 5 | 5 | 4 | 4 | 4 | 4 | 3 | 3 | 2 |
| p38 | 1 | 3 | 2 | 2 | 2 | 1 | 1 | 1 | 1 | 1 |
| p39 | 2 | 4 | 2 | 4 | 3 | 2 | 2 | 2 | 3 | 2 |
| p40 | 2 | 4 | 2 | 3 | 3 | 3 | 4 | 3 | 4 | 3 |

**Table 35.** Averages of the scores that each participant gave to the original and generated infographics they were presented in the survey (this is the first dataset analysed with SPSS)

| Participant | Group | O | G |
|---|---|---|---|
| p1 | 1 | 31.8 | 31.2 |
| p2 | 1 | 28.6 | 31.6 |

| Participant | Group | O | G |
|---|---|---|---|
| p3 | 1 | 22.0 | 24.6 |
| p4 | 1 | 30.0 | 36.4 |
| p5 | 1 | 27.0 | 32.4 |
| p6 | 1 | 39.6 | 40.2 |
| p7 | 1 | 31.4 | 38.8 |
| p8 | 1 | 36.8 | 39.4 |
| p9 | 1 | 30.8 | 32.8 |
| p10 | 1 | 16.2 | 20.2 |
| p11 | 1 | 32.8 | 31.2 |
| p12 | 1 | 34.0 | 35.8 |
| p13 | 1 | 12.8 | 19.4 |
| p14 | 1 | 24.0 | 30.2 |
| p15 | 1 | 32.8 | 32.0 |
| p16 | 1 | 29.2 | 25.8 |
| p17 | 1 | 31.2 | 34.6 |
| p18 | 1 | 30.0 | 30.2 |
| p19 | 1 | 23.0 | 26.0 |
| p20 | 1 | 27.0 | 25.8 |
| p21 | 2 | 21.4 | 20.2 |
| p22 | 2 | 23.8 | 23.6 |
| p23 | 2 | 21.2 | 18.8 |
| p24 | 2 | 36.6 | 32.0 |
| p25 | 2 | 28.4 | 32.2 |
| p26 | 2 | 30.6 | 28.2 |
| p27 | 2 | 27.4 | 31.4 |
| p28 | 2 | 32.4 | 33.8 |
| p29 | 2 | 26.0 | 21.2 |
| p30 | 2 | 25.6 | 26.8 |
| p31 | 2 | 30.6 | 32.2 |
| p32 | 2 | 22.6 | 23.2 |
| p33 | 2 | 19.4 | 29.4 |
| p34 | 2 | 21.2 | 26.6 |
| p35 | 2 | 31.8 | 31.0 |
| p36 | 2 | 29.8 | 32.0 |
| p37 | 2 | 28.4 | 25.2 |
| p38 | 2 | 13.0 | 17.8 |
| p39 | 2 | 21.8 | 30.6 |
| p40 | 2 | 31.0 | 39.2 |
| | Average | 27.40 | 29.40 |
| | Standard deviation | 6.09 | 5.91 |

**Table 36.** Compact table showing the guesses of each participant in group 1, for each of the infographics presented to them in the survey. The values in italics correspond to the real type of the infographics for group 1. A red cell background indicates that the participant failed in their guess. A green cell background and bold font indicates that the participant succeeded in their guess.

| Participant | #1 | #2 | #3 | #4 | #5 | #6 | #7 | #8 | #9 | #10 |
|---|---|---|---|---|---|---|---|---|---|---|
| | *O* | *O* | *O* | *G* | *G* | *O* | *G* | *G* | *G* | *O* |
| p1 | G | **O** | G | O | **G** | G | O | **G** | O | G |

| Participant | #1 | #2 | #3 | #4 | #5 | #6 | #7 | #8 | #9 | #10 |
|---|---|---|---|---|---|---|---|---|---|---|
|  | *O* | *O* | *O* | *G* | *G* | *O* | *G* | *G* | *G* | *O* |
| p2 | G | G | G | O | **G** | G | O | **G** | **G** | G |
| p3 | G | **O** | G | **G** | **G** | G | **G** | G | O | G |
| p4 | G | G | G | **G** | **G** | G | **G** | G | O | G |
| p5 | G | G | **O** | O | O | G | O | **G** | O | G |
| p6 | **O** | **O** | G | O | O | **O** | O | **G** | **G** | **O** |
| p7 | G | G | **O** | O | O | G | **G** | **G** | O | G |
| p8 | G | **O** | **O** | O | **G** | G | O | **G** | **G** | G |
| p9 | G | **O** | G | O | **G** | **O** | O | **G** | **G** | G |
| p10 | **O** | **O** | G | O | **G** | G | O | **G** | **G** | G |
| p11 | **O** | **O** | G | **G** | O | G | O | **G** | **G** | G |
| p12 | G | G | G | O | O | **O** | O | O | **G** | G |
| p13 | G | **O** | G | O | O | **O** | **G** | **G** | **G** | G |
| p14 | G | G | **O** | **G** | O | G | O | O | O | **O** |
| p15 | G | G | G | **G** | **G** | G | **G** | **G** | **G** | G |
| p16 | G | **O** | G | O | O | **O** | O | **G** | **G** | G |
| p17 | G | **O** | **O** | **G** | O | **O** | **G** | **G** | O | G |
| p18 | **O** | G | G | O | **G** | G | **G** | **G** | **G** | G |
| p19 | **O** | **O** | G | O | O | G | **G** | O | O | G |
| p20 | G | G | G | O | **G** | G | **G** | O | O | G |

**Table 37.** Compact table showing the guesses of each participant in group 2, for each of the infographics presented to them in the survey. The values in italics correspond to the real type of the infographics for group 2. A red cell background indicates that the participant failed in their guess. A green cell background and bold font indicates that the participant succeeded in their guess.

| Participant | #1 | #2 | #3 | #4 | #5 | #6 | #7 | #8 | #9 | #10 |
|---|---|---|---|---|---|---|---|---|---|---|
|  | *G* | *G* | *G* | *O* | *O* | *G* | *O* | *O* | *O* | *G* |
| p21 | **G** | O | **G** | **O** | **O** | O | G | G | **O** | O |
| p22 | **G** | O | **G** | **O** | **O** | O | G | G | G | **G** |
| p23 | **G** | O | O | G | **O** | O | G | G | **O** | **G** |
| p24 | O | **G** | **G** | **O** | G | **G** | O | G | G | **G** |
| p25 | O | **G** | O | **O** | G | **G** | O | **O** | G | O |
| p26 | **G** | O | O | **O** | G | **G** | O | G | **O** | **G** |
| p27 | **G** | O | **G** | **O** | G | **G** | G | G | G | **G** |
| p28 | **G** | O | **G** | **O** | **O** | O | G | G | **O** | **G** |
| p29 | **G** | O | **G** | **O** | **O** | **G** | O | G | G | **G** |
| p30 | **G** | **G** | O | **O** | **O** | O | G | G | **O** | **G** |
| p31 | **G** | O | **G** | **O** | G | **G** | G | G | G | **G** |
| p32 | O | O | **G** | **O** | **O** | **G** | G | **O** | **O** | **G** |
| p33 | **G** | **G** | **G** | **O** | G | **G** | G | G | G | **G** |
| p34 | **G** | **G** | O | **O** | G | O | O | G | G | **G** |
| p35 | O | O | **G** | **O** | **O** | **G** | O | G | **O** | O |
| p36 | **G** | **G** | **G** | G | G | **G** | G | G | G | **G** |
| p37 | O | O | O | G | G | **G** | O | G | G | **G** |
| p38 | **G** | **G** | **G** | **O** | **O** | **O** | O | **O** | **O** | **G** |
| p39 | **G** | **G** | **G** | **O** | G | **G** | **O** | **O** | G | **G** |
| p40 | **G** | **G** | **G** | G | G | **O** | **O** | G | **O** | **G** |

**Table 38.** The same data about the guesses presented in Table 36 and Table 37, now presented as 400 individual decision moments. This is the second dataset analysed with SPSS.

| Point | Participant | Infographic | Group | Type | Guess | Success |
|---|---|---|---|---|---|---|
| 1 | 1 | 1 | 1 | O | G | Fail |
| 2 | 1 | 2 | 1 | O | O | Success |
| 3 | 1 | 3 | 1 | O | G | Fail |
| 4 | 1 | 4 | 1 | G | O | Fail |
| 5 | 1 | 5 | 1 | G | G | Success |
| 6 | 1 | 6 | 1 | O | G | Fail |
| 7 | 1 | 7 | 1 | G | O | Fail |
| 8 | 1 | 8 | 1 | G | G | Success |
| 9 | 1 | 9 | 1 | G | O | Fail |
| 10 | 1 | 10 | 1 | O | G | Fail |
| 11 | 2 | 1 | 1 | O | G | Fail |
| 12 | 2 | 2 | 1 | O | G | Fail |
| 13 | 2 | 3 | 1 | O | G | Fail |
| 14 | 2 | 4 | 1 | G | O | Fail |
| 15 | 2 | 5 | 1 | G | G | Success |
| 16 | 2 | 6 | 1 | O | G | Fail |
| 17 | 2 | 7 | 1 | G | O | Fail |
| 18 | 2 | 8 | 1 | G | G | Success |
| 19 | 2 | 9 | 1 | G | G | Success |
| 20 | 2 | 10 | 1 | O | G | Fail |
| 21 | 3 | 1 | 1 | O | G | Fail |
| 22 | 3 | 2 | 1 | O | O | Success |
| 23 | 3 | 3 | 1 | O | G | Fail |
| 24 | 3 | 4 | 1 | G | G | Success |
| 25 | 3 | 5 | 1 | G | G | Success |
| 26 | 3 | 6 | 1 | O | G | Fail |
| 27 | 3 | 7 | 1 | G | G | Success |
| 28 | 3 | 8 | 1 | G | G | Success |
| 29 | 3 | 9 | 1 | G | O | Fail |
| 30 | 3 | 10 | 1 | O | G | Fail |
| 31 | 4 | 1 | 1 | O | G | Fail |
| 32 | 4 | 2 | 1 | O | G | Fail |
| 33 | 4 | 3 | 1 | O | G | Fail |
| 34 | 4 | 4 | 1 | G | G | Success |
| 35 | 4 | 5 | 1 | G | G | Success |
| 36 | 4 | 6 | 1 | O | G | Fail |
| 37 | 4 | 7 | 1 | G | G | Success |
| 38 | 4 | 8 | 1 | G | G | Success |
| 39 | 4 | 9 | 1 | G | O | Fail |
| 40 | 4 | 10 | 1 | O | G | Fail |
| 41 | 5 | 1 | 1 | O | G | Fail |
| 42 | 5 | 2 | 1 | O | G | Fail |
| 43 | 5 | 3 | 1 | O | O | Success |
| 44 | 5 | 4 | 1 | G | O | Fail |
| 45 | 5 | 5 | 1 | G | O | Fail |
| 46 | 5 | 6 | 1 | O | G | Fail |

| Point | Participant | Infographic | Group | Type | Guess | Success |
|---|---|---|---|---|---|---|
| 47 | 5 | 7 | 1 | G | O | Fail |
| 48 | 5 | 8 | 1 | G | G | Success |
| 49 | 5 | 9 | 1 | G | O | Fail |
| 50 | 5 | 10 | 1 | O | G | Fail |
| 51 | 6 | 1 | 1 | O | O | Success |
| 52 | 6 | 2 | 1 | O | O | Success |
| 53 | 6 | 3 | 1 | O | G | Fail |
| 54 | 6 | 4 | 1 | G | O | Fail |
| 55 | 6 | 5 | 1 | G | O | Fail |
| 56 | 6 | 6 | 1 | O | O | Success |
| 57 | 6 | 7 | 1 | G | O | Fail |
| 58 | 6 | 8 | 1 | G | G | Success |
| 59 | 6 | 9 | 1 | G | G | Success |
| 60 | 6 | 10 | 1 | O | O | Success |
| 61 | 7 | 1 | 1 | O | G | Fail |
| 62 | 7 | 2 | 1 | O | G | Fail |
| 63 | 7 | 3 | 1 | O | O | Success |
| 64 | 7 | 4 | 1 | G | O | Fail |
| 65 | 7 | 5 | 1 | G | O | Fail |
| 66 | 7 | 6 | 1 | O | G | Fail |
| 67 | 7 | 7 | 1 | G | G | Success |
| 68 | 7 | 8 | 1 | G | G | Success |
| 69 | 7 | 9 | 1 | G | O | Fail |
| 70 | 7 | 10 | 1 | O | G | Fail |
| 71 | 8 | 1 | 1 | O | G | Fail |
| 72 | 8 | 2 | 1 | O | O | Success |
| 73 | 8 | 3 | 1 | O | O | Success |
| 74 | 8 | 4 | 1 | G | O | Fail |
| 75 | 8 | 5 | 1 | G | G | Success |
| 76 | 8 | 6 | 1 | O | G | Fail |
| 77 | 8 | 7 | 1 | G | O | Fail |
| 78 | 8 | 8 | 1 | G | G | Success |
| 79 | 8 | 9 | 1 | G | G | Success |
| 80 | 8 | 10 | 1 | O | G | Fail |
| 81 | 9 | 1 | 1 | O | G | Fail |
| 82 | 9 | 2 | 1 | O | O | Success |
| 83 | 9 | 3 | 1 | O | G | Fail |
| 84 | 9 | 4 | 1 | G | O | Fail |
| 85 | 9 | 5 | 1 | G | G | Success |
| 86 | 9 | 6 | 1 | O | O | Success |
| 87 | 9 | 7 | 1 | G | O | Fail |
| 88 | 9 | 8 | 1 | G | G | Success |
| 89 | 9 | 9 | 1 | G | G | Success |
| 90 | 9 | 10 | 1 | O | G | Fail |
| 91 | 10 | 1 | 1 | O | O | Success |
| 92 | 10 | 2 | 1 | O | O | Success |
| 93 | 10 | 3 | 1 | O | G | Fail |
| 94 | 10 | 4 | 1 | G | O | Fail |
| 95 | 10 | 5 | 1 | G | G | Success |

| Point | Participant | Infographic | Group | Type | Guess | Success |
|---|---|---|---|---|---|---|
| 96 | 10 | 6 | 1 | O | G | Fail |
| 97 | 10 | 7 | 1 | G | O | Fail |
| 98 | 10 | 8 | 1 | G | G | Success |
| 99 | 10 | 9 | 1 | G | G | Success |
| 100 | 10 | 10 | 1 | O | G | Fail |
| 101 | 11 | 1 | 1 | O | O | Success |
| 102 | 11 | 2 | 1 | O | O | Success |
| 103 | 11 | 3 | 1 | O | G | Fail |
| 104 | 11 | 4 | 1 | G | G | Success |
| 105 | 11 | 5 | 1 | G | O | Fail |
| 106 | 11 | 6 | 1 | O | G | Fail |
| 107 | 11 | 7 | 1 | G | O | Fail |
| 108 | 11 | 8 | 1 | G | G | Success |
| 109 | 11 | 9 | 1 | G | G | Success |
| 110 | 11 | 10 | 1 | O | G | Fail |
| 111 | 12 | 1 | 1 | O | G | Fail |
| 112 | 12 | 2 | 1 | O | G | Fail |
| 113 | 12 | 3 | 1 | O | G | Fail |
| 114 | 12 | 4 | 1 | G | O | Fail |
| 115 | 12 | 5 | 1 | G | O | Fail |
| 116 | 12 | 6 | 1 | O | O | Success |
| 117 | 12 | 7 | 1 | G | O | Fail |
| 118 | 12 | 8 | 1 | G | O | Fail |
| 119 | 12 | 9 | 1 | G | G | Success |
| 120 | 12 | 10 | 1 | O | G | Fail |
| 121 | 13 | 1 | 1 | O | G | Fail |
| 122 | 13 | 2 | 1 | O | O | Success |
| 123 | 13 | 3 | 1 | O | G | Fail |
| 124 | 13 | 4 | 1 | G | O | Fail |
| 125 | 13 | 5 | 1 | G | O | Fail |
| 126 | 13 | 6 | 1 | O | O | Success |
| 127 | 13 | 7 | 1 | G | G | Success |
| 128 | 13 | 8 | 1 | G | G | Success |
| 129 | 13 | 9 | 1 | G | G | Success |
| 130 | 13 | 10 | 1 | O | G | Fail |
| 131 | 14 | 1 | 1 | O | G | Fail |
| 132 | 14 | 2 | 1 | O | G | Fail |
| 133 | 14 | 3 | 1 | O | O | Success |
| 134 | 14 | 4 | 1 | G | G | Success |
| 135 | 14 | 5 | 1 | G | O | Fail |
| 136 | 14 | 6 | 1 | O | G | Fail |
| 137 | 14 | 7 | 1 | G | O | Fail |
| 138 | 14 | 8 | 1 | G | O | Fail |
| 139 | 14 | 9 | 1 | G | O | Fail |
| 140 | 14 | 10 | 1 | O | O | Success |
| 141 | 15 | 1 | 1 | O | G | Fail |
| 142 | 15 | 2 | 1 | O | G | Fail |
| 143 | 15 | 3 | 1 | O | G | Fail |
| 144 | 15 | 4 | 1 | G | G | Success |

| Point | Participant | Infographic | Group | Type | Guess | Success |
|---|---|---|---|---|---|---|
| 145 | 15 | 5 | 1 | G | G | Success |
| 146 | 15 | 6 | 1 | O | G | Fail |
| 147 | 15 | 7 | 1 | G | G | Success |
| 148 | 15 | 8 | 1 | G | G | Success |
| 149 | 15 | 9 | 1 | G | G | Success |
| 150 | 15 | 10 | 1 | O | G | Fail |
| 151 | 16 | 1 | 1 | O | G | Fail |
| 152 | 16 | 2 | 1 | O | O | Success |
| 153 | 16 | 3 | 1 | O | G | Fail |
| 154 | 16 | 4 | 1 | G | O | Fail |
| 155 | 16 | 5 | 1 | G | O | Fail |
| 156 | 16 | 6 | 1 | O | O | Success |
| 157 | 16 | 7 | 1 | G | O | Fail |
| 158 | 16 | 8 | 1 | G | G | Success |
| 159 | 16 | 9 | 1 | G | G | Success |
| 160 | 16 | 10 | 1 | O | G | Fail |
| 161 | 17 | 1 | 1 | O | G | Fail |
| 162 | 17 | 2 | 1 | O | O | Success |
| 163 | 17 | 3 | 1 | O | O | Success |
| 164 | 17 | 4 | 1 | G | G | Success |
| 165 | 17 | 5 | 1 | G | O | Fail |
| 166 | 17 | 6 | 1 | O | O | Success |
| 167 | 17 | 7 | 1 | G | G | Success |
| 168 | 17 | 8 | 1 | G | G | Success |
| 169 | 17 | 9 | 1 | G | O | Fail |
| 170 | 17 | 10 | 1 | O | G | Fail |
| 171 | 18 | 1 | 1 | O | O | Success |
| 172 | 18 | 2 | 1 | O | G | Fail |
| 173 | 18 | 3 | 1 | O | G | Fail |
| 174 | 18 | 4 | 1 | G | O | Fail |
| 175 | 18 | 5 | 1 | G | G | Success |
| 176 | 18 | 6 | 1 | O | G | Fail |
| 177 | 18 | 7 | 1 | G | G | Success |
| 178 | 18 | 8 | 1 | G | G | Success |
| 179 | 18 | 9 | 1 | G | G | Success |
| 180 | 18 | 10 | 1 | O | G | Fail |
| 181 | 19 | 1 | 1 | O | O | Success |
| 182 | 19 | 2 | 1 | O | O | Success |
| 183 | 19 | 3 | 1 | O | G | Fail |
| 184 | 19 | 4 | 1 | G | O | Fail |
| 185 | 19 | 5 | 1 | G | O | Fail |
| 186 | 19 | 6 | 1 | O | G | Fail |
| 187 | 19 | 7 | 1 | G | G | Success |
| 188 | 19 | 8 | 1 | G | O | Fail |
| 189 | 19 | 9 | 1 | G | O | Fail |
| 190 | 19 | 10 | 1 | O | G | Fail |
| 191 | 20 | 1 | 1 | O | G | Fail |
| 192 | 20 | 2 | 1 | O | G | Fail |
| 193 | 20 | 3 | 1 | O | G | Fail |

| Point | Participant | Infographic | Group | Type | Guess | Success |
|---|---|---|---|---|---|---|
| 194 | 20 | 4 | 1 | G | O | Fail |
| 195 | 20 | 5 | 1 | G | G | Success |
| 196 | 20 | 6 | 1 | O | O | Success |
| 197 | 20 | 7 | 1 | G | G | Success |
| 198 | 20 | 8 | 1 | G | G | Success |
| 199 | 20 | 9 | 1 | G | O | Fail |
| 200 | 20 | 10 | 1 | O | G | Fail |
| 201 | 21 | 1 | 2 | G | G | Success |
| 202 | 21 | 2 | 2 | G | O | Fail |
| 203 | 21 | 3 | 2 | G | G | Success |
| 204 | 21 | 4 | 2 | O | O | Success |
| 205 | 21 | 5 | 2 | O | O | Success |
| 206 | 21 | 6 | 2 | G | O | Fail |
| 207 | 21 | 7 | 2 | O | G | Fail |
| 208 | 21 | 8 | 2 | O | G | Fail |
| 209 | 21 | 9 | 2 | O | O | Success |
| 210 | 21 | 10 | 2 | G | O | Fail |
| 211 | 22 | 1 | 2 | G | G | Success |
| 212 | 22 | 2 | 2 | G | O | Fail |
| 213 | 22 | 3 | 2 | G | G | Success |
| 214 | 22 | 4 | 2 | O | O | Success |
| 215 | 22 | 5 | 2 | O | O | Success |
| 216 | 22 | 6 | 2 | G | O | Fail |
| 217 | 22 | 7 | 2 | O | G | Fail |
| 218 | 22 | 8 | 2 | O | G | Fail |
| 219 | 22 | 9 | 2 | O | G | Fail |
| 220 | 22 | 10 | 2 | G | G | Success |
| 221 | 23 | 1 | 2 | G | G | Success |
| 222 | 23 | 2 | 2 | G | O | Fail |
| 223 | 23 | 3 | 2 | G | O | Fail |
| 224 | 23 | 4 | 2 | O | G | Fail |
| 225 | 23 | 5 | 2 | O | O | Success |
| 226 | 23 | 6 | 2 | G | O | Fail |
| 227 | 23 | 7 | 2 | O | G | Fail |
| 228 | 23 | 8 | 2 | O | G | Fail |
| 229 | 23 | 9 | 2 | O | O | Success |
| 230 | 23 | 10 | 2 | G | G | Success |
| 231 | 24 | 1 | 2 | G | O | Fail |
| 232 | 24 | 2 | 2 | G | G | Success |
| 233 | 24 | 3 | 2 | G | G | Success |
| 234 | 24 | 4 | 2 | O | O | Success |
| 235 | 24 | 5 | 2 | O | G | Fail |
| 236 | 24 | 6 | 2 | G | G | Success |
| 237 | 24 | 7 | 2 | O | O | Success |
| 238 | 24 | 8 | 2 | O | G | Fail |
| 239 | 24 | 9 | 2 | O | G | Fail |
| 240 | 24 | 10 | 2 | G | G | Success |
| 241 | 25 | 1 | 2 | G | O | Fail |
| 242 | 25 | 2 | 2 | G | G | Success |

| Point | Participant | Infographic | Group | Type | Guess | Success |
|---|---|---|---|---|---|---|
| 243 | 25 | 3 | 2 | G | O | Fail |
| 244 | 25 | 4 | 2 | O | O | Success |
| 245 | 25 | 5 | 2 | O | G | Fail |
| 246 | 25 | 6 | 2 | G | G | Success |
| 247 | 25 | 7 | 2 | O | O | Success |
| 248 | 25 | 8 | 2 | O | O | Success |
| 249 | 25 | 9 | 2 | O | G | Fail |
| 250 | 25 | 10 | 2 | G | O | Fail |
| 251 | 26 | 1 | 2 | G | G | Success |
| 252 | 26 | 2 | 2 | G | O | Fail |
| 253 | 26 | 3 | 2 | G | O | Fail |
| 254 | 26 | 4 | 2 | O | O | Success |
| 255 | 26 | 5 | 2 | O | G | Fail |
| 256 | 26 | 6 | 2 | G | G | Success |
| 257 | 26 | 7 | 2 | O | O | Success |
| 258 | 26 | 8 | 2 | O | G | Fail |
| 259 | 26 | 9 | 2 | O | O | Success |
| 260 | 26 | 10 | 2 | G | G | Success |
| 261 | 27 | 1 | 2 | G | G | Success |
| 262 | 27 | 2 | 2 | G | O | Fail |
| 263 | 27 | 3 | 2 | G | G | Success |
| 264 | 27 | 4 | 2 | O | O | Success |
| 265 | 27 | 5 | 2 | O | G | Fail |
| 266 | 27 | 6 | 2 | G | G | Success |
| 267 | 27 | 7 | 2 | O | G | Fail |
| 268 | 27 | 8 | 2 | O | G | Fail |
| 269 | 27 | 9 | 2 | O | G | Fail |
| 270 | 27 | 10 | 2 | G | G | Success |
| 271 | 28 | 1 | 2 | G | G | Success |
| 272 | 28 | 2 | 2 | G | O | Fail |
| 273 | 28 | 3 | 2 | G | G | Success |
| 274 | 28 | 4 | 2 | O | O | Success |
| 275 | 28 | 5 | 2 | O | O | Success |
| 276 | 28 | 6 | 2 | G | O | Fail |
| 277 | 28 | 7 | 2 | O | G | Fail |
| 278 | 28 | 8 | 2 | O | G | Fail |
| 279 | 28 | 9 | 2 | O | O | Success |
| 280 | 28 | 10 | 2 | G | G | Success |
| 281 | 29 | 1 | 2 | G | G | Success |
| 282 | 29 | 2 | 2 | G | O | Fail |
| 283 | 29 | 3 | 2 | G | G | Success |
| 284 | 29 | 4 | 2 | O | O | Success |
| 285 | 29 | 5 | 2 | O | O | Success |
| 286 | 29 | 6 | 2 | G | G | Success |
| 287 | 29 | 7 | 2 | O | O | Success |
| 288 | 29 | 8 | 2 | O | G | Fail |
| 289 | 29 | 9 | 2 | O | G | Fail |
| 290 | 29 | 10 | 2 | G | G | Success |
| 291 | 30 | 1 | 2 | G | G | Success |

| Point | Participant | Infographic | Group | Type | Guess | Success |
|---|---|---|---|---|---|---|
| 292 | 30 | 2 | 2 | G | G | Success |
| 293 | 30 | 3 | 2 | G | O | Fail |
| 294 | 30 | 4 | 2 | O | O | Success |
| 295 | 30 | 5 | 2 | O | O | Success |
| 296 | 30 | 6 | 2 | G | O | Fail |
| 297 | 30 | 7 | 2 | O | G | Fail |
| 298 | 30 | 8 | 2 | O | G | Fail |
| 299 | 30 | 9 | 2 | O | O | Success |
| 300 | 30 | 10 | 2 | G | G | Success |
| 301 | 31 | 1 | 2 | G | G | Success |
| 302 | 31 | 2 | 2 | G | O | Fail |
| 303 | 31 | 3 | 2 | G | G | Success |
| 304 | 31 | 4 | 2 | O | O | Success |
| 305 | 31 | 5 | 2 | O | G | Fail |
| 306 | 31 | 6 | 2 | G | G | Success |
| 307 | 31 | 7 | 2 | O | G | Fail |
| 308 | 31 | 8 | 2 | O | G | Fail |
| 309 | 31 | 9 | 2 | O | G | Fail |
| 310 | 31 | 10 | 2 | G | G | Success |
| 311 | 32 | 1 | 2 | G | O | Fail |
| 312 | 32 | 2 | 2 | G | O | Fail |
| 313 | 32 | 3 | 2 | G | G | Success |
| 314 | 32 | 4 | 2 | O | O | Success |
| 315 | 32 | 5 | 2 | O | O | Success |
| 316 | 32 | 6 | 2 | G | G | Success |
| 317 | 32 | 7 | 2 | O | G | Fail |
| 318 | 32 | 8 | 2 | O | O | Success |
| 319 | 32 | 9 | 2 | O | O | Success |
| 320 | 32 | 10 | 2 | G | G | Success |
| 321 | 33 | 1 | 2 | G | G | Success |
| 322 | 33 | 2 | 2 | G | G | Success |
| 323 | 33 | 3 | 2 | G | G | Success |
| 324 | 33 | 4 | 2 | O | O | Success |
| 325 | 33 | 5 | 2 | O | G | Fail |
| 326 | 33 | 6 | 2 | G | G | Success |
| 327 | 33 | 7 | 2 | O | G | Fail |
| 328 | 33 | 8 | 2 | O | G | Fail |
| 329 | 33 | 9 | 2 | O | G | Fail |
| 330 | 33 | 10 | 2 | G | G | Success |
| 331 | 34 | 1 | 2 | G | G | Success |
| 332 | 34 | 2 | 2 | G | G | Success |
| 333 | 34 | 3 | 2 | G | O | Fail |
| 334 | 34 | 4 | 2 | O | O | Success |
| 335 | 34 | 5 | 2 | O | G | Fail |
| 336 | 34 | 6 | 2 | G | O | Fail |
| 337 | 34 | 7 | 2 | O | O | Success |
| 338 | 34 | 8 | 2 | O | G | Fail |
| 339 | 34 | 9 | 2 | O | G | Fail |
| 340 | 34 | 10 | 2 | G | G | Success |

| Point | Participant | Infographic | Group | Type | Guess | Success |
|---|---|---|---|---|---|---|
| 341 | 35 | 1 | 2 | G | O | Fail |
| 342 | 35 | 2 | 2 | G | O | Fail |
| 343 | 35 | 3 | 2 | G | G | Success |
| 344 | 35 | 4 | 2 | O | O | Success |
| 345 | 35 | 5 | 2 | O | O | Success |
| 346 | 35 | 6 | 2 | G | G | Success |
| 347 | 35 | 7 | 2 | O | O | Success |
| 348 | 35 | 8 | 2 | O | G | Fail |
| 349 | 35 | 9 | 2 | O | O | Success |
| 350 | 35 | 10 | 2 | G | O | Fail |
| 351 | 36 | 1 | 2 | G | G | Success |
| 352 | 36 | 2 | 2 | G | G | Success |
| 353 | 36 | 3 | 2 | G | G | Success |
| 354 | 36 | 4 | 2 | O | G | Fail |
| 355 | 36 | 5 | 2 | O | G | Fail |
| 356 | 36 | 6 | 2 | G | G | Success |
| 357 | 36 | 7 | 2 | O | G | Fail |
| 358 | 36 | 8 | 2 | O | G | Fail |
| 359 | 36 | 9 | 2 | O | G | Fail |
| 360 | 36 | 10 | 2 | G | G | Success |
| 361 | 37 | 1 | 2 | G | O | Fail |
| 362 | 37 | 2 | 2 | G | O | Fail |
| 363 | 37 | 3 | 2 | G | O | Fail |
| 364 | 37 | 4 | 2 | O | G | Fail |
| 365 | 37 | 5 | 2 | O | G | Fail |
| 366 | 37 | 6 | 2 | G | G | Success |
| 367 | 37 | 7 | 2 | O | O | Success |
| 368 | 37 | 8 | 2 | O | G | Fail |
| 369 | 37 | 9 | 2 | O | G | Fail |
| 370 | 37 | 10 | 2 | G | G | Success |
| 371 | 38 | 1 | 2 | G | G | Success |
| 372 | 38 | 2 | 2 | G | G | Success |
| 373 | 38 | 3 | 2 | G | G | Success |
| 374 | 38 | 4 | 2 | O | O | Success |
| 375 | 38 | 5 | 2 | O | O | Success |
| 376 | 38 | 6 | 2 | G | O | Fail |
| 377 | 38 | 7 | 2 | O | O | Success |
| 378 | 38 | 8 | 2 | O | O | Success |
| 379 | 38 | 9 | 2 | O | O | Success |
| 380 | 38 | 10 | 2 | G | G | Success |
| 381 | 39 | 1 | 2 | G | G | Success |
| 382 | 39 | 2 | 2 | G | G | Success |
| 383 | 39 | 3 | 2 | G | G | Success |
| 384 | 39 | 4 | 2 | O | O | Success |
| 385 | 39 | 5 | 2 | O | G | Fail |
| 386 | 39 | 6 | 2 | G | G | Success |
| 387 | 39 | 7 | 2 | O | O | Success |
| 388 | 39 | 8 | 2 | O | O | Success |
| 389 | 39 | 9 | 2 | O | G | Fail |

| Point | Participant | Infographic | Group | Type | Guess | Success |
|-------|-------------|-------------|-------|------|-------|---------|
| 390 | 39 | 10 | 2 | G | G | Success |
| 391 | 40 | 1 | 2 | G | G | Success |
| 392 | 40 | 2 | 2 | G | G | Success |
| 393 | 40 | 3 | 2 | G | G | Success |
| 394 | 40 | 4 | 2 | O | G | Fail |
| 395 | 40 | 5 | 2 | O | G | Fail |
| 396 | 40 | 6 | 2 | G | O | Fail |
| 397 | 40 | 7 | 2 | O | O | Success |
| 398 | 40 | 8 | 2 | O | G | Fail |
| 399 | 40 | 9 | 2 | O | O | Success |
| 400 | 40 | 10 | 2 | G | G | Success |